\documentclass[aps,pra,twocolumn,superscriptaddress,showpacs,10pt]{revtex4-1}
\usepackage{amsmath,amssymb,graphicx,color}
\usepackage{epstopdf}
\begin{document}

\title{Maximal coin-position entanglement and non-Hermitian skin effect in discrete-time quantum walks}

\author{Ding Cheng}
\affiliation{College of Physics and Electronic Science, Hubei Normal University, Huangshi 435002, China}

\author{Yi Li}
\affiliation{State Key Laboratory of Quantum Optics Technologies and Devices,
	Institute of Laser Spectroscopy, Shanxi University, Taiyuan, Shanxi 030006, China}

\author{Hao Zhao}
\altaffiliation{zhaohao@hbnu.edu.cn}
\affiliation{College of Physics and Electronic Science, Hubei Normal University, Huangshi 435002, China}

\author{Haijun Kang}
\affiliation{Science School, Qingdao University of Technology, Qingdao 266520, China}

\author{Cui Kong}
\affiliation{College of Physics and Electronic Science, Hubei Normal University, Huangshi 435002, China}
	
\author{Jiguo Wang}
\affiliation{College of Physics and Electronic Science, Hubei Normal University, Huangshi 435002, China}

\author{Feng Mei}
\affiliation{State Key Laboratory of Quantum Optics Technologies and Devices, Institute of Laser Spectroscopy, Shanxi University, Taiyuan, Shanxi 030006, China}
\affiliation{Collaborative Innovation Center of Extreme Optics, Shanxi University, Taiyuan, Shanxi 030006, China}

\author{Chuanjia Shan}
\affiliation{College of Physics and Electronic Science, Hubei Normal University, Huangshi 435002, China}
	
\author{Jibing Liu}
\affiliation{College of Physics and Electronic Science, Hubei Normal University, Huangshi 435002, China}

\begin{abstract}
 A distinctive feature of non-Hermitian systems is the skin effect, which has recently attracted widespread attention. Quantum walks provide a powerful platform for investigating the underlying mechanisms of the non-Hermitian skin effect. Additionally, hybrid entanglement generation in quantum walks is recognized as another crucial property. However, experimentally investigating the influence of skin effect on the evolution of entanglement dynamics in the non-Hermitian systems remains a challenge. In this paper, we present a flexible implementation of 20-step discrete-time quantum walks using an optimized time-multiplexed loop configuration. By optimizing the coin parameter, we achieve maximal coin-position entanglement after 20-step quantum walks. Moreover, we experimentally measure the polarization-averaged growth rates and the evolution of coin-position entanglement under specific coin and loss parameters. We observe the asymmetric Lyapunov exponent profiles and the entanglement suppression induced by the non-Hermitian skin effect. Interestingly, this entanglement suppression weakens with increasing coin parameters and enhances with increasing loss parameters and evolution steps. Our results demonstrate the potential of quantum walks as a powerful platform for investigating hybrid entanglement properties and skin effects in non-Hermitian systems.
\end{abstract}

\maketitle
\section{Introduction}
Quantum walks (QWs) extend classical random walk processes into the quantum realm through coherent superposition of position states. In QWs, the quantum walker's wave function undergoes coherent spreading across lattice sites under the unitary evolution, leading to interference between multiple propagation paths. These transport mechanisms exhibit ballistic spreading characteristics and result in a faster spread of the walkers' positions compared to the classical case\cite{pearson1905problem,aharonov1993quantum}. The unique space-time evolution has significant advantages in the general-purpose quantum computers\cite{childs2009universal,childs2013universal}, quantum simulators\cite{rudner2009topological,weidemann2022topological}, and search algorithms\cite{shenvi2003quantum,potovcek2009optimized,qu2022deterministic}. \par

In the discrete-time QWs (DTQWs), this typically refers to a coin-based quantum walk\cite{qiang2024quantum}, namely flipping a coin and moving in the direction determined by the result of the coin flip. The walker's coin and position degrees of freedom can be entangled with each other\cite{carneiro2005entanglement,abal2006quantum}. The entanglement here differs from its original definition between multiple parties, which is defined among different degrees of freedom within a single particle and called hybrid coin-position entanglement\cite{tao2021experimental,zhang2022maximal}. The entanglement fluctuates at each step, gradually stabilizing to an asymptotic value. This stabilization depends on factors such as the coin operation, the shift operation, the initial state, and on-site loss. The resulting entanglement is typically not a maximally entangled state. Thus, the generation of the maximal coin-position entanglement and its related research has attracted widespread attention\cite{wang2018dynamic,tao2021experimental,naves2022enhancing,zhang2022maximal,fang2023maximal}. 
Moreover, DTQWs also provide a platform for studying the fundamental mechanisms of non-Hermitian physics, including exceptional points\cite{xiao2021observation}, topological phase transitions\cite{xiao2017observation,wang2019simulating,lin2022topological}, and non-Hermitian skin effects (NHSE)\cite{longhi2019probing,xiao2020non,lin2022observation,lin2022topological}. The NHSE refers to the phenomenon in which bulk eigenstates of a non-Hermitian system are exponentially localized at its boundaries. In DTQWs, the polarization-dependent on-site loss parameters typically induce an asymmetric hopping amplitude as a source of non-Hermiticity\cite{lin2022observation,kawabata2023entanglement}. The loss parameters and the effective coupling between the coin and position states can realize NHSE experimentally. This effect has led to exciting applications, including light funneling\cite{weidemann2020topological}, topological sensors\cite{mcdonald2020exponentially}, and topological amplification\cite{wanjura2020topological}. In addition, non-Hermitian theories offer a profound understanding of the dynamics of quantum correlations and entanglement among particles in open quantum systems\cite{wanjura2020topological,chen2022quantum}. The suppression of entanglement and reduction of von Neumann entropy induced by NHSE have been explored with theoretical studies in open condensed matter\cite{kawabata2023entanglement} and experimentally confirmed in the non-Hermitian photonic lattices system\cite{gao2024quantum}. However, experimental investigations into the influence of NHSE on entanglement dynamics in non-Hermitian DTQW systems remain challenging.\par

To date, DTQWs have been realized in a variety of physical systems, including superconducting qubits\cite{flurin2017observing,ramasesh2017direct}, nuclear magnetic resonance\cite{ryan2005experimental}, trapped atoms \cite{karski2009quantum}, trapped ions\cite{schmitz2009quantum,zahringer2010realization}, integrated photonic circuits\cite{sansoni2012two,crespi2013anderson}, or optical systems\cite{goyal2013implementing,giordani2019experimental,schreiber2010photons,schreiber2011decoherence,schreiber20122d,wang2018dynamic,tao2021experimental,zhang2022maximal,lin2023manipulating,xue2015experimental,xiao2021observation,zhan2017detecting,xiao2017observation,wang2019simulating,lin2022topological,xiao2020non,lin2022observation}. Photons are an excellent carrier of quantum state due to their low transmission loss in free-space and optical fiber channels. Therefore, optical systems have made remarkable progress in realizing DTQWs with a variety of well-established technologies, including orbital angular momentum (OAM) \cite{goyal2013implementing,giordani2019experimental}, time multiplexing \cite{lin2022observation,schreiber2010photons,schreiber2011decoherence,schreiber20122d,wang2018dynamic,tao2021experimental,zhang2022maximal,lin2022observation,lin2023manipulating}, and spatial displacers \cite{xue2015experimental,xiao2021observation,zhan2017detecting,xiao2017observation,wang2019simulating,xiao2020non,lin2022topological}. Optical systems that use spatial displacers or OAM may suffer from larger size, limited scalability, and stability issues, making it quite challenging to study the NHSE's effect on long-term coin-position entanglement. Time-multiplexed optical systems can implement a time-bin encoded quantum state by mapping the position state onto the time domain. This configuration is scalable in terms of the number of accessible steps and positions in Hilbert space, offering more flexibility in manipulating the walker's internal degree of freedom. Introducing dynamical disorder into the coin or shift operations in time-multiplexed optical systems can significantly enhance the coin-position entanglement and drive it towards maximal entanglement regardless of initial states\cite{wang2018dynamic,tao2021experimental,zhang2022maximal}. Nevertheless, the dynamical disorder inevitably induce Anderson localization, which fundamentally differs from NHSE. Alternatively, maximal entanglement can also be obtained with a fixed initial state and specific coin operation in time-multiplexed optical systems \cite{carneiro2005entanglement,abal2006quantum}. The previous experimental studies, such as Lin et al.  \cite{lin2022observation}, implemented a nonunitary DTQW using a time-multiplexed optical loop and observed up to 10 steps of the quantum walk. Their results demonstrated signatures of the NHSE through asymmetric profiles of the polarization-averaged growth rate. However, there have been no experimental reports on how the NHSE influences the dynamics of hybrid maximally entangled states in non-Hermitian DTQW systems.\par

In this paper, we demonstrate the generation of maximal coin-position entanglement over 20 steps and the entanglement suppression induced by the NHSE in a photonic DTQW system. In particular, various DTQW scenarios are investigated by changing the coin parameters, loss parameters, and initial states. The Lyapunov exponent and von Neumann entropy under the different scenarios were numerically analyzed. Meanwhile, we numerically and experimentally presented the polarization-resolved photon distributions and polarization-averaged growth rates after the 20-step DTQWs. Then, we chose specific coin parameters and an asymmetric initial state to observe the evolution of entanglement and delocalization as the number of walking steps increases. The results show that the entanglement can be enhanced by choosing appropriate coin parameters and initial states. In addition, we provide the first experimental observation of the entanglement suppression phenomenon induced by NHSE during the evolution of hybrid entanglement in DTQWs. Finally, we demonstrate that the suppression weakens with increasing coin parameters and enhances with increasing loss parameters and evolution steps. The presented results provide an efficient way to study the entanglement evolution and non-Hermitian properties in DTQWs systems.
\par

\section{Theoretical analysis}
In a one-dimensional photonic DTQW, a photon (walker) moves left or right along a straight line based on a coin operation that manipulates its polarization\cite{wang2018dynamic}. The internal degree of freedom of the photon is represented by its polarization state, where the horizontal polarization $|H \rangle=\left ( 1,0\right ) ^{T} $ and vertical polarization $|V \rangle=\left ( 0,1\right ) ^{T} $. The walker performs the spatial shift based on its internal state, and the resulting position is represented by the integer value $x\in Z$. Thus, the coin-based QW consists of two separate spaces: the coin space
$H_{C}$ and the position space $H_{S}$. $H_{C}$  is a
two-dimensional Hilbert space spanned by $|H \rangle $ and $|V \rangle$. $H_{S}$ is an infinite dimensional Hilbert space spanned by orthogonal vectors $\left |x  \right \rangle$, where $x$ corresponds to the possible positions of the walker.
\par

The evolution operator of the DTQW can be defined by  
 \begin{equation}
	\hat{U}=\hat{S} \left [ \hat{I}_{p} \otimes \hat{L}\left ( \gamma  \right ) \hat{C} \left (\theta  \right )\right ] ,
\end{equation} 
where $\hat{I}_{p} $ is the identity operator in space $H_{S}$. $ \hat{C} \left (\theta  \right )$ is the quantum coin operator in space $H_{C}$, which manipulates the polarization state of a photon via a half-wave plate (HWP). The matrix representation of  $ \hat{C} \left (\theta  \right )$ is given by
\begin{equation}
	\hat{C} \left ( \theta  \right ) =\left ( \begin{matrix}
		\cos \theta & \sin \theta \\
		\sin \theta  &- \cos \theta
	\end{matrix} \right ) ,
\end{equation} 
where  $\theta $ is the rotation angle of the HWP relative to one of its optical axes. When the coin parameter $\theta $ is varied, it generates different superposition states of the polarization states $|H \rangle$ and  $|V \rangle$. Specifically, when $\theta=\pi/4 $, the quantum walk becomes a Hadamard quantum walk, where the states $|H \rangle$ and $|V \rangle$ are in an equal superposition. \par

In this one-dimensional photonic QW, polarization-dependent losses are described by a non-unitary loss operation $\hat{L}\left ( \gamma  \right ) $, with its matrix representation given by
\begin{equation}
	\hat{L} \left ( \gamma  \right ) = {\textstyle \sum_{x}^{}}|x \rangle\langle x|\otimes \left (|H \rangle\langle H|+e^{-\gamma } |V \rangle\langle V| \right ),
\end{equation} 
where $\gamma $ is the loss parameter that indicates loss imbalance between the H-polarized and V-polarized photon. By changing the parameters $\gamma $ and  $\theta $ in a controlled way, we can generate a variety of DTQW scenarios. The shift operation $\hat{S} $ moves the photon's position $x$ to $x+1$ if it is in the $|H \rangle $ state, or to $x-1$ if it is in the $|V \rangle $ state, which can be expressed as
\begin{equation}
	\hat{S}= {\textstyle \sum_{x}^{}}|x-1 \rangle\langle x|\otimes |V \rangle\langle V|+|x+1 \rangle\langle x|\otimes |H \rangle\langle H|.
\end{equation} \par

The operator $\hat{S}$ is implemented within the optical feedback loop, where horizontally polarized photons travel along a longer path, while vertically polarized photons travel along a shorter path in time-multiplexed optical systems. Photons from these two paths are coherently recombined at the output and fed back into the optical feedback loop for the next shift operation.\par

We consider that the initial state is a localized state at the original position, expressed as
\begin{equation}
	|\psi_{0} \left ( 0 \right )  \rangle=|0 \rangle \otimes \left [  a_{0}\left ( 0 \right )  \right|H \rangle+
	b_{0}\left ( 0 \right )|V \rangle ],
\end{equation} 
with the complex coefficients satisfying $ \left | a_{0}\left ( 0 \right ) \right | ^{2} +\left | b_{0}\left ( 0 \right ) \right | ^{2}=1 $. The photon’s wave function at the $t$-step evolution can be expressed as  $|\psi_{x} \left ( t \right )  \rangle=\hat{U}^{t}|\psi_{0} \left ( 0 \right )  \rangle$. The system is a discrete-time simulation of non-unitary evolution driven by a non-Hermitian effective Hamiltonian $\hat{H}_{\mathsf{eff}}=i\ln_{}{\hat{U}}$. Thus, the normalized photon’s wave function after $t$-steps evolves as\cite{longhi2023phase,xiao2020non,kawabata2023entanglement,lin2022observation}:
\begin{equation}
	\begin{split}|\psi_{x} \left ( t \right )  \rangle &=\frac{e^{{-i\hat{H}_{\mathsf{eff} }t} }|\psi_{0} \left ( 0 \right )  \rangle}{\left \| e^{{-i\hat{H}_{\mathsf{eff} }t} }|\psi_{0} \left ( 0 \right )  \rangle \right \| } \\&= {\textstyle \sum_{x}^{}}|x \rangle \otimes \left [  a_{x}\left ( t \right )  \right|H \rangle+
		b_{x}\left ( t \right )|V \rangle ]
	\end{split},
\end{equation} 
with $x=-t,-t+2,\dots ,t-2,t$. The complex coefficients satisfy $ {\textstyle \sum_{x}^{}}\left | a_{x}\left ( t \right ) \right | ^{2} +\left | b_{x}\left ( t \right ) \right | ^{2}=1 $ due to the normalized condition.\par 

The properties of $\hat{U}$ manifest in the eigenstates and energy spectrum of $\hat{H}_{\mathsf{eff} } $. The quantum walk is governed by $\hat{U}$, which is influenced by the interplay between the effective coupling of polarization and position states and the polarization-dependent loss parameter. This interplay will give rise to the accumulation of eigenstates at the boundaries, a phenomenon known as the NHSE\cite{xiao2020non,lin2022observation,yao2018edge,yokomizo2019non,imura2019generalized,li2025observation}. A key consequence of the NHSE is that the bulk bands of the system exhibit significant differences between open boundary conditions (OBC) and periodic boundary conditions (PBC)\cite{longhi2019probing,hatano1996localization}. When the NHSE is present, the energy spectrum under PBC forms one or more closed loops enclosing a non-vanishing area. In contrast, under OBC, the energy spectrum consists of a set of open arcs located within the interior of the PBC loci. In the absence of the NHSE, the energy spectrum under both PBC and OBC coincide on a straight line along the real axis, and the distinction between the two boundary conditions vanishes. \par

The NHSE also influences bulk dynamics, leaving distinctive signatures in the Lyapunov exponent\cite{longhi2019probing,lin2022observation,gao2024quantum}. The Lyapunov exponent in bulk wave dynamics can generally reveal the presence of the non-Hermitian skin effect, which is defined as
\begin{equation}
	\lambda \left ( v \right ) =\lim_{t \to \infty} \frac{\log_{}{\left |\psi_{x=vt} \left ( t \right )   \right | } }{t}, 
\end{equation} 
where $v $ is the shift velocity. When the shift velocity $v $ is set to 0, $\lambda\left ( 0  \right )$ converges to an asymptotic value, known as the Lyapunov exponent $\lambda$, as $t\to \infty $. The Lyapunov exponent $\lambda$ describes the asymptotic growth rate of light intensity at $x=0$, reflecting the shifting behavior of the photon’s wave function as it propagates through the lattice \cite{longhi2019probing,gao2024quantum}. A nonzero Lyapunov exponent $\lambda$ indicates that the wave function exhibits unidirectional diffusion induced by NHSE in the photonic lattice. In contrast, when the Lyapunov exponent $\lambda$ is zero, the wave packet exhibits unitary diffusion in the photonic lattice. In addition, the location of the peak of the Lyapunov exponent $\lambda \left ( v  \right )$ also shows how the wave function propagates along the lattice \cite{longhi2019probing,lin2022observation}. Specifically, in the presence of NHSE, the peak of $\lambda \left ( v  \right )$ appears at finite shift velocity $v $, exhibiting an asymmetric profile around $v =0 $. While in the absence of NHSE, the peak of $\lambda \left ( v  \right )$ remains at $v =0 $ and the profile is symmetric. \par

The coherent action of  $\hat{S} $ and $ \hat{C}$ leads to coin-position entanglement of the walker, which can be quantified by the von Neumann entropy\cite{wang2018dynamic,zhang2022maximal}. The von Neumann entropy is defined as $S_{\mathrm{ E}} \left ( \rho _{c}  \right ) =-\mathsf{Tr}\left (\rho _{c}\log_{2}{\rho _{c}}   \right )  $, where $\rho _{c} =\mathsf{Tr_{p}} \left [ |\psi_{x} \left ( t \right )  \rangle\langle \psi_{x} \left ( t \right )  | \right ] $ is the reduced density matrix of coin state obtained by the partial trace over position. Using Eq.(6), we obtain the reduced density matrix $\rho _{c}$ as
\begin{equation}
	\rho _{c}  =\left ( \begin{matrix}
		\alpha \left ( t \right ) &\chi  \left ( t \right )  \\
		\chi ^{\ast }  \left ( t \right ) &\beta \left ( t \right ) 
	\end{matrix} \right ) ,
\end{equation}
with $\alpha \left ( t \right ) = {\textstyle \sum_{x}^{}} \left | a_{x} \left ( t \right )  \right |^{2} $, $\beta  \left ( t \right ) = {\textstyle \sum_{x}^{}} \left | b_{x} \left ( t \right )  \right |^{2} $, $\chi  \left ( t \right ) = {\textstyle \sum_{x}^{}} a_{x} \left ( t \right ) b_{x}^{\ast }\left ( t \right ) $. The evolution operator $\hat{U} $ is purely real in the $\left \{ \left | H \right \rangle , \left | V \right \rangle \right \}$ basis. When the initial state is $ \left | 0\right \rangle \otimes \left | H \right \rangle $ or $ \left | 0\right \rangle \otimes \left | V \right \rangle $, $ a_{x}(t)$ and $b_{x}(t)$ both are also real, which can simplifies the off-diagonal term to $ \chi  \left ( t \right ) = {\textstyle \sum_{x}^{}} a_{x} \left ( t \right ) b_{x}\left ( t \right ) $. Then the von Neumann entropy $S_{\mathrm{ E}} \left ( \rho _{c}  \right )$ is calculated by 
\begin{equation}
	S_{\mathrm{ E}} \left ( \rho _{c}  \right )=-\lambda _{1} \log_{2}{\lambda _{1} } -\lambda _{2} \log_{2}{\lambda _{2} } ,
\end{equation}
where $\lambda _{1}$ and $\lambda _{2}$ are the eigenvalues of matrix  $\rho _{c}$  expressed as
\begin{equation}
	\lambda _{1,2} =\frac{1\pm\sqrt{1-4\left [ \alpha \left ( t \right ) \beta \left ( t \right ) -\left | \chi \left ( t \right )  \right | ^{2}  \right ] }  }{2} .
\end{equation}\par

\begin{figure*}[ht!]
	\begin{minipage}{0.33\textwidth}
		\centering
		\includegraphics[width=\linewidth]{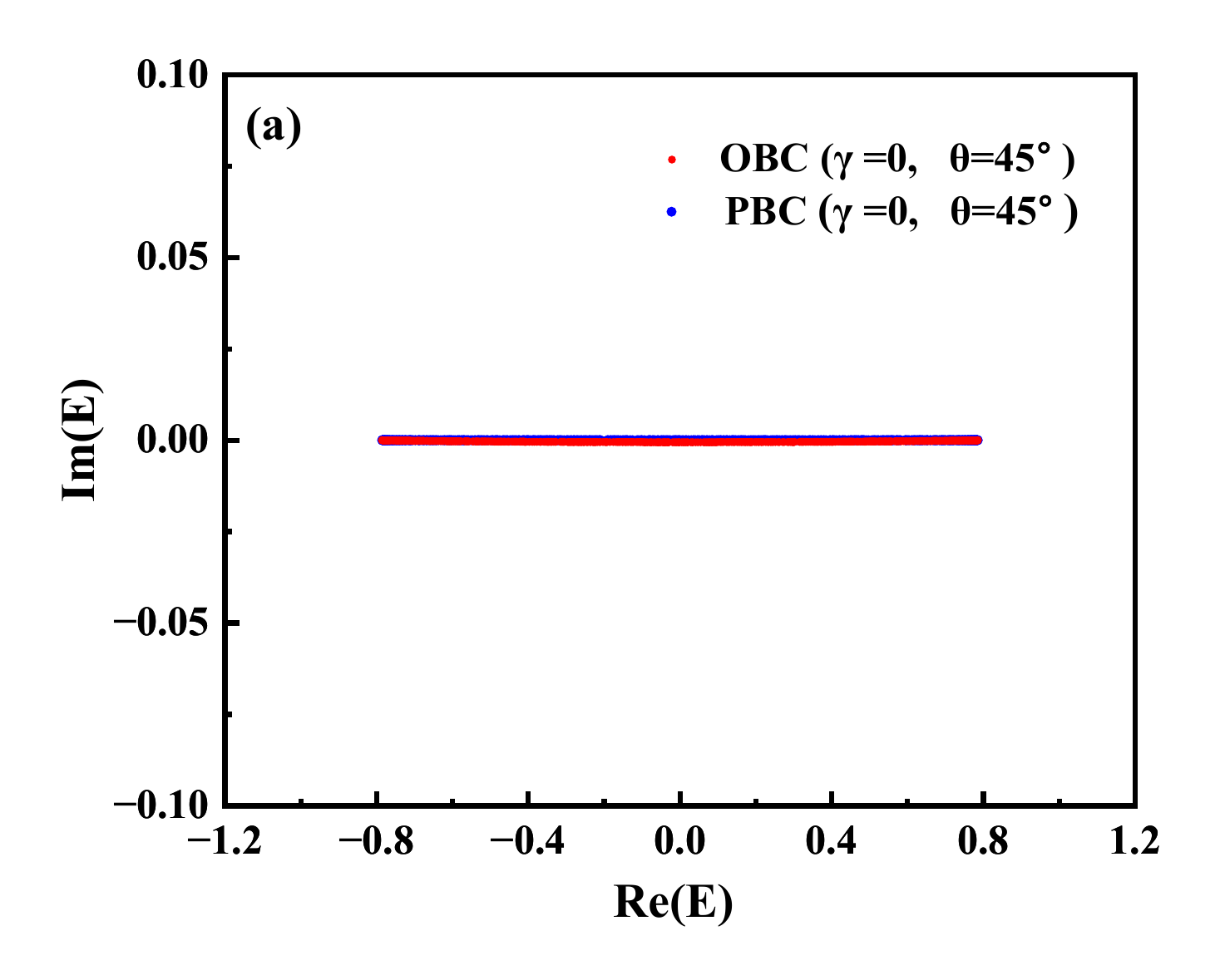}
	\end{minipage}%
	\begin{minipage}{0.33\textwidth}
		\centering
		\includegraphics[width=\linewidth]{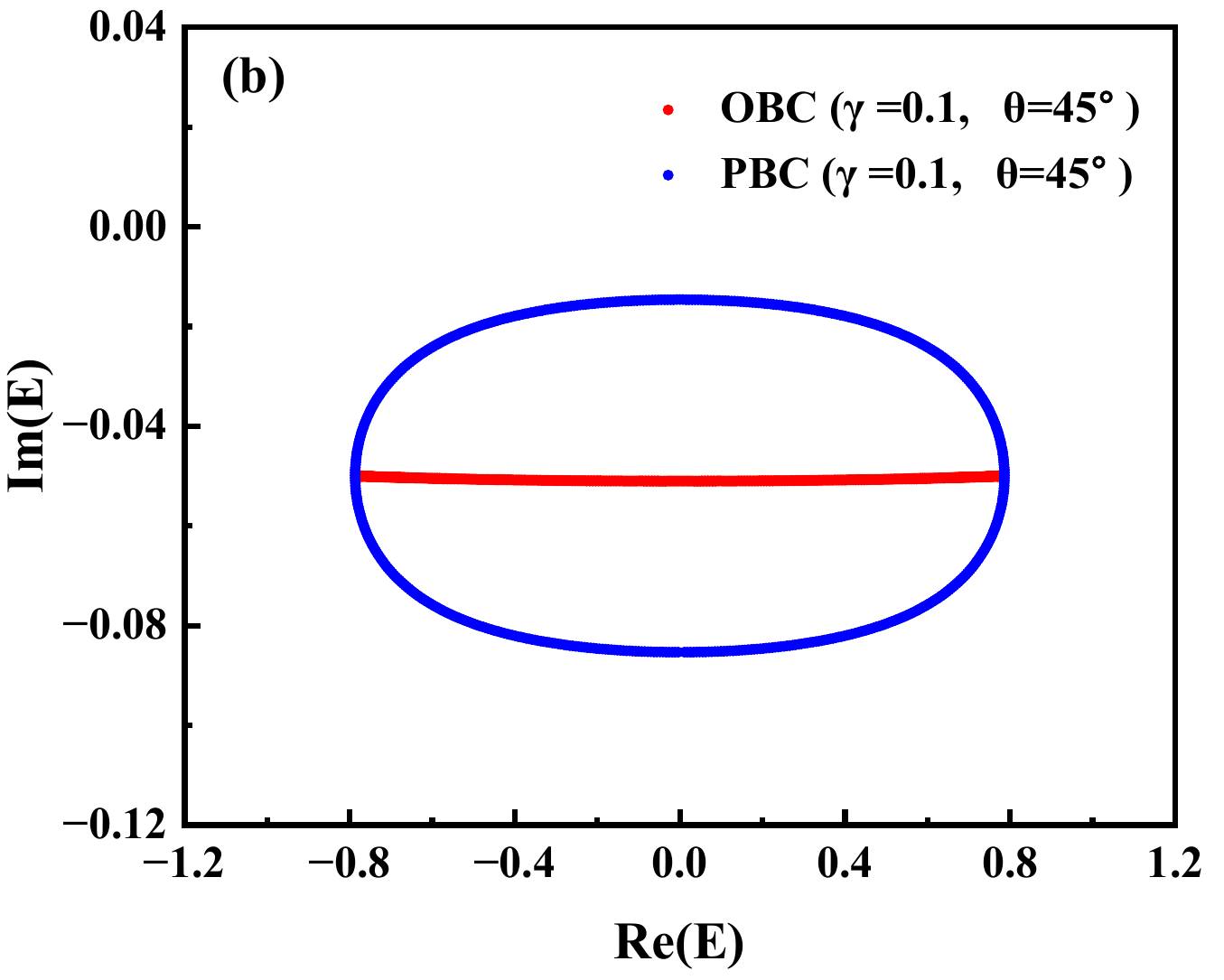}
	\end{minipage}%
		\begin{minipage}{0.34\textwidth}
		\centering
		\includegraphics[width=\linewidth]{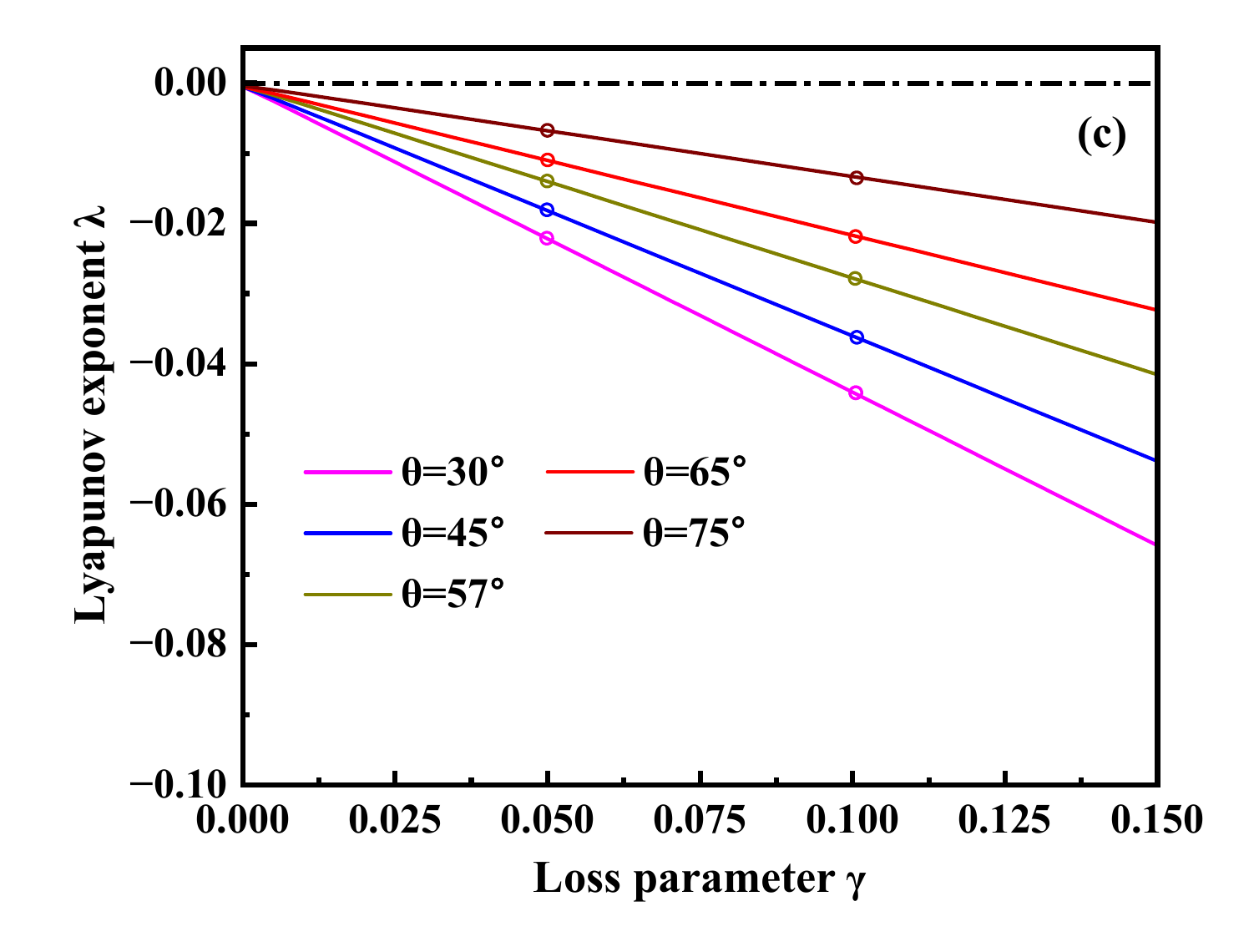}
	\end{minipage}%
	
	\begin{minipage}{0.34\textwidth}
		\centering
		\includegraphics[width=\linewidth]{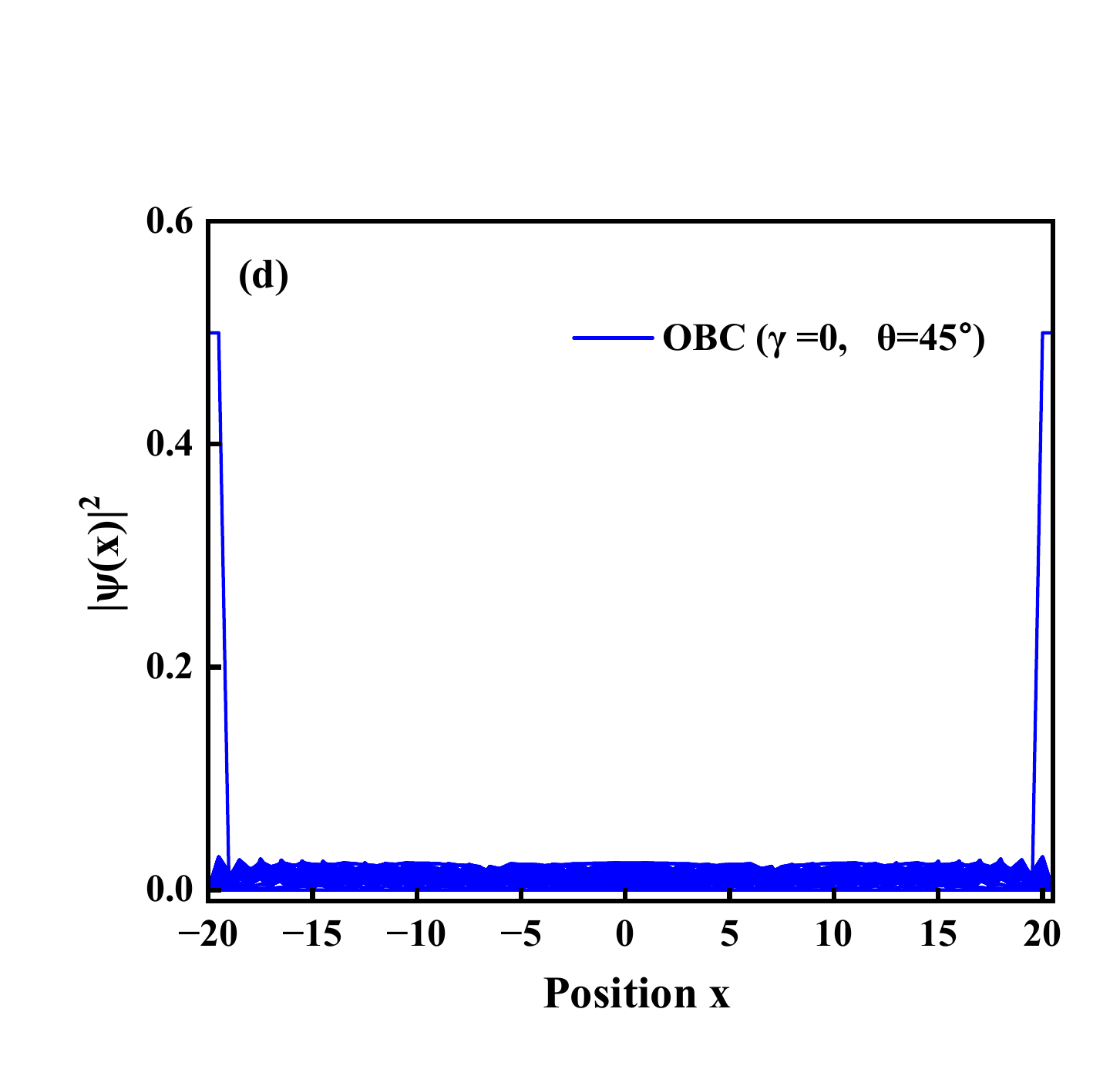}
	\end{minipage}
		\begin{minipage}{0.32\textwidth}
		\centering
		\includegraphics[width=\linewidth]{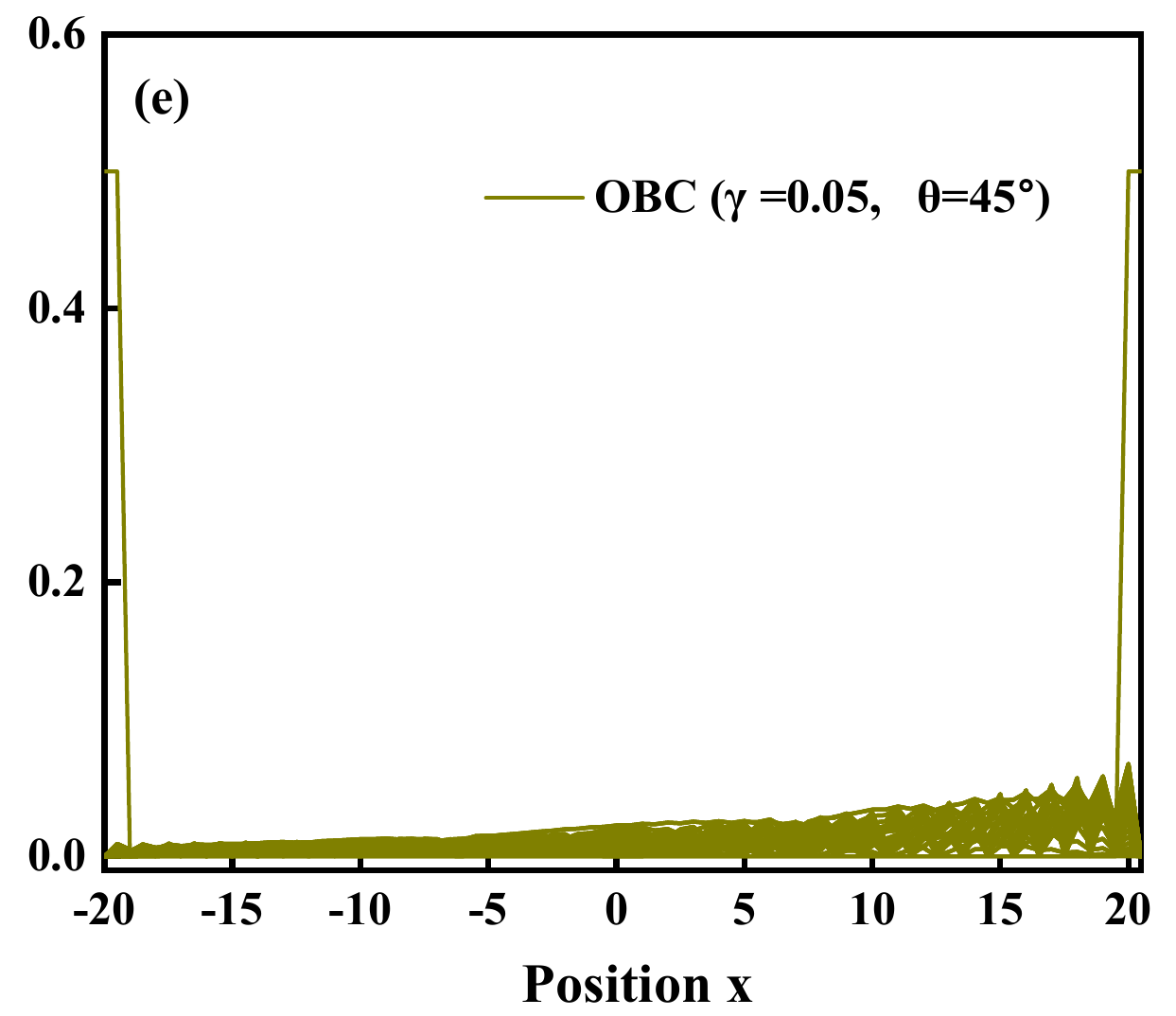}
	\end{minipage}
			\begin{minipage}{0.32\textwidth}
		\centering
		\includegraphics[width=\linewidth]{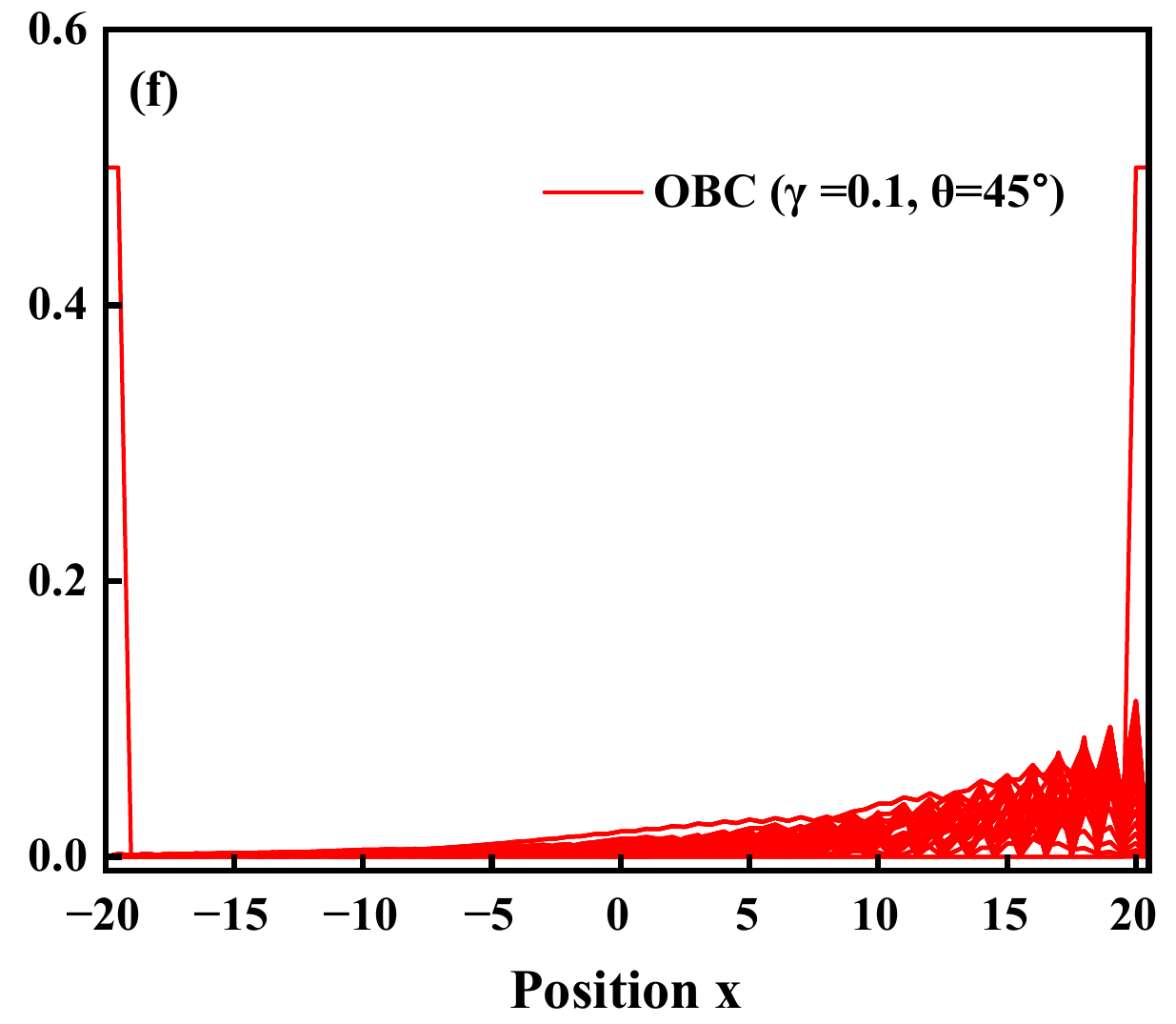}
	\end{minipage}
	
	\begin{minipage}{1\textwidth}
		\centering
		\caption{(a) The eigen spectra of the Hermitian DTQW for parameter values $\gamma=0$, $\theta=45^{\circ}$ under PBC and OBC. (b) The eigen spectra of the non-Hermitian DTQW for parameter values $\gamma=0.1$, $\theta=45^{\circ}$ under PBC and OBC. (c) The Lyapunov exponent $\lambda$ in the system as a function of the loss parameter $\gamma$ for different coin operation parameters $\theta$. (d)-(f) The spatial distribution of eigen wave functions in the DTQW for different loss parameters $\gamma$ at the coin operation parameter $\theta=45^{\circ}$.}
	\end{minipage}
\end{figure*}
\par
\begin{figure}[ht!]
	\centering\includegraphics[width=\linewidth]{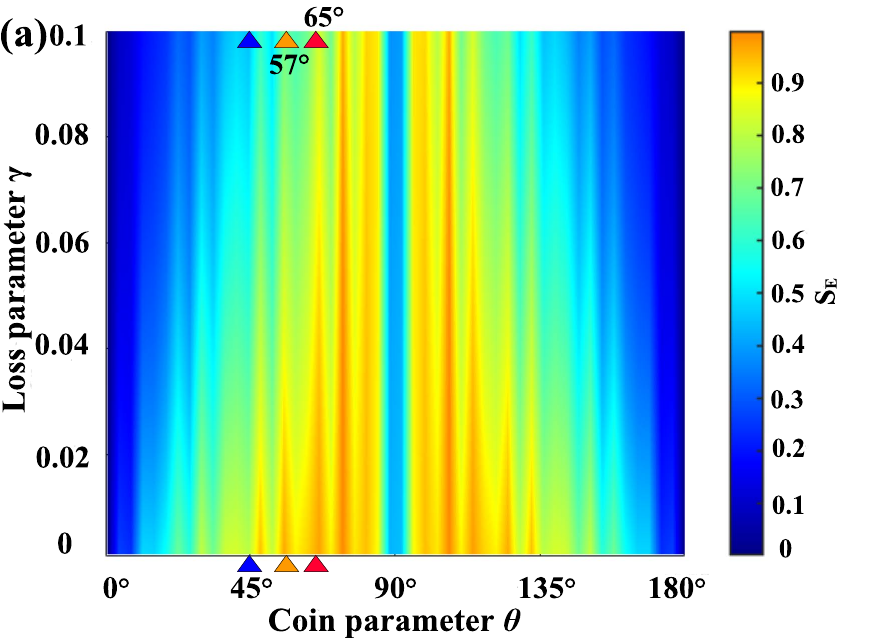}
	\centering\includegraphics[width=\linewidth]{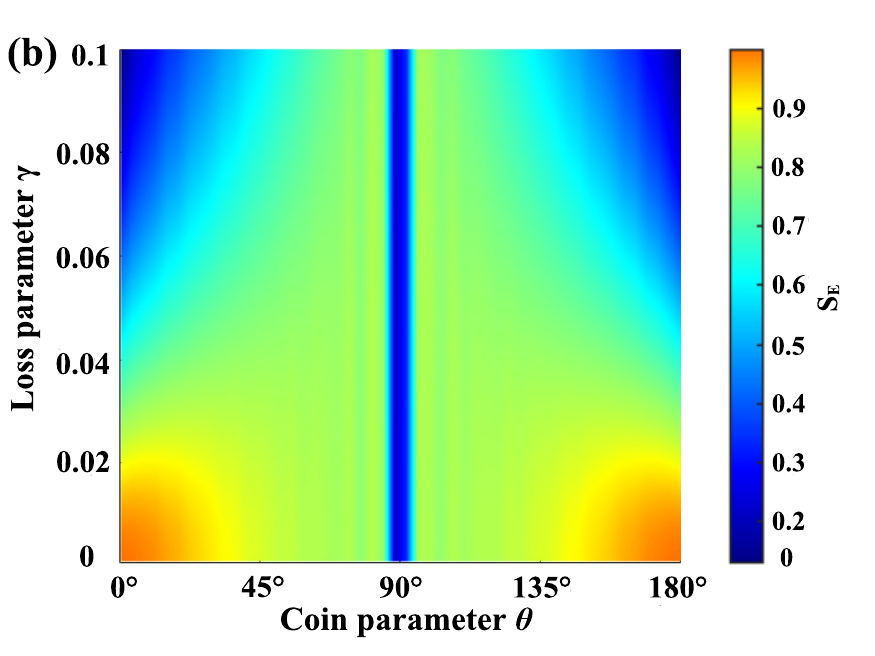}
	\centering
	\caption{The von Neumann entropy $S_{\mathrm{ E}}$ of the coin-position entanglement as a function of the coin operation parameter $\theta $ and the loss parameter $\gamma$ for the 20-step quantum walk, considering different initial states. (a) The initial state is $|\psi_{0} \left ( 0 \right )  \rangle=|0 \rangle \otimes|H \rangle  $, where $a_{0}\left ( 0 \right )=1$ and $b_{0}\left ( 0 \right )=0$. (b) The initial state is $|\psi_{0} \left ( 0 \right )  \rangle= |0 \rangle \otimes \left ( |H \rangle+i|V \rangle \right )  /\sqrt{2}$, where $a_{0}\left ( 0 \right )=1/\sqrt{2}$ and $b_{0}\left ( 0 \right )=i/\sqrt{2}$.}
\end{figure}\par

The coin-position state is separable when $S_{\mathrm{ E}} = 0$ and maximally entangled when $S_{\mathrm{ E}} = 1$. In addition to studying the entanglement properties of the walker, the inverse participation ratio ($\mathrm{IPR}$) can be used to quantify its localization (delocalization) properties\cite{evers2000fluctuations,buarque2019aperiodic,lin2022topological}. It can be expressed as $\mathrm{IPR}\left ( t \right ) ={\textstyle \sum_{x}\left [P \left ( t,x \right )   \right ]^{2}  } $, where $ P \left ( t,x \right ) =\left | a_{x}\left ( t \right )   \right | ^{2} +\left | b_{x}\left ( t \right )   \right | ^{2}$ represents the probability of finding the walker at site $x$ at time $t$. When the photon's wave function $|\psi_{x} \left ( t \right )  \rangle$ is highly concentrated at a few positions $x$, $\mathrm{IPR}\left ( t \right )$ is close to 1, indicating a localized state. Conversely, when $|\psi_{x} \left ( t \right )  \rangle$ is uniformly distributed over an $N$-site lattice, $\mathrm{IPR}\left ( t \right )$ is approximately $1/N$, which corresponds to a delocalized state. Although the NHSE significantly influences the walker's localization properties, we primarily focus here on its impact on entanglement. \par

Fig. 1(a) and Fig. 1(b) show the numerically calculated eigen spectra for $\theta$=45° in the Hermitian($\gamma=0$) and non-Hermitian($\gamma=0.1$) system, respectively. The red dots represent the eigenvalues under OBC, while the blue dots represent the eigenvalues under PBC. When the loss parameter $\gamma$ is set to 0.1, the energy spectrum for PBC forms a closed loop and encloses a non-vanishing area, whereas the energy spectrum for OBC forms a set of open arcs, indicating the presence of the NHSE. In contrast, in the lossless case ($\gamma = 0$), the energy spectrum under both PBC and OBC coincide on a straight line along the real axis, indicating the absence of the NHSE.\par

Fig. 1(c) shows the numerically calculated Lyapunov exponent $\lambda$ of the system as a function of the loss parameter $\gamma$ at different values of the coin operation parameter $\theta$, using Eq. (7). Here, the coin operation parameter $\theta$ varies continuously within the range ($0$, $\pi/2$). For convenience, we consider five specific values of $\theta$ =$30^{\circ}$(magenta line), $45^{\circ}$(blue line), $57^{\circ}$(green line), $65^{\circ}$(red line), and $75^{\circ}$(wine line). It can be observed that when $\gamma=0$, the Lyapunov exponent is equal to zero for all coin parameters, implying that the wave packet exhibits unitary diffusion and no skin effects occur in the system. However, when $\gamma>0$, the Lyapunov exponent becomes non-zero, implying that the wave packet exhibits unidirectional diffusion and the skin effects occur in the system. Notably, the amplitude of the Lyapunov exponent $\lambda$ gradually increases as the loss parameter $\gamma$ increases. Additionally, the amplitude of the Lyapunov exponent $\lambda$ decreases as the coin parameter $\theta$ increases. This suggests that as the coin parameters increase further, the shift velocity of the wave function towards the boundary decreases.\par
   
Fig. 1(d)-(f) shows the numerically calculated eigenstates of the Hermitian system for $\gamma=0$, $\theta$=45°, and the non-Hermitian system for $\gamma=0.05 $ and $0.1$, $\theta$=45°. The blue solid line represents the eigenwave functions of the Hermitian systems under OBC, while the yellow and red solid lines correspond to those of the non-Hermitian system under OBC. It can be observed that the eigenstates of the Hermitian system are extended over the position space. When the loss parameter $\gamma$ increases to 0.05, the eigenstates gradually localize at the boundaries, which is a characteristic feature of the NHSE. As $\gamma$ increases to 0.1, the eigenstates localize more rapidly at the boundaries, exhibiting a more pronounced manifestation of the NHSE.\par

For a fixed initial state, as the number of walk steps increases, the coin-position entanglement approaches an asymptotically stable value. This asymptotic value generally cannot reach its maximum in the Hadamard QW\cite{carneiro2005entanglement,abal2006quantum}. However, this situation can be significantly altered in non-Hadamard DTQW systems. Moreover, we know that the NHSE is not dependent on the specifics of the initial state \cite{kawabata2023entanglement}. Nevertheless, the coin-position entanglement in non-Hermitian DTQW systems exhibits significantly different behaviors for different initial states. Fig. 2(a) shows numerically calculated von Neumann entropy $S_{\mathrm{ E}}$ for the 20-step quantum walk as a function of the coin operation parameter $\theta$ and the loss parameter $\gamma$, with the initial state $|0 \rangle \otimes |H \rangle$, using Eq. (9). Here, we consider the coin operation parameter $\theta$ in the range (0°, 90°) and the loss parameter $\gamma$ in the interval (0, 0.1). When the coin parameters $\theta $ are chosen as 45°, 57°, and 65°, and the loss parameter $\gamma$ is set to 0, the von Neumann entropy for the 20-step quantum walk is 0.858, 0.982, and 0.996, respectively. Thus, in non-Hadamard DTQW systems, the von Neumann entropy can be optimized to approach its maximum value for a specific small range of coin operation parameters. Interestingly, the von Neumann entropy $S_{\mathrm{ E}}$ decreases as the loss parameter increases. As the coin operation parameters increase, the decrease in entanglement entropy slows down. This also indicates that within some specific ranges of coin parameters, the system’s entanglement can effectively resist the impact of the polarization-dependent loss.\par 

Fig. 2(b) shows the numerically calculated von Neumann entropy $S_{\mathrm{ E}}$ for the 20-step quantum walk as a function of the coin parameter $\theta$ and the loss parameter $\gamma$, with the initial state $|0 \rangle \otimes \left ( |H \rangle+i|V \rangle \right )  /\sqrt{2} $, using Eq. (9). As the coin parameter $\theta $ is gradually reduced from $45^{\circ }$ to $0^{\circ }$, the von Neumann entropy increases continuously from 0.875 to 1. This indicates that in non-Hadamard DTQW systems, the coin-position entanglement can be optimized to approach its maximum value when the coin operation parameter is close to $0^{\circ }$. Specifically, when $S_{\mathrm{ E}}$ exceeds 0.9, the loss parameter $\gamma$ and the coin parameter $\theta$ are approximately bounded by $0< \gamma< 0.02$ and $0<\theta <  34.7^{\circ}$. This behavior can be understood as follows: when $\gamma=0$ and $\theta = 0^\circ$, the coin and loss operator do not alter the polarization components, the walker's state is equivalent to a Bell-like state, representing maximal entanglement between the position and polarization degrees of freedom. In contrast, a nonzero loss parameter $\gamma$ introduces an amplitude imbalance between the $|H\rangle$ and $ |V\rangle$ components. As $\gamma$ increases, the attenuation of the $|V\rangle$ components gradually drives the system towards a separable state. Consequently, the coin-position entanglement entropy decreases monotonically with the loss parameter. Similarly, as $\theta$ increases, the coin operator progressively mixes the $|H\rangle$ and $|V\rangle$ components, resulting in the $|H\rangle$ and $ |V\rangle$ components partially overlapping at each position. This overlap reduces their distinguishability, thereby weakening the coin-position correlation and leading to a gradual decrease in $S_{\mathrm{E}}$. It can also be intuitively seen that $S_{\mathrm{ E}}$ decreases as the loss parameter increases, and this decrease slows down as the coin parameter increases. With the theoretical analysis established, we now turn to the experimental setup and the analysis of the results. \par

\section{Experimental demonstrations}

\begin{figure*}[ht!]
	\centering
	\includegraphics[width=12cm]{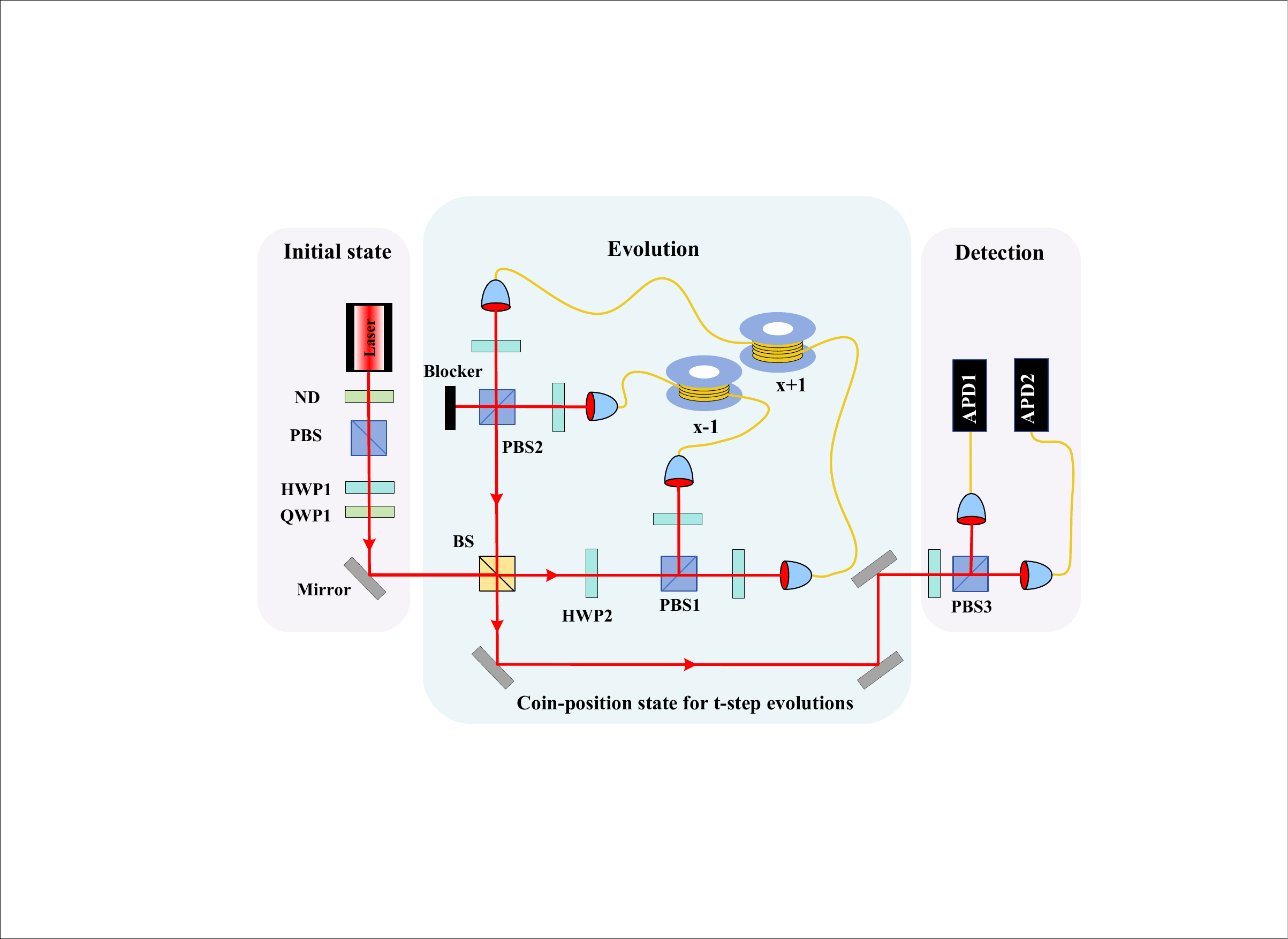} %
	\begin{minipage}{1\textwidth}
		\caption{Experimental setup of the one-dimensional photonic DTQW. HWP: half-wave plate; PBS: polarization beam splitter; QWP: quarter-wave plate; ND: neutral density filters; BS: 90/10 beam splitter; APD: avalanche photodiode; Setup dimensions: 1.5 m in free space and 31 m (30 m) in fiber.}
	\end{minipage}
\end{figure*}\par

The experimental setup of the one-dimensional photonic DTQWs is shown in Fig. 3. A pulsed laser provides a photon source with a pulse width of 88 ps, a wavelength of 805 nm, and a repetition rate of 125 kHz. Neutral density filters (ND) attenuate the pulsed laser to the single-photon level. The initial state is prepared at position $|x=0  \rangle $ using HWP1 and a quarter-wave plate (QWP1). The quantum coin operation is implemented by another HWP2 to tune the coin parameter $\theta $. The shift operation is implemented by the optical fiber feedback loop, which consists of two polarization beam splitters (PBS1 and PBS2) and two optical fibers of different lengths. When photons pass through the PBS1, H-polarized photons enter the long fiber loop with a transmission time of 155 ns, while V-polarized photons enter the short fiber loop with a transmission time of 150 ns, which is 5 ns shorter than the transmission time of H-polarized photons. The resulting temporal difference of 5 ns between both polarization components corresponds to a spatial displacement of one step (either $x+1$ or $x-1$). The two paths are coherently recombined by PBS2, and the photon is sent back to PBS1 for the next step, such that the walker's position is mapped to the time domain. \par

In our experiment, the system's overall losses are caused by the losses of various optical components in the experimental setup. The photons are initially split by a 90:10 beam splitter (BS) and coupled into the fiber loop with a probability of 0.1. Subsequently, the photon undergoes a complete round trip and is coupled into the fiber loop through the same BS with probability 0.9. Another 0.1 probability is transmitted for the coin-position state measurement. Taking into account the photon loss introduced by the BS, two PBS, long and short fibers, half-wave plates, and other optical components in the fiber loop, the overall probability that a photon successfully undergoes a complete round trip (without being lost or detected) is about 0.61 per step. The overall photon detection efficiency, with an avalanche photodiode (APD) efficiency of 0.6 and an optical fiber coupler efficiency of 0.83, is about 0.498. To control the loss parameter $\gamma$ accurately, it is necessary to balance the inherent losses of the two polarizations in the long and short fiber loops. By measuring the transmission count rate, it is found that the per-step loss for H-polarized photons is about $3\%$ lower than that for V-polarized photons. Without additional adjustment, this would result in an inherent imbalance between the two polarizations. To compensate for this difference, an additional loss channel for the H-polarized photons is introduced in the long-path fiber loop. Specifically, a pair of HWPs is inserted into the long and short fiber loops, respectively. The angle of the HWP in the long fiber loop is adjusted so that part of the H-polarized component is rotated and leaks out through PBS2, which is blocked by a blocker\cite{lin2022observation}. In this way, the per-step loss for both polarizations can be balanced, ensuring that the loss parameter $\gamma$ is effectively set to zero. Subsequently, the angle of the HWP in the short fiber loop can be adjusted so that part of the V-polarized photons is rotated and leaked out through PBS2, which is blocked by the blocker. This introduces a tunable polarization-dependent loss and achieves the desired value of the loss parameter $\gamma$. 
\par

\begin{figure*}[ht!]
	\begin{minipage}{0.34\textwidth}
		\centering\includegraphics[width=\linewidth]{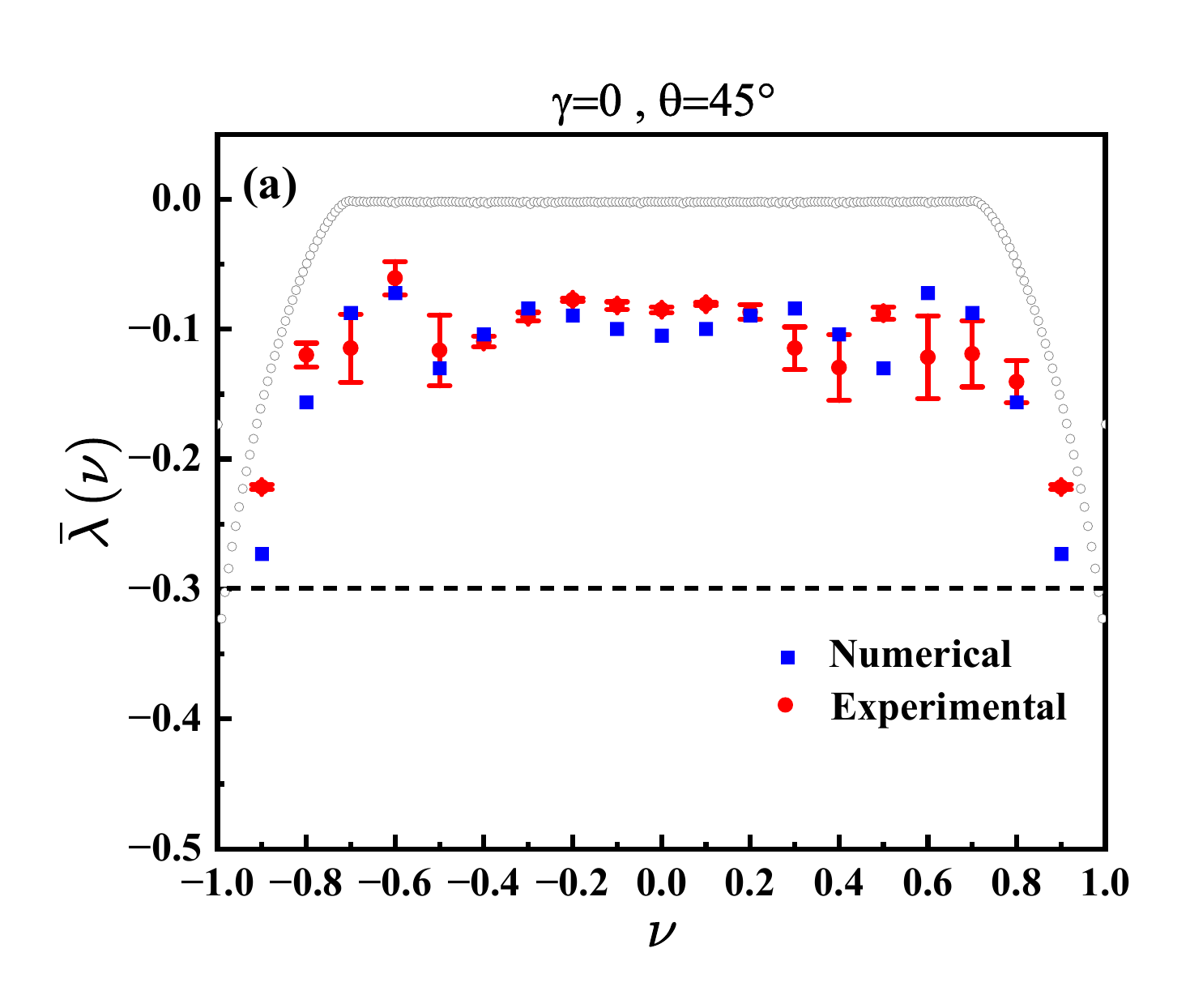}
	\end{minipage}%
	\begin{minipage}{0.315\textwidth}
		\centering\includegraphics[width=\linewidth]{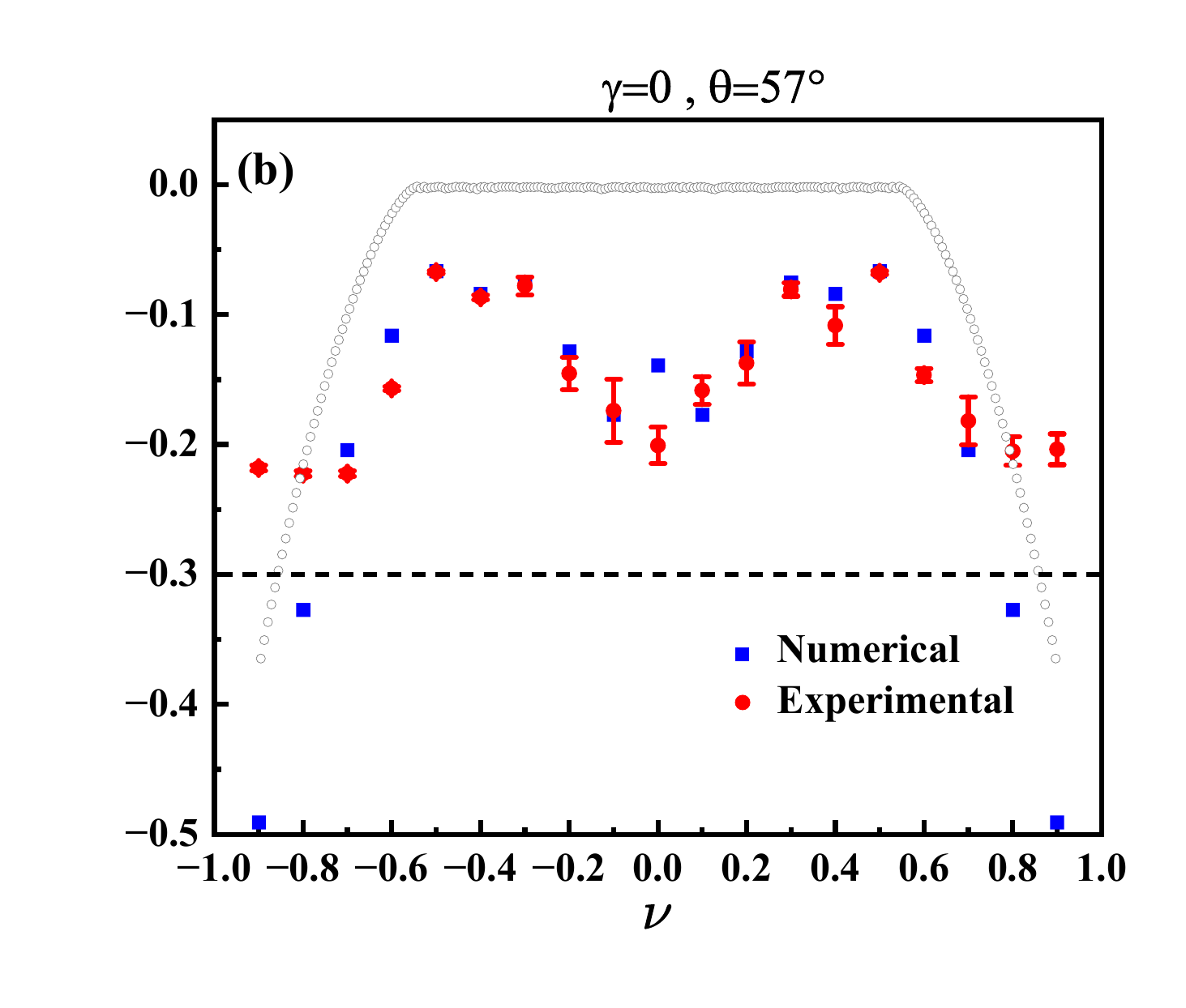}
	\end{minipage}%
	\begin{minipage}{0.315\textwidth}
		\centering\includegraphics[width=\linewidth]{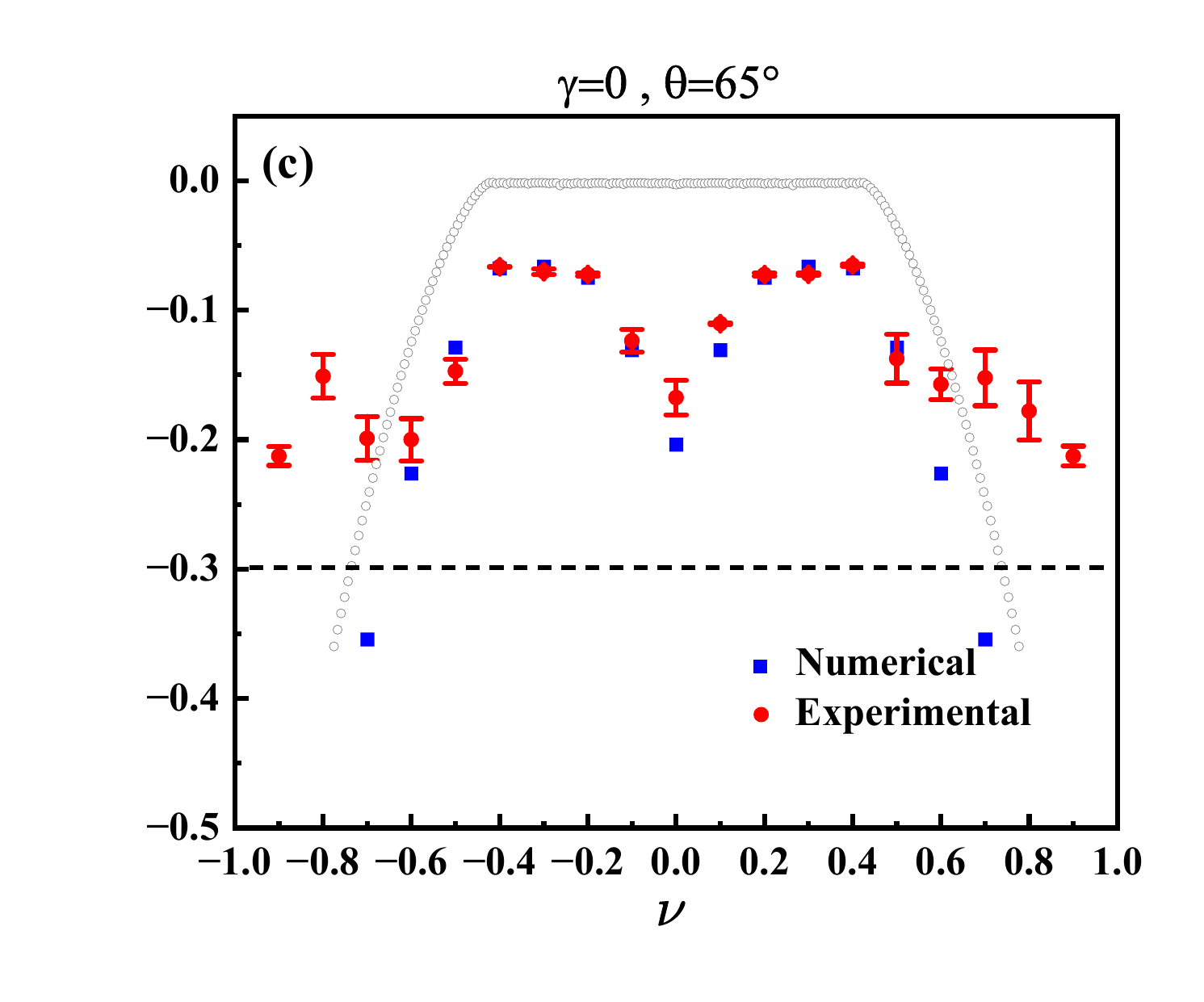}
	\end{minipage}%
	
	\begin{minipage}{0.345\textwidth}
		\centering\includegraphics[width=\linewidth]{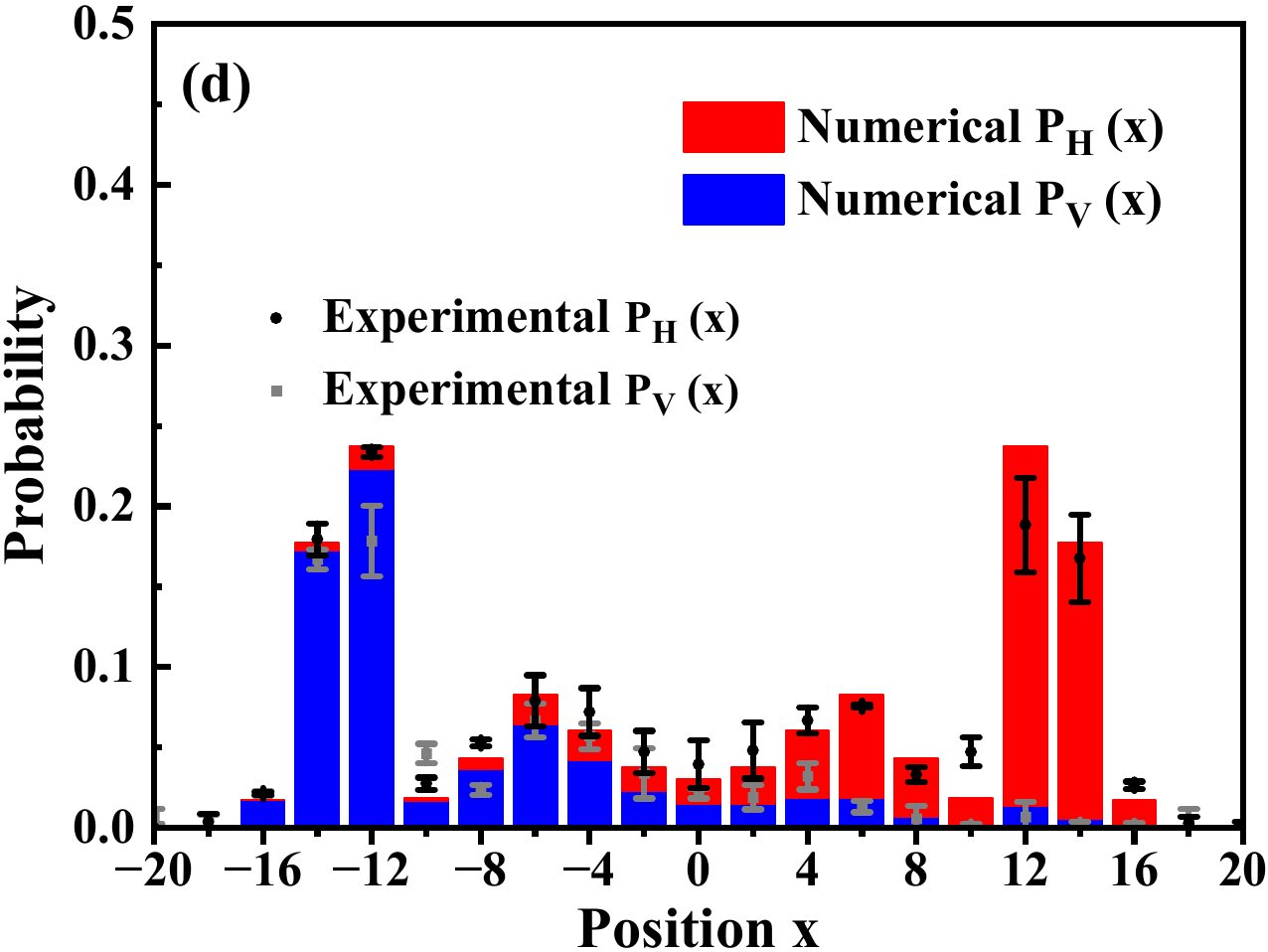}
	\end{minipage}%
	\begin{minipage}{0.3\textwidth}
		\centering\includegraphics[width=\linewidth]{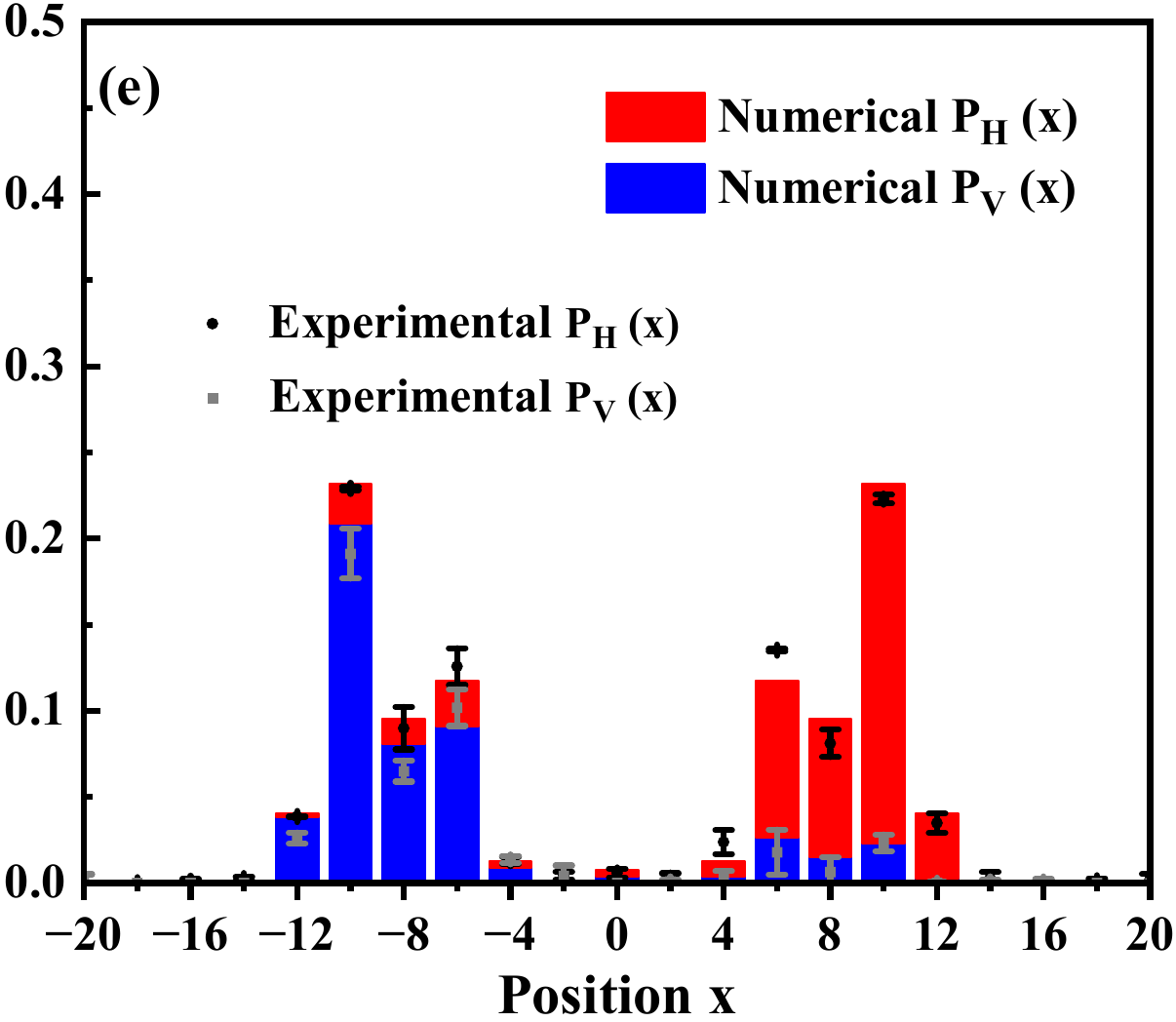}
	\end{minipage}%
	\begin{minipage}{0.3\textwidth}
		\centering\includegraphics[width=\linewidth]{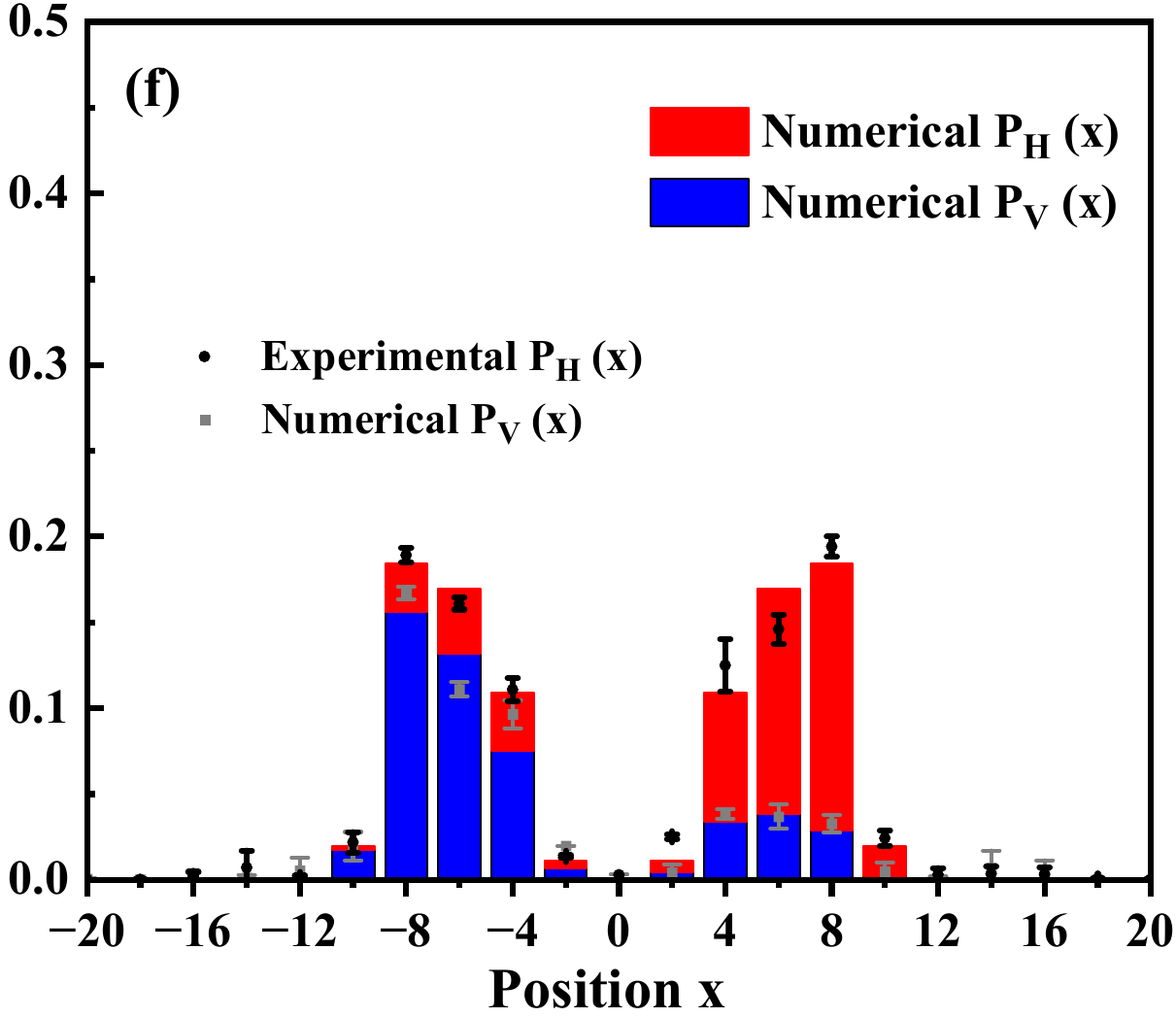}
	\end{minipage}%
	
	\begin{minipage}{1\textwidth}
		\centering
		\caption{(a)-(c) Experimental and numerical polarization-averaged growth rates $\bar{\lambda } \left ( v  \right ) $  for the 20-step quantum walk versus the shift velocity $v $, with the loss parameter $\gamma=0 $ and the coin parameter $\theta$=45°, $\theta$=57°, $\theta$=65°. Red dots with error bars represent the experimental data, and blue squares correspond to numerical simulations. Black dashed lines mark the reliability threshold, beyond which experimental data become unreliable due to photon loss. Gray circles are the numerical simulation results for the 2000-step quantum walk. (d)-(f) Experimental and numerical horizontally (vertically) polarized photon distribution for the 20-step quantum walk with the initial state $|0 \rangle \otimes |H \rangle $($|0 \rangle \otimes |V \rangle $) versus the position $x$, with the loss parameter $\gamma=0 $ and the coin parameter $\theta$=45°, $\theta$=57°, $\theta$=65°. The red (top) and blue (bottom) bars are the numerical results for the horizontal and vertical-polarization photon distributions, respectively. Gray dots denote the experimental values of the vertical-polarization photon distribution, while the black dots represent the total polarization-resolved distributions. The error bars account for statistical uncertainties in photon number counting. }
	\end{minipage}
\end{figure*}\par

\begin{figure*}[ht!]
	\begin{minipage}{0.33\textwidth}
		\centering\includegraphics[width=\linewidth]{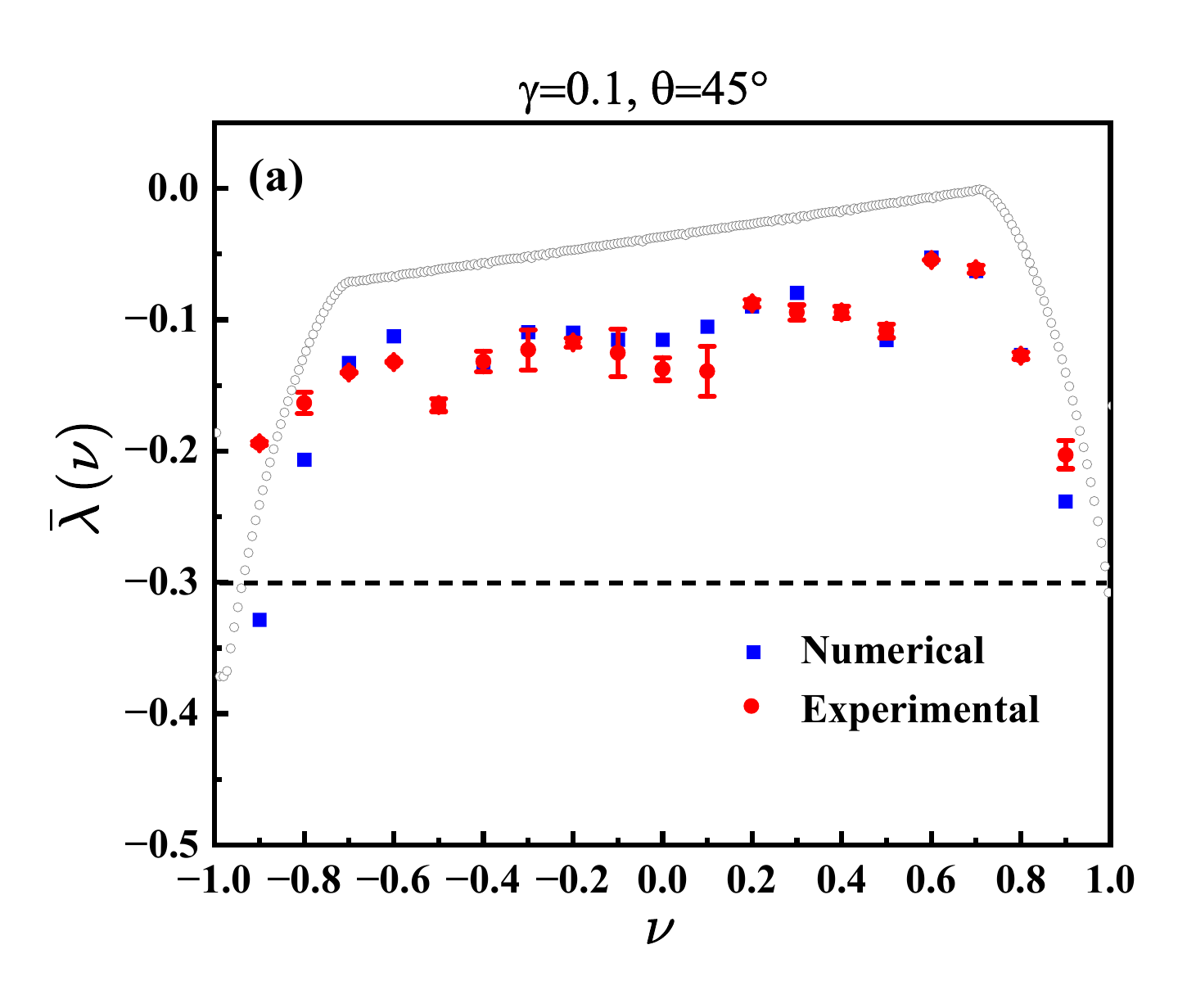}
	\end{minipage}%
	\begin{minipage}{0.31\textwidth}
		\centering\includegraphics[width=\linewidth]{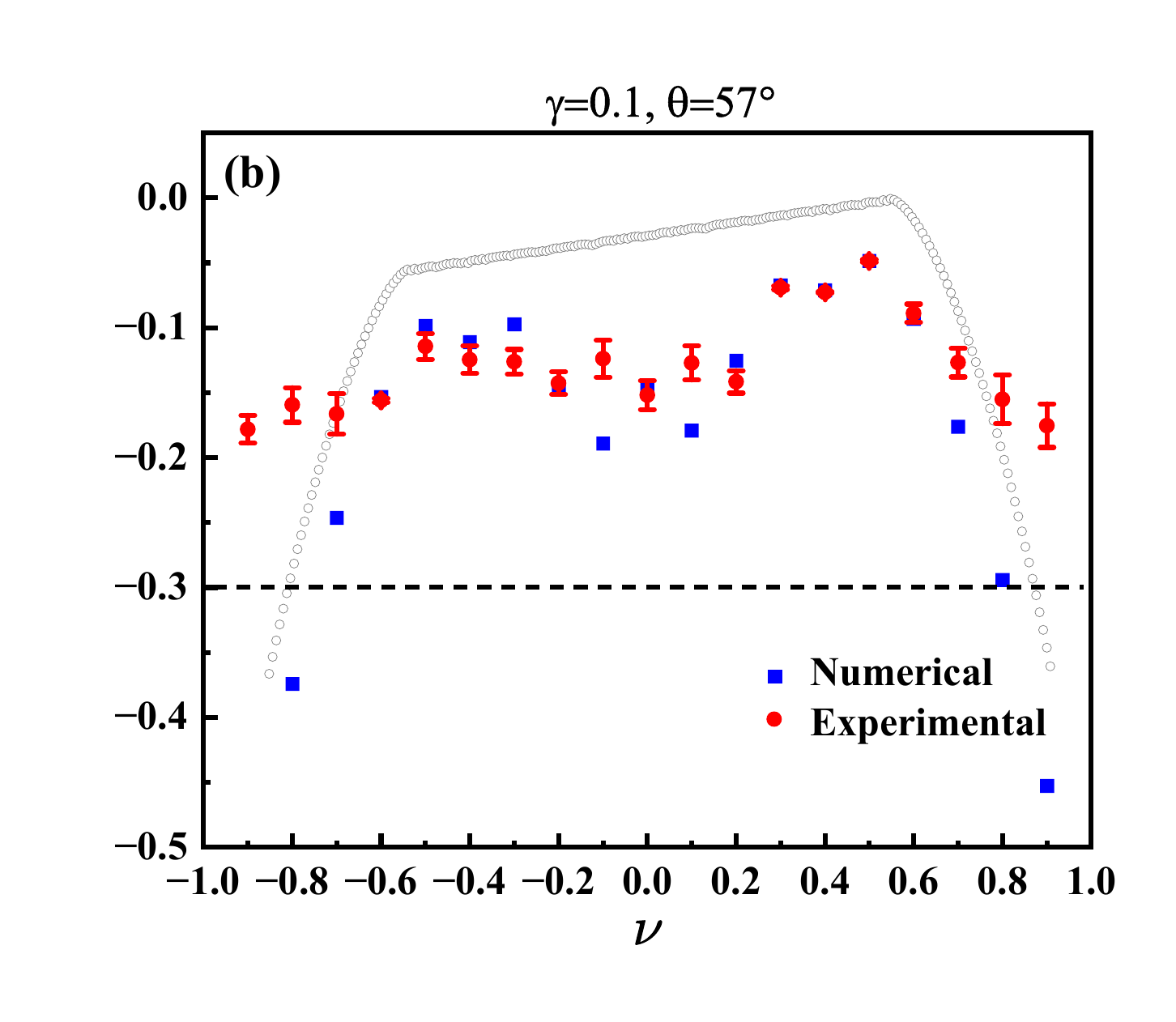}
	\end{minipage}%
	\begin{minipage}{0.31\textwidth}
		\centering\includegraphics[width=\linewidth]{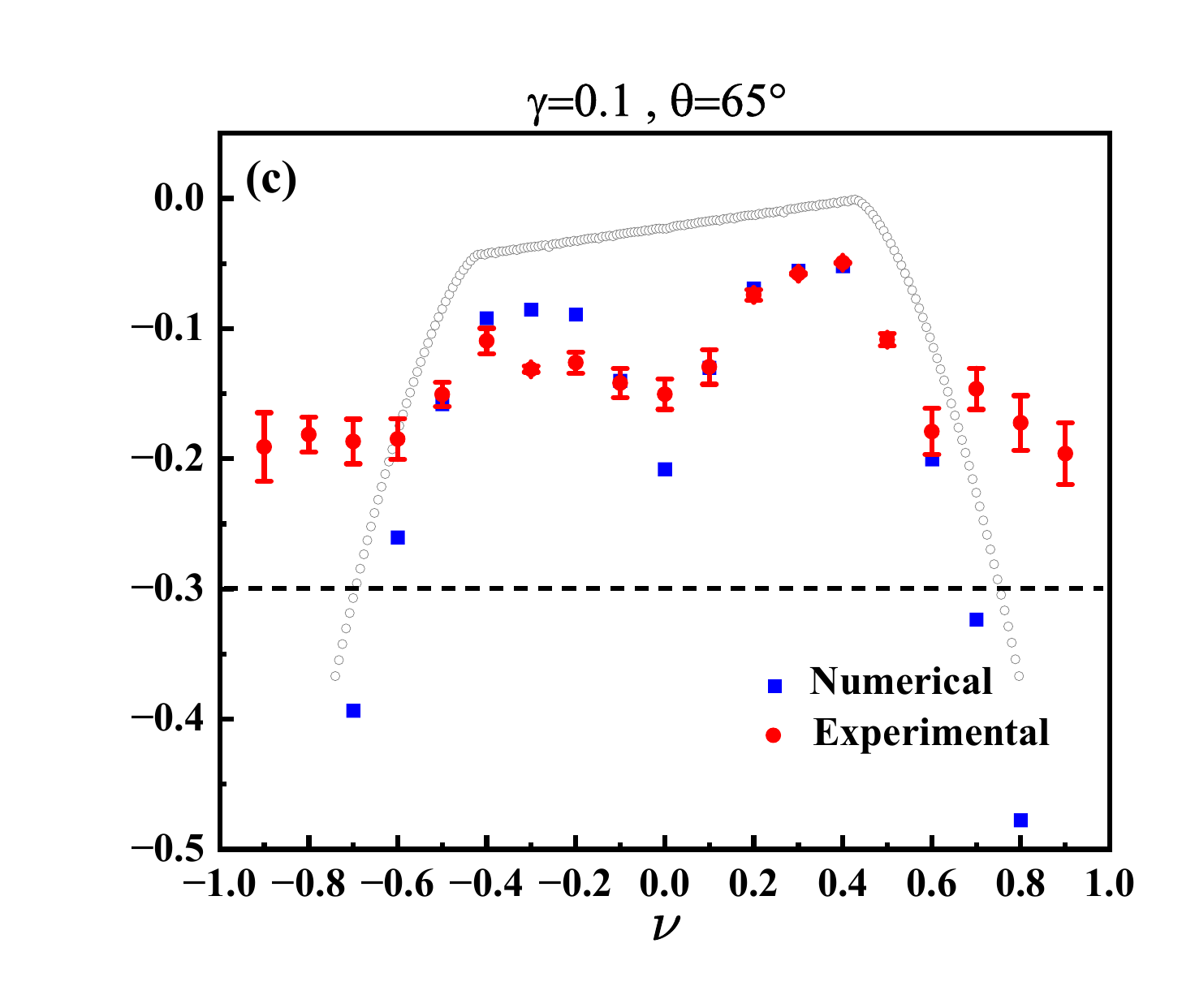}
	\end{minipage}%
	
	\begin{minipage}{0.32\textwidth}
		\centering\includegraphics[width=\linewidth]{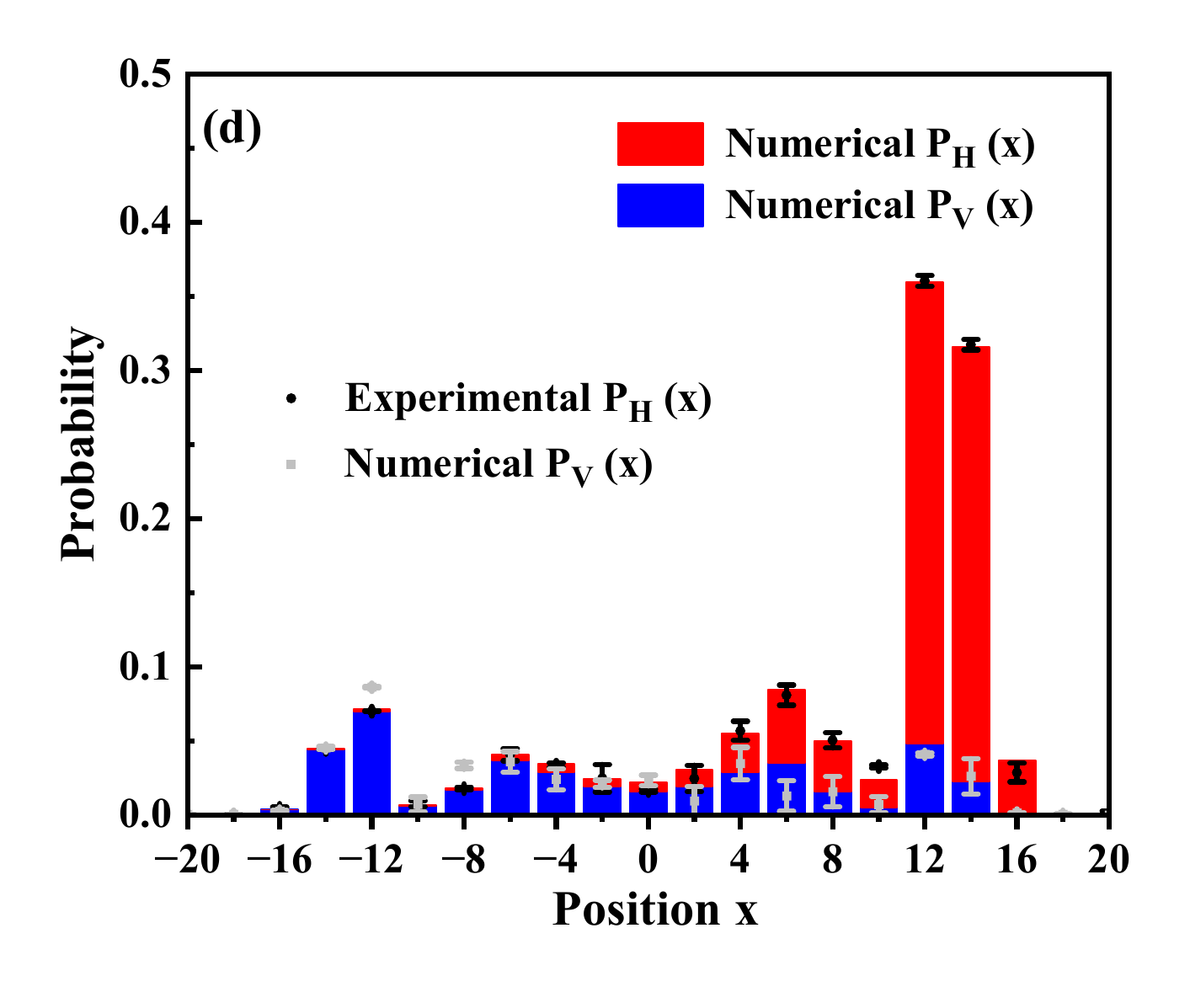}
	\end{minipage}%
	\begin{minipage}{0.3\textwidth}
		\centering\includegraphics[width=\linewidth]{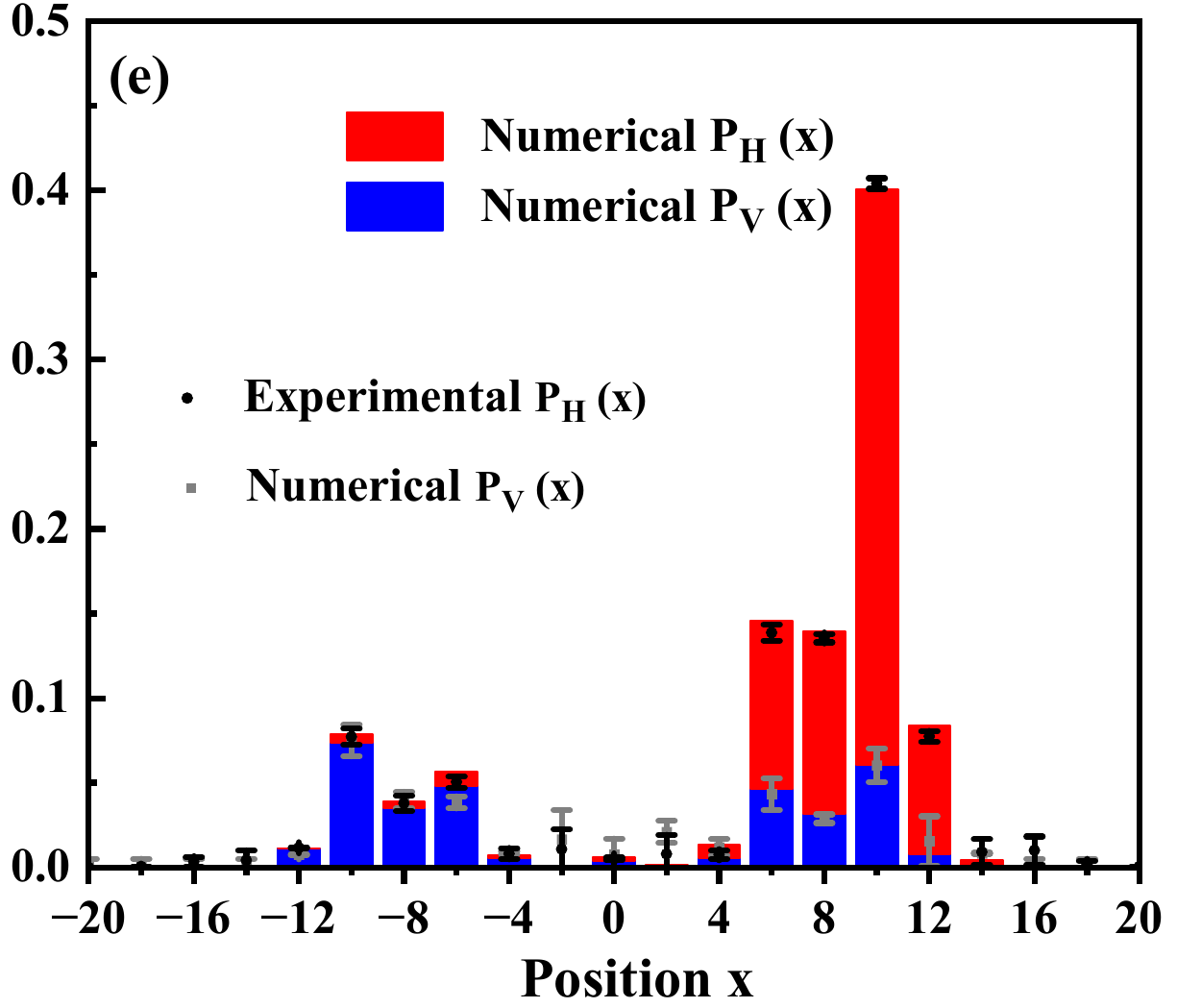}
	\end{minipage}%
	\begin{minipage}{0.3\textwidth}
		\centering\includegraphics[width=\linewidth]{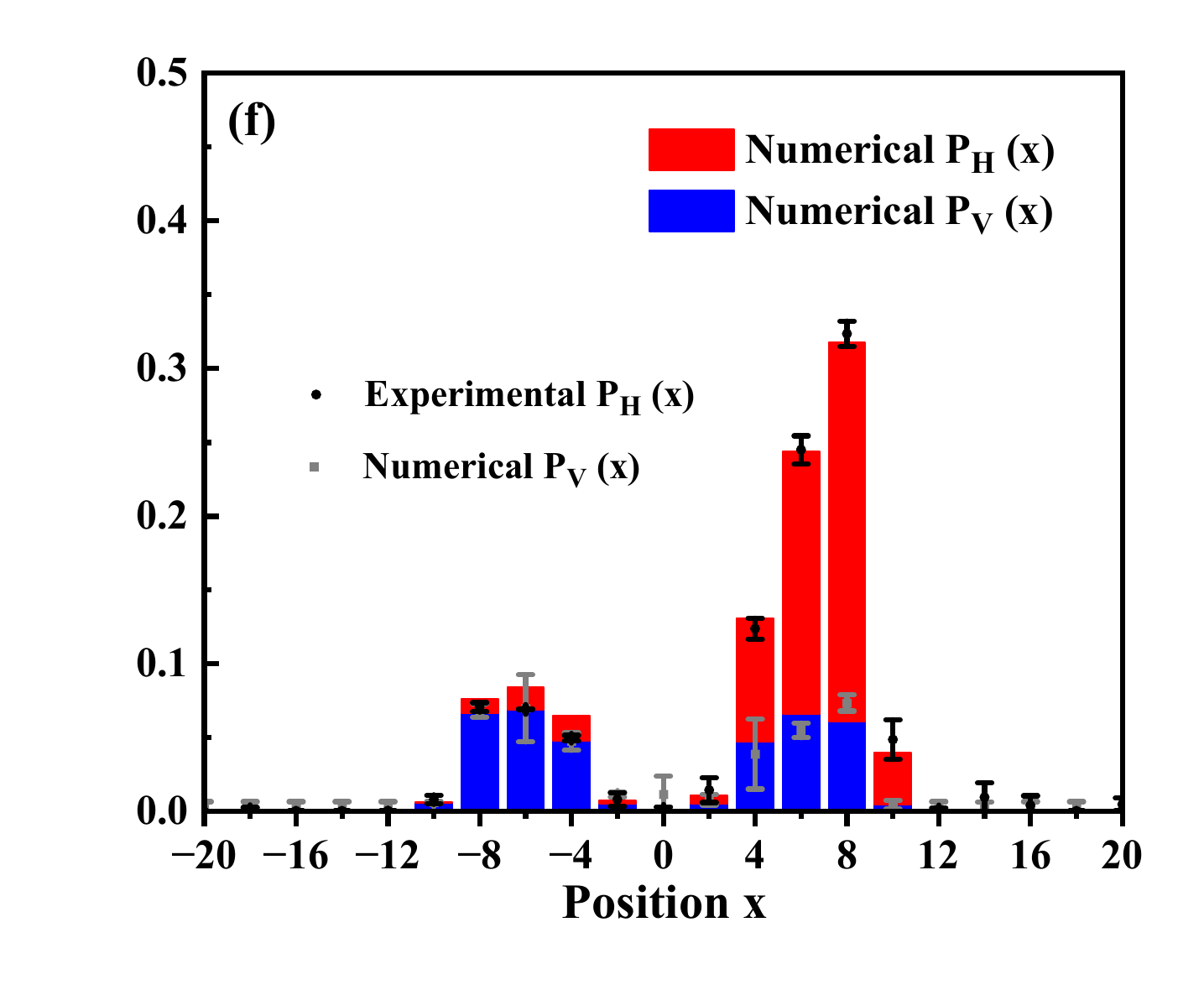}
	\end{minipage}%
	
	\begin{minipage}{1\textwidth}
		\centering
		\caption{(a)-(c) Experimental and numerical polarization-averaged growth rates $\bar{\lambda } \left ( v \right ) $  for the 20-step quantum walk versus the shift velocity $v $, with the loss parameter $\gamma=0.1 $ and the coin parameter $\theta$=45°, $\theta$=57°, $\theta$=65°. Gray circles are the numerical simulation results for the 2000-step quantum walk. (d)-(f) Experimental and numerical horizontally (vertically) polarized photon distribution for the 20-step quantum walk with the initial state $|0 \rangle \otimes |H \rangle $($|0 \rangle \otimes |V \rangle $) versus the position $x$, with the loss parameter $\gamma=0.1 $ and the coin parameter $\theta$=45°, $\theta$=57°, $\theta$=65°. The red (top) and blue (bottom) bars are the numerical results for the horizontal and vertical-polarization photon distributions, respectively. Gray dots denote the experimental values of the vertical-polarization photon distribution, while the black dots represent the total polarization-resolved distributions. The error bars account for statistical uncertainties in photon number counting.}
	\end{minipage}
\end{figure*}\par

\begin{figure*}[ht!]
	\begin{minipage}{0.315\textwidth}
		\centering\includegraphics[width=\linewidth]{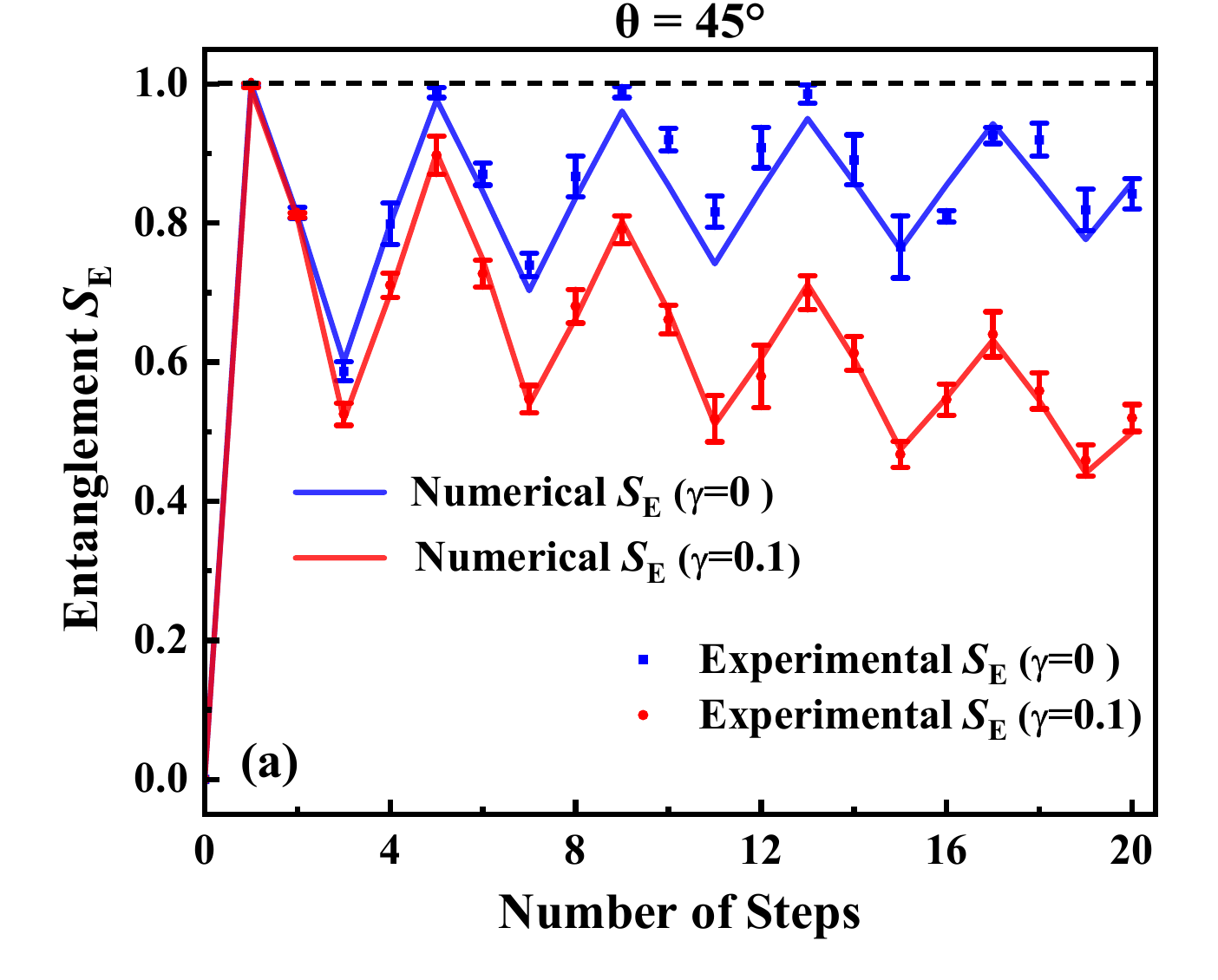}
	\end{minipage}%
	\hspace{0.1em}
	\begin{minipage}{0.3\textwidth}
		\centering\includegraphics[width=\linewidth]{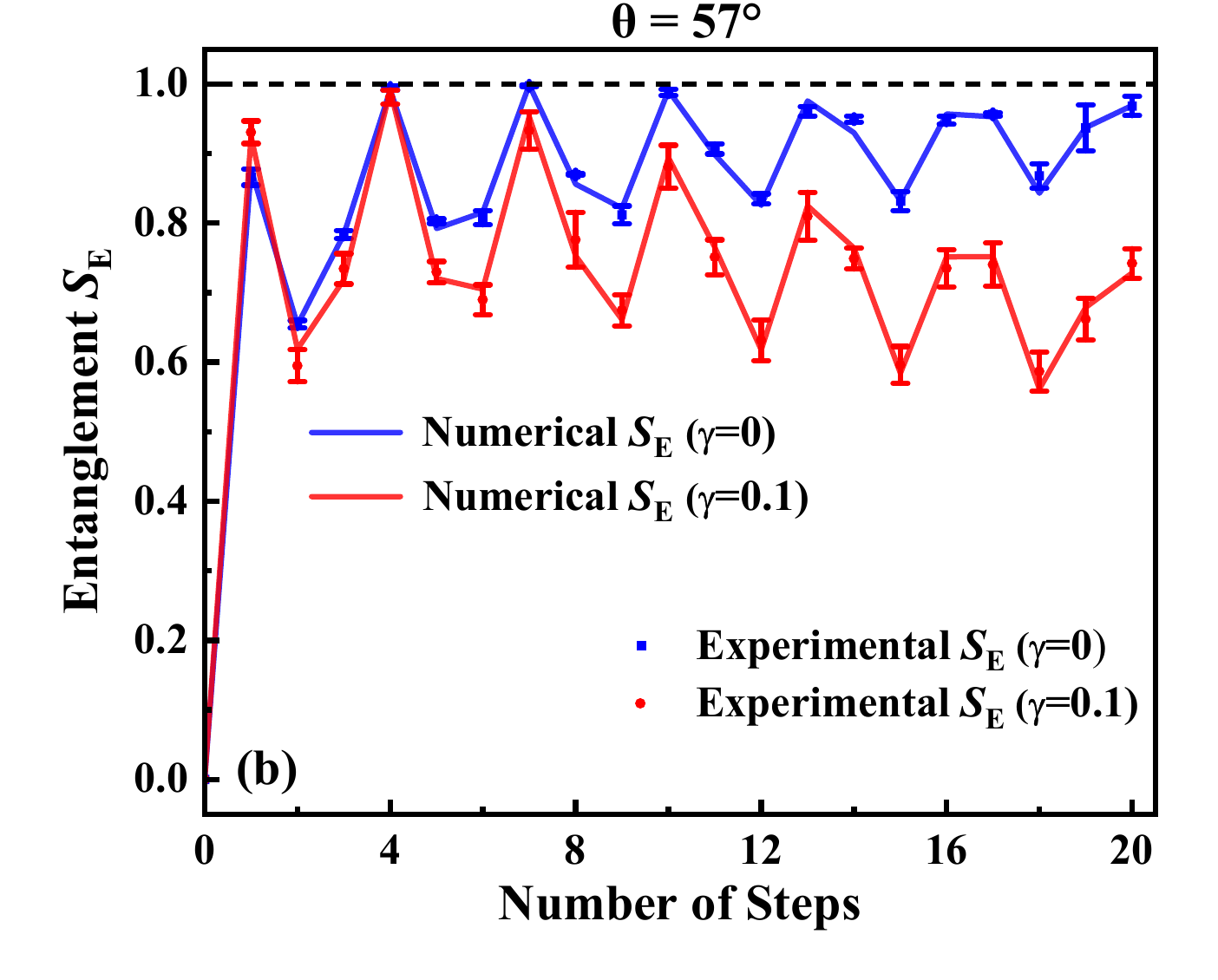}
	\end{minipage}%
	\hspace{0.1em}
	\begin{minipage}{0.3\textwidth}
		\centering\includegraphics[width=\linewidth]{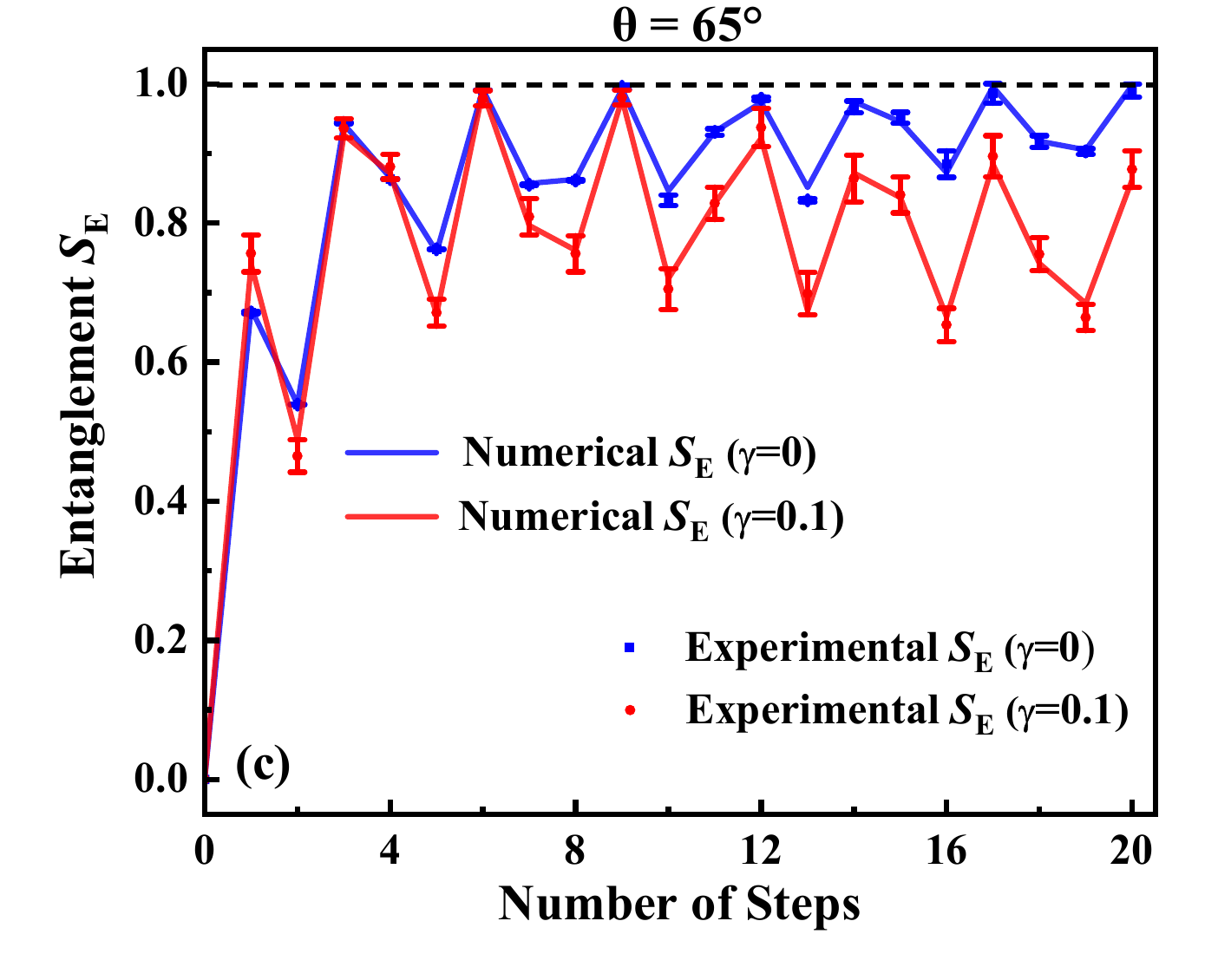}
	\end{minipage}%
	\hspace{0.1em}
	\begin{minipage}{0.31\textwidth}
		\centering\includegraphics[width=\linewidth]{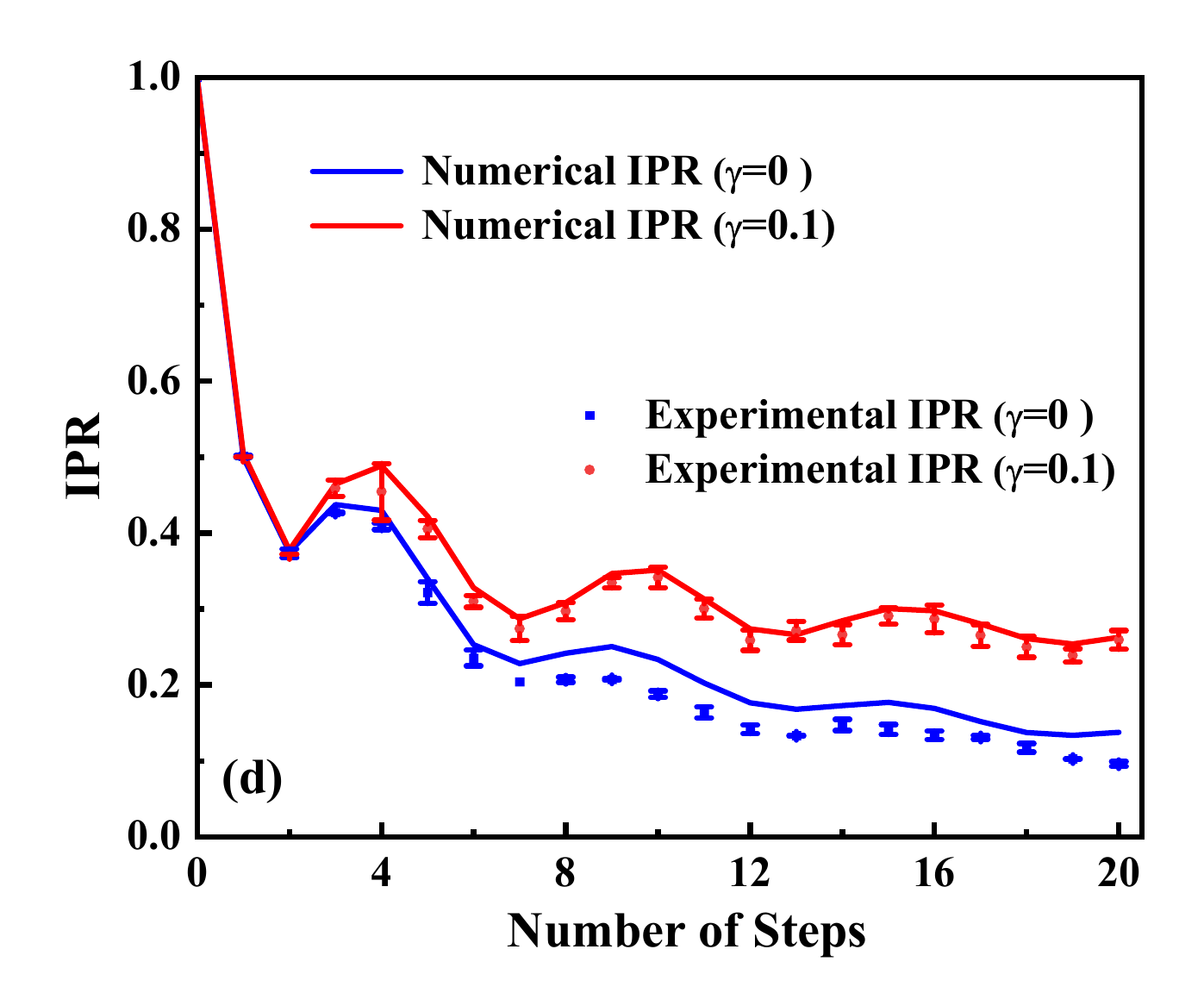}
	\end{minipage}%
	\hspace{0.1em}
	\begin{minipage}{0.3\textwidth}
		\centering\includegraphics[width=\linewidth]{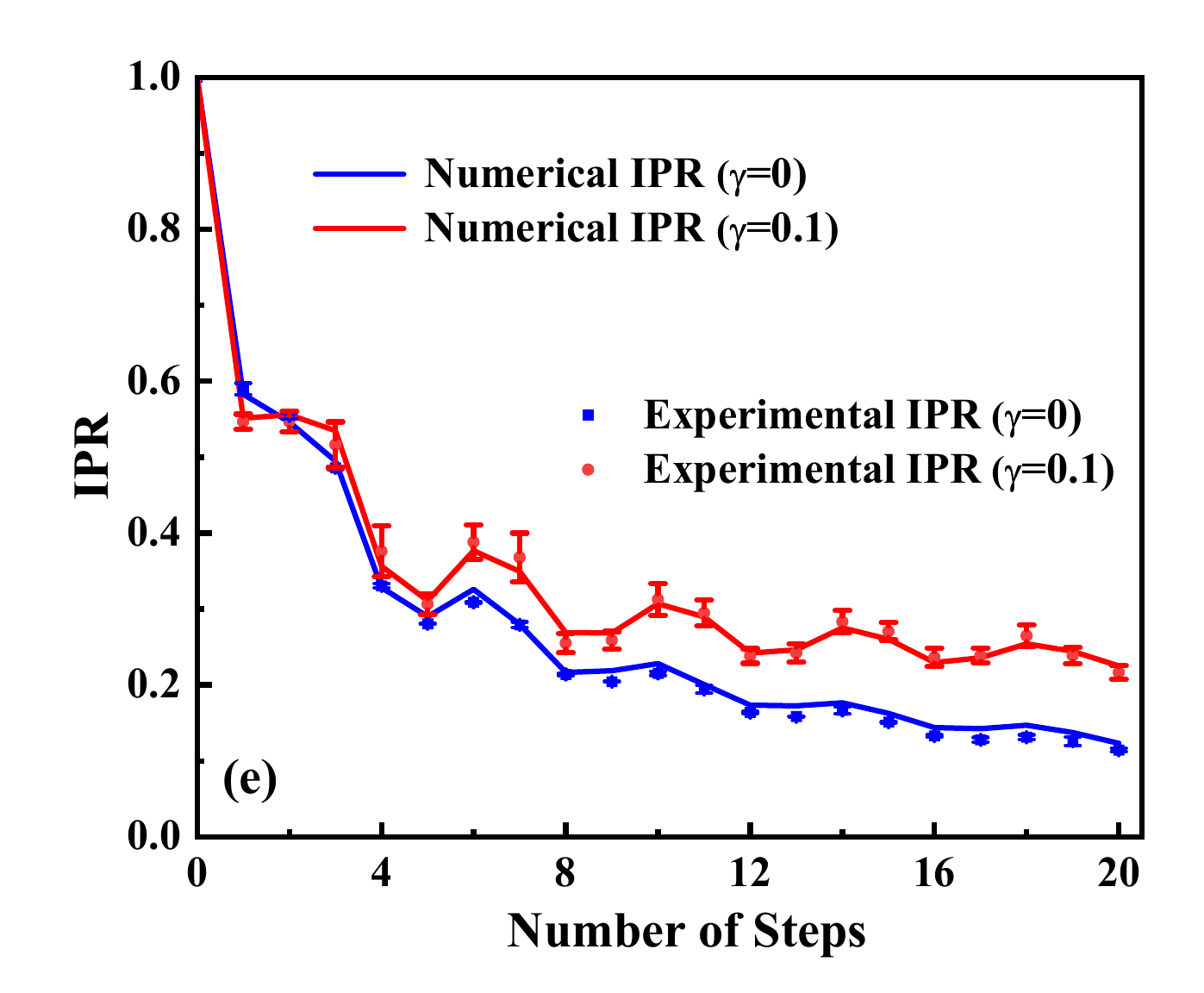}
	\end{minipage}%
	\hspace{0.1em}
	\begin{minipage}{0.3\textwidth}
		\centering\includegraphics[width=\linewidth]{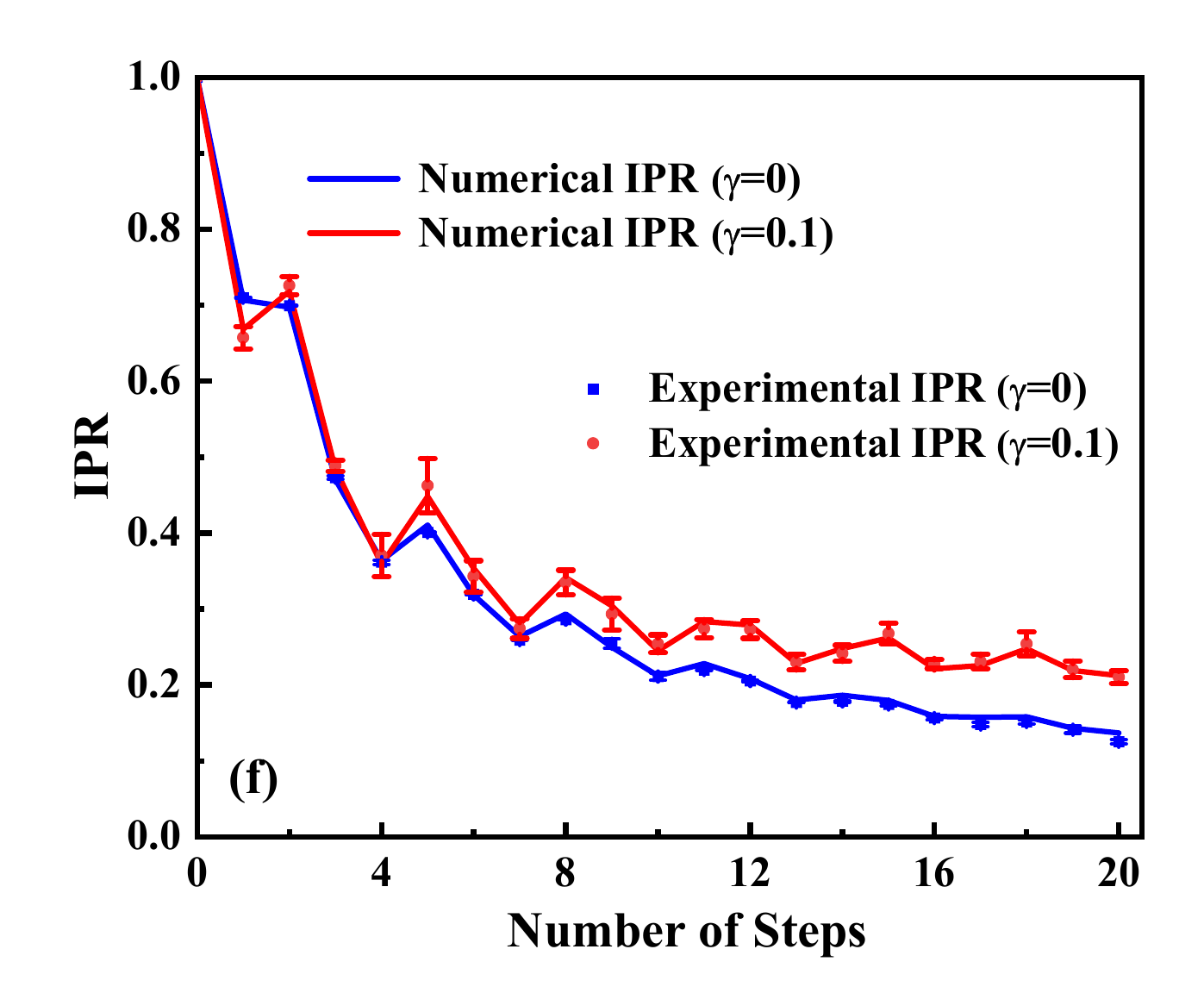}
	\end{minipage}%
	
	\begin{minipage}{1\textwidth}
		\centering
		\caption{(a)-(c) Experimental and numerical entanglement entropy $S_{\mathrm{E}} $ for the quantum walk versus the evolution steps, with the coin parameter $\theta$=45°, $\theta$=57°, and $\theta$=65°. (d)-(f) Experimental and numerical $\mathrm{IPR}$ for the quantum walk versus the evolution steps, with the coin parameter $\theta$=45°, $\theta$=57°, and $\theta$=65°. For the loss parameter $\gamma=0$, blue squares with error bars represent experimental data, and blue lines denote numerical simulations. For the loss parameter $\gamma=0.1$, red dots with error bars represent the experimental data, and red lines denote numerical simulations.}
	\end{minipage}
\end{figure*}\par

A series of consecutive experiments are conducted to characterize the walker’s probability distributions, with each experiment yielding at most one detection event within a specific time window. These detection events are recorded by a computer through a time-to-digital converter interface. After $t$ steps of evolution, photon wave packets are distributed across $t + 1$ time windows, corresponding with the positions $x=-t,-t+2,\dots,t-2,t$. Photons at position $x$ are sent to a detection terminal consisting of a PBS3, APD1, and APD2. The PBS3 separates the horizontal and vertical polarization components, which are detected by APD1 and APD2, respectively. APD1 records the counts $N_{V} \left ( t,x \right ) $ of vertically polarized photons, while APD2 records the counts $N_{H} \left ( t,x \right ) $ of horizontally polarized photons. Therefore, the polarization-resolved photon distribution probabilities, $P_{H}\left ( x \right ) $ and $P_{V}\left ( x \right ) $, are obtained by counting the number of detected photons in the $|H \rangle$ and $|V \rangle$ states at each position, and then normalized by dividing by the total photon count across all positions at a given time step. This configuration performs a projection measurement in the $ \left |H \right \rangle $ and $ \left |V \right \rangle $ polarization basis, corresponding to measurement of the observable $ \sigma _{z} $. An HWP set at $22.5^{\circ } $ is inserted before PBS3 in the detection terminal. The HWP rotates the polarization basis from $\left \{ |H\rangle, |V\rangle \right \} $ to $\left \{ |+\rangle, |-\rangle \right \}$, where $\left | + \right \rangle = \frac{1}{\sqrt{2}}\big(\left | H \right \rangle + \left | V \right \rangle\big)$ and $\left | - \right \rangle = \frac{1}{\sqrt{2}}\big(\left | H \right \rangle - \left | V \right \rangle\big)$. The PBS3 then separates the $\left |+ \right \rangle $ and $\left |- \right \rangle $ components, which are detected by APD1 and APD2, respectively. APD1 records the counts $N_{+}(t,x)$ for photons polarized along $+45^\circ$, while APD2 records $N_{-}(t,x)$ for photons polarized along $-45^\circ$. Therefore, the normalized $P_{+}\left ( x \right ) $ and $P_{-}\left ( x \right ) $ are obtained by counting the number of detected photons in the $|+ \rangle$ and $|- \rangle$ states at each position. This configuration performs a projection measurement in the $ \left |+ \right \rangle $ and $ \left |- \right \rangle $ polarization basis, corresponding to measurement of the observable $ \sigma _{x} $. These two detection configurations enable the measurement of polarization-resolved probability distributions and the reconstruction of the reduced density matrix of the coin state. As the number of walking steps increases, photon loss accumulates, leading to an increase in the overall loss of the system. In order to obtain a reliable statistical distribution of walkers with more steps, hours of ensemble measurements are required. Therefore, it is required to maintain the system's stability and minimize external influences such as ambient light pollution and wind interference. 
 \par

To observe the NHSE, we measured the polarization-averaged growth rate $\bar{\lambda } \left ( v  \right ) =\frac{1}{2}\left ( \lambda _{H} \left ( v   \right ) +\lambda _{V} \left ( v   \right ) \right )$ in our experiment. Here, $\bar{\lambda } \left ( v  \right )$  enables us to qualitatively capture the distinctive features of the Lyapunov exponent using 20-step DTQW. In Eq. (7), the polarization-resolved growth rates are defined as $\lambda _{i}\left ( v   \right )=\frac{1}{t} \log_{}{\left | \psi_{x}^{i}\left ( t \right )   \right | }(i=H, V) $, where $ \psi_{x}^{i}\left ( t \right )  =\langle i|\otimes \langle x|\psi _{x}\left ( t\right )  \rangle\otimes|i \rangle $. To construct $ \psi_{x}^{i}\left ( t\right )$, the initial state is chosen as $|0 \rangle \otimes |i \rangle $, and the probability distribution of photons in the polarization state $|i \rangle$ at position $|x \rangle$ is measured projectively.
Fig. 4(a)-(c) shows experimental and numerical polarization-averaged growth rates $\bar{\lambda } \left ( v  \right )$ for the 20-step quantum walk as functions of the shift velocity $v $. Here, the loss parameter $\gamma$ is chosen as 0, and the coin parameter $\theta $ is selected as 45°, 57°, and 65°, respectively. Gray circles are the numerical simulation results for the 2000-step quantum walk, which is approximated as the Lyapunov exponent of the system. As can be seen from the gray circles and red data dots, $\bar{\lambda } \left ( v   \right )$ with a relatively small number of steps ($t$ = 20) was able to capture approximately the significant features of the Lyapunov exponent. Apparently, $\bar{\lambda } \left ( v  \right )$ shows a symmetric profile with respect to its peak at  $v =0 $, and there is no NHSE. Such a profile directly originates from the propagation of probability in the bulk.\par

Then, we perform projection measurements on the polarization bases $|H \rangle$ and $|V \rangle$.  As shown in Fig. 4(d)-(f), the horizontally (vertically) polarized photon distributions for the 20-step quantum walk with the initial state $|0 \rangle \otimes |H \rangle$($|0 \rangle \otimes |V \rangle $) are presented. It can be seen that these photon position distributions clearly demonstrate the non-Gaussian behavior of the DTQW. To compare the experimental and numerical distributions, we used the fidelity defined as $F= {\textstyle \sum_{x}^{}}\left (  \sqrt{P_{H}^{\mathrm{Exp} }(x) P_{H}^{\mathrm{Num}}}(x) +\sqrt{P_{V}^{\mathrm{Exp} }(x) P_{V}^{\mathrm{Num}}}(x)  \right )$, which ranges from 0 (complete mismatch) to 1 (identical distributions). The fidelity for the 20-step quantum walk at $\theta$=45° is calculated as $F=0.972\pm0.002$. The experimental results confirm that the 20-step quantum walk remains nearly coherent. The polarization-resolved photon distribution also exhibits a symmetric profile. The symmetric probability propagation does not lead to population accumulation at the boundaries. This observation confirms that when the loss parameter $\gamma$ is 0, there is no skin effect in the DTQW. \par

Fig. 5(a)-(c) also shows experimental and numerical polarization-averaged growth rates $\bar{\lambda } \left ( v  \right )$ for the 20-step quantum walk as functions of the shift velocity $v $. Here, the loss parameter $\gamma$ is chosen as 0.1, and the coin parameter $\theta $ is selected as 45°, 57°, and 65°, respectively. As seen from the gray circles and red data points, $\bar{\lambda } \left ( v   \right )$ can qualitatively capture the significant features of the Lyapunov exponent. As shown in Fig. 5(a), the maximum peak of $\bar{\lambda } \left ( v  \right )$ with the coin parameter $\theta$ =45° appears at shift velocity $v =0.6 $; in Fig. 5(b), the maximum peak with $\theta$=57° appeared at $v =0.5 $; and in Fig. 5(c), the maximum peak with $\theta$=65° appeared at $v =0.4 $.  Apparently, $\bar{\lambda } \left ( v  \right )$ shows an asymmetric profile. This observation confirms the presence of the NHSE when the loss parameter $\gamma$ is 0.1. In addition, with the same loss parameter, the shift velocity $v $ corresponding to the maximum peak of $\bar{\lambda } \left ( v  \right )$ is gradually moved towards the origin $v =0$ as the coin parameter $\theta$ increases. This behavior can be attributed to the directional propagation of probability in the bulk. As shown in Fig. 5(d)-(f), the horizontally (vertically) polarized photon distributions for the 20-step quantum walk with the initial state $|0 \rangle \otimes |H \rangle$($|0 \rangle \otimes |V \rangle $) are presented. The polarization-resolved photon distribution also exhibits an asymmetric profile. The maximum peak of photon distribution appeared at position $x=12$ for $\theta$ = 45°, at $x=10$ for $\theta$ = 57°, and at $x=8$ for $\theta$ = 65°. This localization of the photon population is a clear signature of the NHSE-induced eigenstate localization. This also means that increasing the coin parameter leads to a more uniform distribution of the photon population for a given loss parameter. \par
\begin{figure*}[ht!]
	\begin{minipage}{0.345\textwidth}
		\centering\includegraphics[width=\linewidth]{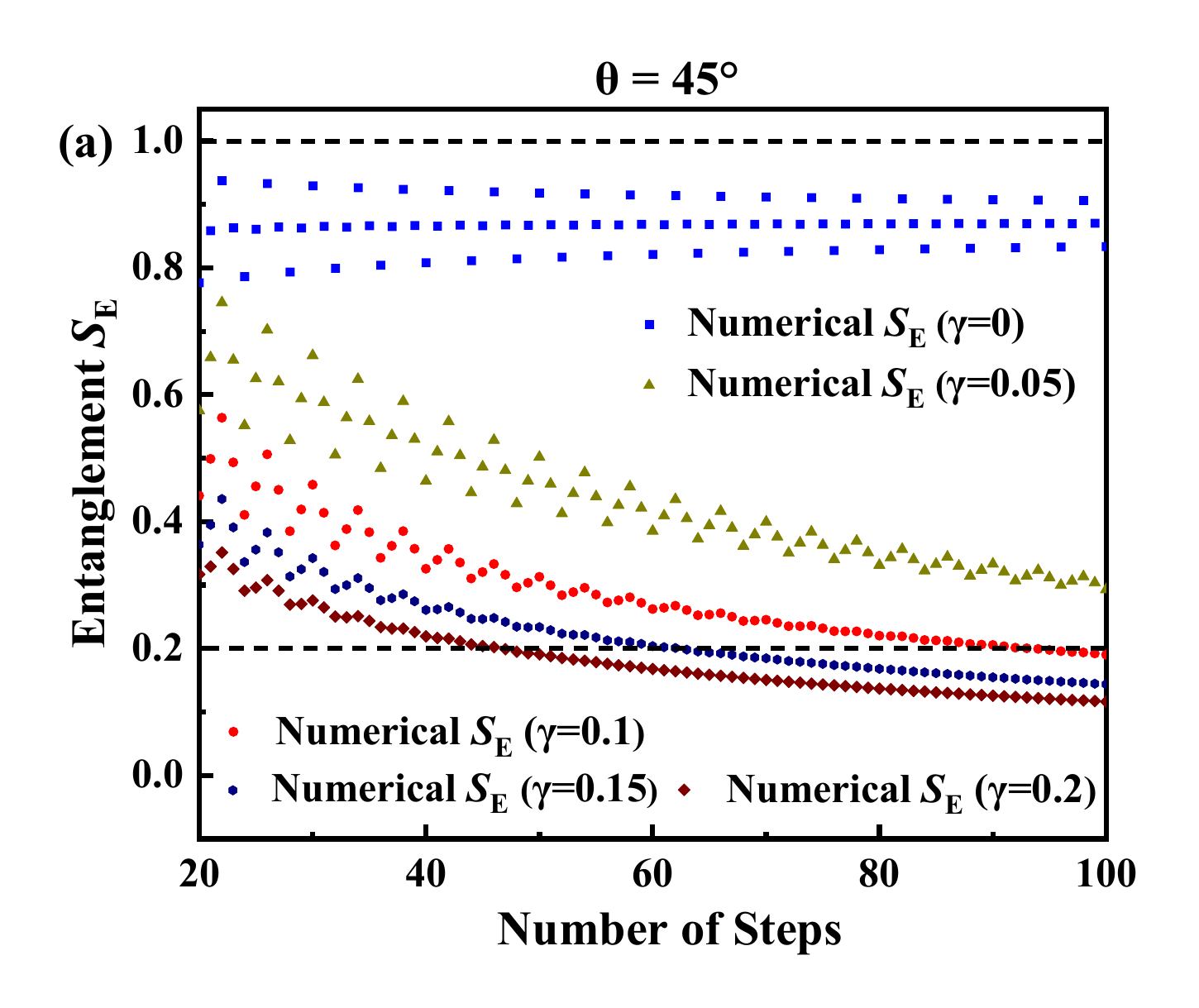}
	\end{minipage}%
	\begin{minipage}{0.325\textwidth}
		\centering\includegraphics[width=\linewidth]{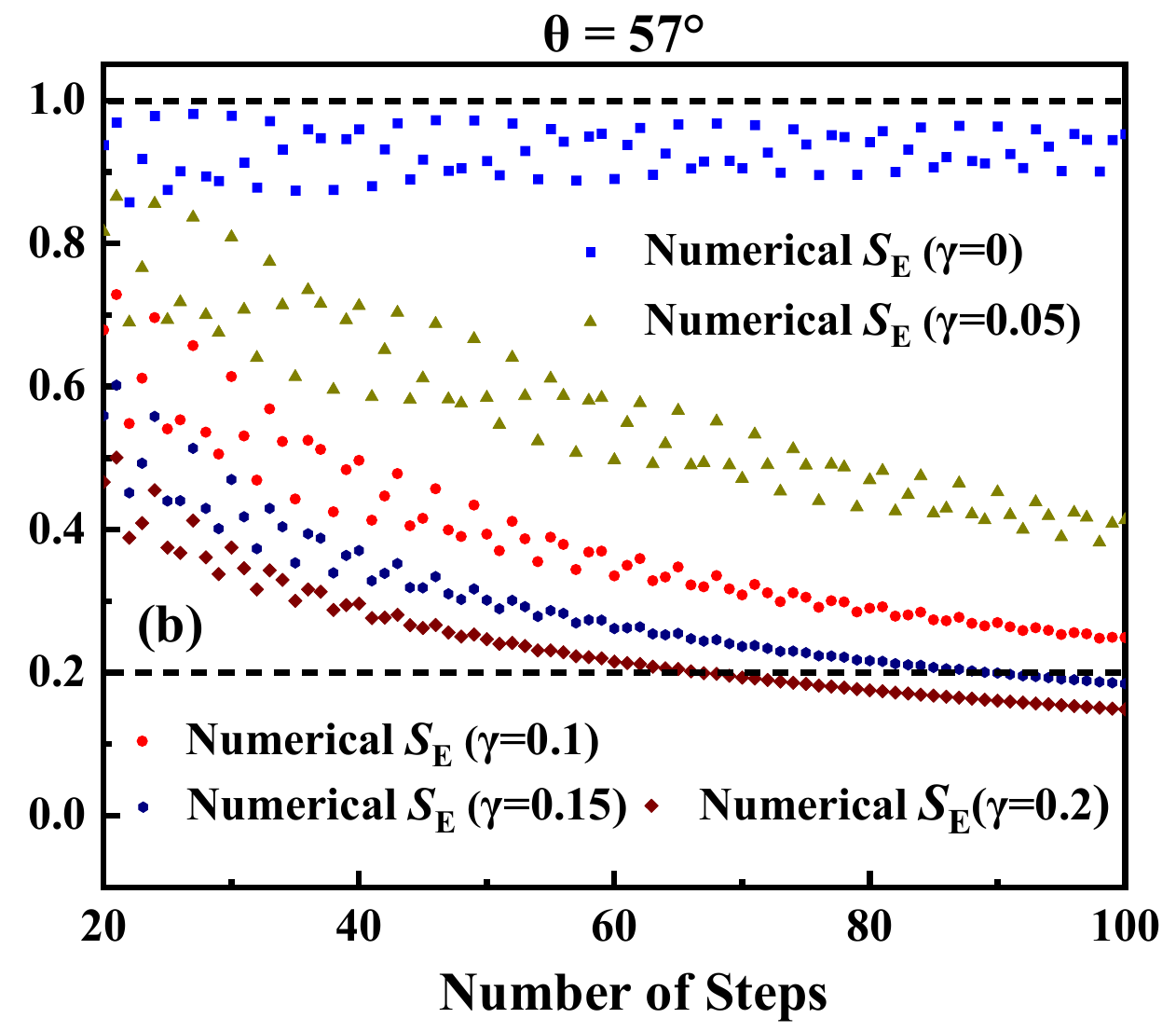}
	\end{minipage}%
	\begin{minipage}{0.33\textwidth}
		\centering\includegraphics[width=\linewidth]{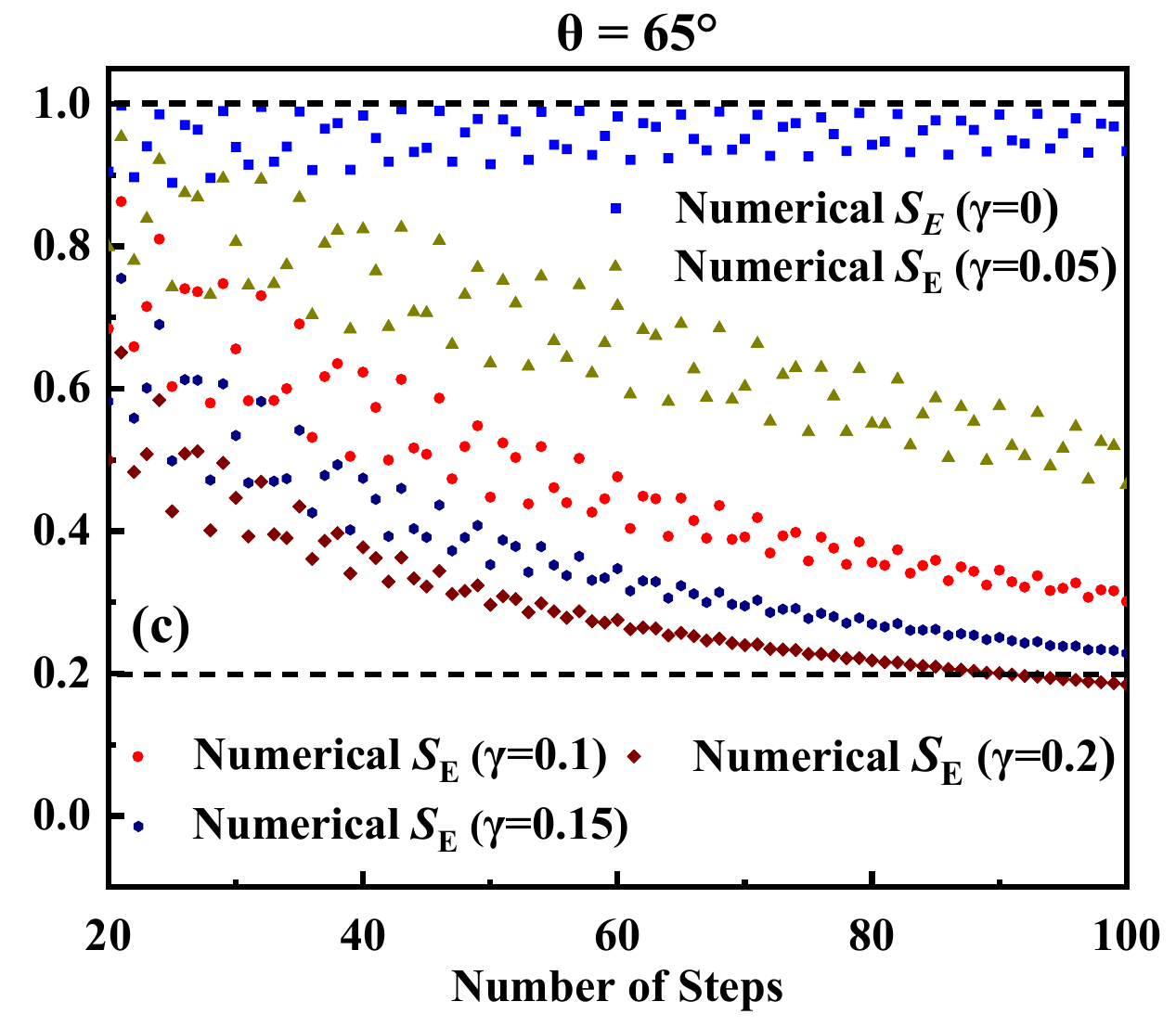}
	\end{minipage}%
	\hspace{0.1em}		
	\begin{minipage}{1\textwidth}
		\centering
		\caption{(a)-(c) Numerical entanglement entropy $S_{\mathrm{E}} $ for the quantum walk versus the extended evolution steps, with the coin parameter $\theta$=45°, $\theta$=57°, and $\theta$=65°. The blue, yellow, red, navy, and wine lines represent numerical simulations for the loss parameter $\gamma=0$, $\gamma=0.05$, $\gamma=0.1$, $\gamma=0.15$, $\gamma=0.2$, respectively.}
	\end{minipage}
\end{figure*}\par
To determine the entanglement entropy experimentally, we reconstruct the complete reduced density matrix $\rho_c $ of the coin state after $t$ steps. Reconstructing $\rho_c $ requires measuring both its diagonal and off-diagonal elements. To obtain the diagonal elements, we perform a projection measurement of the observable $\sigma _{z}$. After $t$ steps, the polarization-resolved photon distribution $P_{H}\left ( x \right )$ and $P_{V}\left ( x \right ) $ at each position $x$ provide the absolute values of the amplitude coefficients: $ \left | a_{x} \left ( t \right )  \right | =\sqrt{P_{H} \left ( x \right ) } $, $ \left | b_{x} \left ( t \right )  \right | =\sqrt{P_{V} \left (x \right ) } $. The diagonal elements are then calculated by summing over all positions: $ \alpha \left ( t \right )= {\textstyle \sum_{x}^{}} P_{H}\left ( x \right ) $, $ \beta \left ( t \right )= {\textstyle \sum_{x}^{}} P_{V}\left ( x \right ) $. To extract the off-diagonal elements, we perform an additional projection measurement of the observable $\sigma _{x}$. The difference between the measured probabilities $ P_{+}\left ( x \right )$ and $ P_{-}\left ( x \right )$ determines the relative sign between $ a_{x}(t) $ and $ b_{x}(t)$ at each position: $ P_{+}\left ( x \right )- P_{+}\left ( x \right )=2 a_{x}(t) b_{x}(t)$. Meanwhile, the absolute value of the product $ a_{x}(t) b_{x}(t) $ is determined by: $|a_{x}(t)b_{x}(t)|=\sqrt{P_{H}(x)P_{V}(x)} $. The off-diagonal element is then obtained by summing over all positions ${\textstyle \sum_{x}^{}} a_{x}\left ( t \right )b_{x}\left ( t \right )$. By performing polarization-resolved measurements in both the $\sigma _{x}$ and $\sigma _{z}$ bases, we reconstruct the coin state's reduced density matrix $ \rho_c $ after $t$ steps. The resulting matrix is then used to determine its eigenvalues according to Eq. (10), and its eigenvalues are subsequently substituted into Eq. (9) to calculate the entanglement entropy between the coin and position degrees of freedom.

Fig. 6(a)-(c) shows experimental and numerical entanglement entropy $S_{\mathrm{E}} $ for the quantum walk as functions of the evolution steps. Here, the initial state is chosen as  $|0 \rangle \otimes |H \rangle $, and the loss parameter $\gamma$ is chosen as 0 and 0.1, respectively. The coin parameters are selected as 45°, 57°, and 65°, respectively. It can be seen that the experimental values of the von Neumann entropy for the 20-step walker are 0.842 ± 0.021, 0.968 ± 0.014, and 0.990 ± 0.010, respectively. These values are in good agreement with the numerical results. Using the initial state $|0 \rangle \otimes |H \rangle $ and a coin parameter $\theta=65^\circ$, our experiment thus generated near-maximal coin-position entanglement in the 20-step quantum walks. Moreover, the entanglement entropy of the quantum walk with loss parameter $\gamma=0$ consistently exceeds that of the quantum walk with loss parameter $\gamma=0.1$. The entanglement is significantly suppressed when the non-Hermitian loss parameter is introduced. In addition, as the number of walking steps increases, the entanglement suppression becomes more pronounced. It is worth noting that the entanglement entropy $S_{\mathrm{E}}$ decays more slowly with increasing walk steps when $\theta$ = 65° compared to $\theta$ = 45° and $\theta$ = 57°. Therefore, increasing the coin parameter slows the decay of entanglement entropy for a fixed loss parameter. As shown in Fig. 1(c), Fig. 2(a), and Fig. 5(a)-(c), these experimental results confirm that entanglement suppression in non-Hermitian DTQW systems becomes stronger with longer evolution time, and weaker with larger coin parameters.\par

Note that entanglement suppression requires a large number of localized modes. As shown in Fig. 6(d)-(f), experimental and numerical $\mathrm{IPR}$ for the quantum walk as functions of the evolution steps are present. Here, the initial state is chosen as  $|0 \rangle \otimes |H \rangle $, and the loss parameter $\gamma$ is chosen as 0 and 0.1, respectively. The coin parameters are selected as 45°, 57°, and 65°, respectively. It can be observed that there is an overall decay in the $\mathrm{IPR}$ value of the quantum walk as the number of steps increases. This behavior arises from the ballistic spreading of quantum walks, which causes the probability distribution of photons to gradually spread out and become progressively flatter as the number of steps $t$ increases, thereby leading to an overall decrease in IPR. The $\mathrm{IPR}$ of the quantum walk with loss parameter $\gamma=0$ is consistently lower than that with loss parameter $\gamma=0.1$. This difference becomes more pronounced as the number of walking steps increases. Besides, the $\mathrm{IPR}$ for $\gamma=0.1$ decays more rapidly with increasing steps when $\theta=$65° compared to $\theta=45$° and $\theta=$57°. Thus, we show that the NHSE suppresses the delocalization and the entanglement of the coin-position state in non-Hermitian DTQW systems. This suppression also weakens as coin parameters increase. \par 

Due to fiber loop loss, limited detection efficiency, and environmental fluctuations, our current experiments cannot measure photon distribution at longer evolution steps. Therefore, we numerically predict the entanglement dynamics of the coin-position state in the DTQW system for extended evolution steps, considering different coin and loss parameters, as shown in Fig. 7(a)–(c). The initial state is chosen as  $|0 \rangle \otimes |H \rangle $, with the loss parameter  $\gamma$ set to 0, 0.05, 0.1, 0.15, and 0.2, respectively. The results show that entanglement suppression in the non-Hermitian DTQW system becomes more pronounced as the number of walk steps and the loss parameter $\gamma$ increase. Additionally, we observe that $S_{\mathrm{E}}$ decays more slowly with increasing steps for $\theta$=65° compared to $\theta$=45° and $\theta$=57°. These results further confirm that entanglement suppression induced by NHSE weakens with increasing coin parameters and enhances with increasing loss parameters and evolution steps.\par 

\section{Conclusion}
In conclusion, we investigate the generation of maximal coin-position entanglement and its suppression induced by NHSE in DTQWs. Our theoretical analysis reveals that under PBC with a nonzero loss parameter, the energy spectrum forms a closed loop, and the eigenstates of the system become localized at the boundary, indicating the existence of NHSE. In addition, the amplitude of the Lyapunov exponent gradually increases with the increase of the loss parameter and decreases with the increase of the coin parameter. Then, we analyzed the dependence of the von Neumann entropy on the loss and coin parameters. We find that, within specific ranges of coin operation parameters, the coin-position entanglement can be optimized to approach its maximum value, effectively resisting the impact of the loss parameter. \par 

Based on theoretical analysis, we experimentally implemented a one-dimensional DTQW using a time-multiplexed fiber loop. We observed polarization-resolved photon distributions for a 20-step quantum walk. The experiment demonstrates the generation of maximal coin-position entanglement for a 20-step quantum walk with a specific coin parameter. In addition, the polarization-averaged growth rate as a function of the shift velocity was experimentally measured. When the loss parameter is set to 0, the polarization-averaged growth rate and photon population distribution show a symmetric profile, indicating the absence of the skin effect in the DTQW. However, when the non-Hermitian loss is introduced, the polarization-averaged growth rate shows an asymmetric profile, and the photon population is accumulated at the boundaries, indicating the presence of the skin effect. Significantly, the shift velocity corresponding to the maximum polarization-averaged growth rate for the fixed loss parameter is gradually moved towards the origin with increasing coin parameters. This asymmetric profile of the polarization-averaged growth rate under loss is similar to observations reported in Ref. \cite{lin2022observation}. \par 

We then studied the dynamic properties of entanglement and delocalization in DTQWs under different coin and loss parameters. Interestingly, as the number of walking steps increases, we observe that in the presence of a loss parameter, the entanglement entropy of the quantum walk gradually decreases, while the IPR gradually increases compared to the lossless case. The entanglement suppression becomes more pronounced as the loss parameter and walking steps increase. In addition, as the coin parameter increases, the entanglement entropy of the quantum walk decays much more slowly. Therefore, our results show that entanglement suppression in the non-Hermitian DTQW system becomes more pronounced as the evolution steps and loss parameter increase, and weakens as the coin parameters increase. To investigate more long-time entanglement dynamics in quantum walks, it is essential to further reduce photon losses in the fiber loop, while simultaneously improving detection efficiency and suppressing environmental noise to increase the number of walk steps. In the future, with further development of photonic quantum state tomography, researchers will be able to reconstruct a more complete density matrix of optical quantum states. This progress will enable our experimental platform to study complex quantum phenomena, including detecting hybrid entanglement properties of multiple degrees of freedom in multi-particle systems and exploring more challenging non-Hermitian open quantum systems. \par

This work was supported by the Scientific and Technological Research Program of the Education Department of Hubei Province under Grant No. B2023139, the Key Laboratory of Photoelectric Conversion Materials and Devices Fund of Hubei province under Grant No. PMD202411 and No. PMD202412, and the Innovation Development Union Fund of Huangshi city under Grant No. 2024AFD010.

\bibliography{sample}

\begin{thebibliography}{54}%
\makeatletter
\providecommand \@ifxundefined [1]{%
 \@ifx{#1\undefined}
}%
\providecommand \@ifnum [1]{%
 \ifnum #1\expandafter \@firstoftwo
 \else \expandafter \@secondoftwo
 \fi
}%
\providecommand \@ifx [1]{%
 \ifx #1\expandafter \@firstoftwo
 \else \expandafter \@secondoftwo
 \fi
}%
\providecommand \natexlab [1]{#1}%
\providecommand \enquote  [1]{``#1''}%
\providecommand \bibnamefont  [1]{#1}%
\providecommand \bibfnamefont [1]{#1}%
\providecommand \citenamefont [1]{#1}%
\providecommand \href@noop [0]{\@secondoftwo}%
\providecommand \href [0]{\begingroup \@sanitize@url \@href}%
\providecommand \@href[1]{\@@startlink{#1}\@@href}%
\providecommand \@@href[1]{\endgroup#1\@@endlink}%
\providecommand \@sanitize@url [0]{\catcode `\\12\catcode `\$12\catcode
  `\&12\catcode `\#12\catcode `\^12\catcode `\_12\catcode `\%12\relax}%
\providecommand \@@startlink[1]{}%
\providecommand \@@endlink[0]{}%
\providecommand \url  [0]{\begingroup\@sanitize@url \@url }%
\providecommand \@url [1]{\endgroup\@href {#1}{\urlprefix }}%
\providecommand \urlprefix  [0]{URL }%
\providecommand \Eprint [0]{\href }%
\providecommand \doibase [0]{http://dx.doi.org/}%
\providecommand \selectlanguage [0]{\@gobble}%
\providecommand \bibinfo  [0]{\@secondoftwo}%
\providecommand \bibfield  [0]{\@secondoftwo}%
\providecommand \translation [1]{[#1]}%
\providecommand \BibitemOpen [0]{}%
\providecommand \bibitemStop [0]{}%
\providecommand \bibitemNoStop [0]{.\EOS\space}%
\providecommand \EOS [0]{\spacefactor3000\relax}%
\providecommand \BibitemShut  [1]{\csname bibitem#1\endcsname}%
\let\auto@bib@innerbib\@empty
\bibitem [{\citenamefont {Pearson}(1905)}]{pearson1905problem}%
  \BibitemOpen
  \bibfield  {author} {\bibinfo {author} {\bibfnamefont {K.}~\bibnamefont
  {Pearson}},\ }\href@noop {} {\bibfield  {journal} {\bibinfo  {journal}
  {Nature}\ }\textbf {\bibinfo {volume} {72}},\ \bibinfo {pages} {342}
  (\bibinfo {year} {1905})}\BibitemShut {NoStop}%
\bibitem [{\citenamefont {Aharonov}\ \emph {et~al.}(1993)\citenamefont
  {Aharonov}, \citenamefont {Davidovich},\ and\ \citenamefont
  {Zagury}}]{aharonov1993quantum}%
  \BibitemOpen
  \bibfield  {author} {\bibinfo {author} {\bibfnamefont {Y.}~\bibnamefont
  {Aharonov}}, \bibinfo {author} {\bibfnamefont {L.}~\bibnamefont
  {Davidovich}}, \ and\ \bibinfo {author} {\bibfnamefont {N.}~\bibnamefont
  {Zagury}},\ }\href@noop {} {\bibfield  {journal} {\bibinfo  {journal} {Phys.
  Rev. A}\ }\textbf {\bibinfo {volume} {48}},\ \bibinfo {pages} {1687}
  (\bibinfo {year} {1993})}\BibitemShut {NoStop}%
\bibitem [{\citenamefont {Childs}(2009)}]{childs2009universal}%
  \BibitemOpen
  \bibfield  {author} {\bibinfo {author} {\bibfnamefont {A.~M.}\ \bibnamefont
  {Childs}},\ }\href@noop {} {\bibfield  {journal} {\bibinfo  {journal} {Phys.
  Rev. Lett.}\ }\textbf {\bibinfo {volume} {102}},\ \bibinfo {pages} {180501}
  (\bibinfo {year} {2009})}\BibitemShut {NoStop}%
\bibitem [{\citenamefont {Childs}\ \emph {et~al.}(2013)\citenamefont {Childs},
  \citenamefont {Gosset},\ and\ \citenamefont {Webb}}]{childs2013universal}%
  \BibitemOpen
  \bibfield  {author} {\bibinfo {author} {\bibfnamefont {A.~M.}\ \bibnamefont
  {Childs}}, \bibinfo {author} {\bibfnamefont {D.}~\bibnamefont {Gosset}}, \
  and\ \bibinfo {author} {\bibfnamefont {Z.}~\bibnamefont {Webb}},\ }\href@noop
  {} {\bibfield  {journal} {\bibinfo  {journal} {Science}\ }\textbf {\bibinfo
  {volume} {339}},\ \bibinfo {pages} {791} (\bibinfo {year}
  {2013})}\BibitemShut {NoStop}%
\bibitem [{\citenamefont {Rudner}\ and\ \citenamefont
  {Levitov}(2009)}]{rudner2009topological}%
  \BibitemOpen
  \bibfield  {author} {\bibinfo {author} {\bibfnamefont {M.~S.}\ \bibnamefont
  {Rudner}}\ and\ \bibinfo {author} {\bibfnamefont {L.}~\bibnamefont
  {Levitov}},\ }\href@noop {} {\bibfield  {journal} {\bibinfo  {journal} {Phys.
  Rev. Lett.}\ }\textbf {\bibinfo {volume} {102}},\ \bibinfo {pages} {065703}
  (\bibinfo {year} {2009})}\BibitemShut {NoStop}%
\bibitem [{\citenamefont {Weidemann}\ \emph {et~al.}(2022)\citenamefont
  {Weidemann}, \citenamefont {Kremer}, \citenamefont {Longhi},\ and\
  \citenamefont {Szameit}}]{weidemann2022topological}%
  \BibitemOpen
  \bibfield  {author} {\bibinfo {author} {\bibfnamefont {S.}~\bibnamefont
  {Weidemann}}, \bibinfo {author} {\bibfnamefont {M.}~\bibnamefont {Kremer}},
  \bibinfo {author} {\bibfnamefont {S.}~\bibnamefont {Longhi}}, \ and\ \bibinfo
  {author} {\bibfnamefont {A.}~\bibnamefont {Szameit}},\ }\href@noop {}
  {\bibfield  {journal} {\bibinfo  {journal} {Nature}\ }\textbf {\bibinfo
  {volume} {601}},\ \bibinfo {pages} {354} (\bibinfo {year}
  {2022})}\BibitemShut {NoStop}%
\bibitem [{\citenamefont {Shenvi}\ \emph {et~al.}(2003)\citenamefont {Shenvi},
  \citenamefont {Kempe},\ and\ \citenamefont {Whaley}}]{shenvi2003quantum}%
  \BibitemOpen
  \bibfield  {author} {\bibinfo {author} {\bibfnamefont {N.}~\bibnamefont
  {Shenvi}}, \bibinfo {author} {\bibfnamefont {J.}~\bibnamefont {Kempe}}, \
  and\ \bibinfo {author} {\bibfnamefont {K.~B.}\ \bibnamefont {Whaley}},\
  }\href@noop {} {\bibfield  {journal} {\bibinfo  {journal} {Phys. Rev. A}\
  }\textbf {\bibinfo {volume} {67}},\ \bibinfo {pages} {052307} (\bibinfo
  {year} {2003})}\BibitemShut {NoStop}%
\bibitem [{\citenamefont {Poto{\v{c}}ek}\ \emph {et~al.}(2009)\citenamefont
  {Poto{\v{c}}ek}, \citenamefont {G{\'a}bris}, \citenamefont {Kiss},\ and\
  \citenamefont {Jex}}]{potovcek2009optimized}%
  \BibitemOpen
  \bibfield  {author} {\bibinfo {author} {\bibfnamefont {V.}~\bibnamefont
  {Poto{\v{c}}ek}}, \bibinfo {author} {\bibfnamefont {A.}~\bibnamefont
  {G{\'a}bris}}, \bibinfo {author} {\bibfnamefont {T.}~\bibnamefont {Kiss}}, \
  and\ \bibinfo {author} {\bibfnamefont {I.}~\bibnamefont {Jex}},\ }\href@noop
  {} {\bibfield  {journal} {\bibinfo  {journal} {Phys. Rev. A}\ }\textbf
  {\bibinfo {volume} {79}},\ \bibinfo {pages} {012325} (\bibinfo {year}
  {2009})}\BibitemShut {NoStop}%
\bibitem [{\citenamefont {Qu}\ \emph {et~al.}(2022)\citenamefont {Qu},
  \citenamefont {Marsh}, \citenamefont {Wang}, \citenamefont {Xiao},
  \citenamefont {Wang},\ and\ \citenamefont {Xue}}]{qu2022deterministic}%
  \BibitemOpen
  \bibfield  {author} {\bibinfo {author} {\bibfnamefont {D.}~\bibnamefont
  {Qu}}, \bibinfo {author} {\bibfnamefont {S.}~\bibnamefont {Marsh}}, \bibinfo
  {author} {\bibfnamefont {K.}~\bibnamefont {Wang}}, \bibinfo {author}
  {\bibfnamefont {L.}~\bibnamefont {Xiao}}, \bibinfo {author} {\bibfnamefont
  {J.}~\bibnamefont {Wang}}, \ and\ \bibinfo {author} {\bibfnamefont
  {P.}~\bibnamefont {Xue}},\ }\href@noop {} {\bibfield  {journal} {\bibinfo
  {journal} {Phys. Rev. Lett.}\ }\textbf {\bibinfo {volume} {128}},\ \bibinfo
  {pages} {050501} (\bibinfo {year} {2022})}\BibitemShut {NoStop}%
\bibitem [{\citenamefont {Qiang}\ \emph {et~al.}(2024)\citenamefont {Qiang},
  \citenamefont {Ma},\ and\ \citenamefont {Song}}]{qiang2024quantum}%
  \BibitemOpen
  \bibfield  {author} {\bibinfo {author} {\bibfnamefont {X.}~\bibnamefont
  {Qiang}}, \bibinfo {author} {\bibfnamefont {S.}~\bibnamefont {Ma}}, \ and\
  \bibinfo {author} {\bibfnamefont {H.}~\bibnamefont {Song}},\ }\href@noop {}
  {\bibfield  {journal} {\bibinfo  {journal} {Intell. Comput.}\ }\textbf
  {\bibinfo {volume} {3}},\ \bibinfo {pages} {0097} (\bibinfo {year}
  {2024})}\BibitemShut {NoStop}%
\bibitem [{\citenamefont {Carneiro}\ \emph {et~al.}(2005)\citenamefont
  {Carneiro}, \citenamefont {Loo}, \citenamefont {Xu}, \citenamefont {Girerd},
  \citenamefont {Kendon},\ and\ \citenamefont
  {Knight}}]{carneiro2005entanglement}%
  \BibitemOpen
  \bibfield  {author} {\bibinfo {author} {\bibfnamefont {I.}~\bibnamefont
  {Carneiro}}, \bibinfo {author} {\bibfnamefont {M.}~\bibnamefont {Loo}},
  \bibinfo {author} {\bibfnamefont {X.}~\bibnamefont {Xu}}, \bibinfo {author}
  {\bibfnamefont {M.}~\bibnamefont {Girerd}}, \bibinfo {author} {\bibfnamefont
  {V.}~\bibnamefont {Kendon}}, \ and\ \bibinfo {author} {\bibfnamefont {P.~L.}\
  \bibnamefont {Knight}},\ }\href@noop {} {\bibfield  {journal} {\bibinfo
  {journal} {New J. Phys.}\ }\textbf {\bibinfo {volume} {7}},\ \bibinfo {pages}
  {156} (\bibinfo {year} {2005})}\BibitemShut {NoStop}%
\bibitem [{\citenamefont {Abal}\ \emph {et~al.}(2006)\citenamefont {Abal},
  \citenamefont {Siri}, \citenamefont {Romanelli},\ and\ \citenamefont
  {Donangelo}}]{abal2006quantum}%
  \BibitemOpen
  \bibfield  {author} {\bibinfo {author} {\bibfnamefont {G.}~\bibnamefont
  {Abal}}, \bibinfo {author} {\bibfnamefont {R.}~\bibnamefont {Siri}}, \bibinfo
  {author} {\bibfnamefont {A.}~\bibnamefont {Romanelli}}, \ and\ \bibinfo
  {author} {\bibfnamefont {R.}~\bibnamefont {Donangelo}},\ }\href@noop {}
  {\bibfield  {journal} {\bibinfo  {journal} {Phys. Rev. A}\ }\textbf {\bibinfo
  {volume} {73}},\ \bibinfo {pages} {042302} (\bibinfo {year}
  {2006})}\BibitemShut {NoStop}%
\bibitem [{\citenamefont {Tao}\ \emph {et~al.}(2021)\citenamefont {Tao},
  \citenamefont {Wang}, \citenamefont {Chen}, \citenamefont {Pan},
  \citenamefont {Yu}, \citenamefont {Chen}, \citenamefont {Xu}, \citenamefont
  {Han}, \citenamefont {Li},\ and\ \citenamefont {Guo}}]{tao2021experimental}%
  \BibitemOpen
  \bibfield  {author} {\bibinfo {author} {\bibfnamefont {S.~J.}\ \bibnamefont
  {Tao}}, \bibinfo {author} {\bibfnamefont {Q.~Q.}\ \bibnamefont {Wang}},
  \bibinfo {author} {\bibfnamefont {Z.}~\bibnamefont {Chen}}, \bibinfo {author}
  {\bibfnamefont {W.~W.}\ \bibnamefont {Pan}}, \bibinfo {author} {\bibfnamefont
  {S.}~\bibnamefont {Yu}}, \bibinfo {author} {\bibfnamefont {G.}~\bibnamefont
  {Chen}}, \bibinfo {author} {\bibfnamefont {X.~Y.}\ \bibnamefont {Xu}},
  \bibinfo {author} {\bibfnamefont {Y.~J.}\ \bibnamefont {Han}}, \bibinfo
  {author} {\bibfnamefont {C.~F.}\ \bibnamefont {Li}}, \ and\ \bibinfo {author}
  {\bibfnamefont {G.~C.}\ \bibnamefont {Guo}},\ }\href@noop {} {\bibfield
  {journal} {\bibinfo  {journal} {Opt. Lett.}\ }\textbf {\bibinfo {volume}
  {46}},\ \bibinfo {pages} {1868} (\bibinfo {year} {2021})}\BibitemShut
  {NoStop}%
\bibitem [{\citenamefont {Zhang}\ \emph {et~al.}(2022)\citenamefont {Zhang},
  \citenamefont {Yang}, \citenamefont {Guo}, \citenamefont {Sun}, \citenamefont
  {Duan}, \citenamefont {Zhou}, \citenamefont {Xie}, \citenamefont {Xu},
  \citenamefont {Gong},\ and\ \citenamefont {Zhu}}]{zhang2022maximal}%
  \BibitemOpen
  \bibfield  {author} {\bibinfo {author} {\bibfnamefont {R.}~\bibnamefont
  {Zhang}}, \bibinfo {author} {\bibfnamefont {R.}~\bibnamefont {Yang}},
  \bibinfo {author} {\bibfnamefont {J.}~\bibnamefont {Guo}}, \bibinfo {author}
  {\bibfnamefont {C.~W.}\ \bibnamefont {Sun}}, \bibinfo {author} {\bibfnamefont
  {J.~C.}\ \bibnamefont {Duan}}, \bibinfo {author} {\bibfnamefont
  {H.}~\bibnamefont {Zhou}}, \bibinfo {author} {\bibfnamefont {Z.}~\bibnamefont
  {Xie}}, \bibinfo {author} {\bibfnamefont {P.}~\bibnamefont {Xu}}, \bibinfo
  {author} {\bibfnamefont {Y.~X.}\ \bibnamefont {Gong}}, \ and\ \bibinfo
  {author} {\bibfnamefont {S.~N.}\ \bibnamefont {Zhu}},\ }\href@noop {}
  {\bibfield  {journal} {\bibinfo  {journal} {Phys. Rev. A}\ }\textbf {\bibinfo
  {volume} {105}},\ \bibinfo {pages} {042216} (\bibinfo {year}
  {2022})}\BibitemShut {NoStop}%
\bibitem [{\citenamefont {Wang}\ \emph {et~al.}(2018)\citenamefont {Wang},
  \citenamefont {Xu}, \citenamefont {Pan}, \citenamefont {Sun}, \citenamefont
  {Xu}, \citenamefont {Chen}, \citenamefont {Han}, \citenamefont {Li},\ and\
  \citenamefont {Guo}}]{wang2018dynamic}%
  \BibitemOpen
  \bibfield  {author} {\bibinfo {author} {\bibfnamefont {Q.~Q.}\ \bibnamefont
  {Wang}}, \bibinfo {author} {\bibfnamefont {X.~Y.}\ \bibnamefont {Xu}},
  \bibinfo {author} {\bibfnamefont {W.~W.}\ \bibnamefont {Pan}}, \bibinfo
  {author} {\bibfnamefont {K.}~\bibnamefont {Sun}}, \bibinfo {author}
  {\bibfnamefont {J.~S.}\ \bibnamefont {Xu}}, \bibinfo {author} {\bibfnamefont
  {G.}~\bibnamefont {Chen}}, \bibinfo {author} {\bibfnamefont {Y.~J.}\
  \bibnamefont {Han}}, \bibinfo {author} {\bibfnamefont {C.~F.}\ \bibnamefont
  {Li}}, \ and\ \bibinfo {author} {\bibfnamefont {G.~C.}\ \bibnamefont {Guo}},\
  }\href@noop {} {\bibfield  {journal} {\bibinfo  {journal} {Optica}\ }\textbf
  {\bibinfo {volume} {5}},\ \bibinfo {pages} {1136} (\bibinfo {year}
  {2018})}\BibitemShut {NoStop}%
\bibitem [{\citenamefont {Naves}\ \emph {et~al.}(2022)\citenamefont {Naves},
  \citenamefont {Pires}, \citenamefont {Soares~Pinto},\ and\ \citenamefont
  {Queir{\'o}s}}]{naves2022enhancing}%
  \BibitemOpen
  \bibfield  {author} {\bibinfo {author} {\bibfnamefont {C.~B.}\ \bibnamefont
  {Naves}}, \bibinfo {author} {\bibfnamefont {M.~A.}\ \bibnamefont {Pires}},
  \bibinfo {author} {\bibfnamefont {D.~O.}\ \bibnamefont {Soares~Pinto}}, \
  and\ \bibinfo {author} {\bibfnamefont {S.~M.~D.}\ \bibnamefont
  {Queir{\'o}s}},\ }\href@noop {} {\bibfield  {journal} {\bibinfo  {journal}
  {Phys. Rev. A}\ }\textbf {\bibinfo {volume} {106}},\ \bibinfo {pages}
  {042408} (\bibinfo {year} {2022})}\BibitemShut {NoStop}%
\bibitem [{\citenamefont {Fang}\ \emph {et~al.}(2023)\citenamefont {Fang},
  \citenamefont {An}, \citenamefont {Zhang}, \citenamefont {Sanders},\ and\
  \citenamefont {Lu}}]{fang2023maximal}%
  \BibitemOpen
  \bibfield  {author} {\bibinfo {author} {\bibfnamefont {X.~X.}\ \bibnamefont
  {Fang}}, \bibinfo {author} {\bibfnamefont {K.}~\bibnamefont {An}}, \bibinfo
  {author} {\bibfnamefont {B.~T.}\ \bibnamefont {Zhang}}, \bibinfo {author}
  {\bibfnamefont {B.~C.}\ \bibnamefont {Sanders}}, \ and\ \bibinfo {author}
  {\bibfnamefont {H.}~\bibnamefont {Lu}},\ }\href@noop {} {\bibfield  {journal}
  {\bibinfo  {journal} {Phys. Rev. A}\ }\textbf {\bibinfo {volume} {107}},\
  \bibinfo {pages} {012433} (\bibinfo {year} {2023})}\BibitemShut {NoStop}%
\bibitem [{\citenamefont {Xiao}\ \emph {et~al.}(2021)\citenamefont {Xiao},
  \citenamefont {Deng}, \citenamefont {Wang}, \citenamefont {Wang},
  \citenamefont {Yi},\ and\ \citenamefont {Xue}}]{xiao2021observation}%
  \BibitemOpen
  \bibfield  {author} {\bibinfo {author} {\bibfnamefont {L.}~\bibnamefont
  {Xiao}}, \bibinfo {author} {\bibfnamefont {T.}~\bibnamefont {Deng}}, \bibinfo
  {author} {\bibfnamefont {K.}~\bibnamefont {Wang}}, \bibinfo {author}
  {\bibfnamefont {Z.}~\bibnamefont {Wang}}, \bibinfo {author} {\bibfnamefont
  {W.}~\bibnamefont {Yi}}, \ and\ \bibinfo {author} {\bibfnamefont
  {P.}~\bibnamefont {Xue}},\ }\href@noop {} {\bibfield  {journal} {\bibinfo
  {journal} {Phys. Rev. Lett.}\ }\textbf {\bibinfo {volume} {126}},\ \bibinfo
  {pages} {230402} (\bibinfo {year} {2021})}\BibitemShut {NoStop}%
\bibitem [{\citenamefont {Xiao}\ \emph {et~al.}(2017)\citenamefont {Xiao},
  \citenamefont {Zhan}, \citenamefont {Bian}, \citenamefont {Wang},
  \citenamefont {Zhang}, \citenamefont {Wang}, \citenamefont {Li},
  \citenamefont {Mochizuki}, \citenamefont {Kim}, \citenamefont {Kawakami}
  \emph {et~al.}}]{xiao2017observation}%
  \BibitemOpen
  \bibfield  {author} {\bibinfo {author} {\bibfnamefont {L.}~\bibnamefont
  {Xiao}}, \bibinfo {author} {\bibfnamefont {X.}~\bibnamefont {Zhan}}, \bibinfo
  {author} {\bibfnamefont {Z.}~\bibnamefont {Bian}}, \bibinfo {author}
  {\bibfnamefont {K.}~\bibnamefont {Wang}}, \bibinfo {author} {\bibfnamefont
  {X.}~\bibnamefont {Zhang}}, \bibinfo {author} {\bibfnamefont
  {X.}~\bibnamefont {Wang}}, \bibinfo {author} {\bibfnamefont {J.}~\bibnamefont
  {Li}}, \bibinfo {author} {\bibfnamefont {K.}~\bibnamefont {Mochizuki}},
  \bibinfo {author} {\bibfnamefont {D.}~\bibnamefont {Kim}}, \bibinfo {author}
  {\bibfnamefont {N.}~\bibnamefont {Kawakami}},  \emph {et~al.},\ }\href@noop
  {} {\bibfield  {journal} {\bibinfo  {journal} {Nat. Phys.}\ }\textbf
  {\bibinfo {volume} {13}},\ \bibinfo {pages} {1117} (\bibinfo {year}
  {2017})}\BibitemShut {NoStop}%
\bibitem [{\citenamefont {Wang}\ \emph {et~al.}(2019)\citenamefont {Wang},
  \citenamefont {Qiu}, \citenamefont {Xiao}, \citenamefont {Zhan},
  \citenamefont {Bian}, \citenamefont {Yi},\ and\ \citenamefont
  {Xue}}]{wang2019simulating}%
  \BibitemOpen
  \bibfield  {author} {\bibinfo {author} {\bibfnamefont {K.}~\bibnamefont
  {Wang}}, \bibinfo {author} {\bibfnamefont {X.}~\bibnamefont {Qiu}}, \bibinfo
  {author} {\bibfnamefont {L.}~\bibnamefont {Xiao}}, \bibinfo {author}
  {\bibfnamefont {X.}~\bibnamefont {Zhan}}, \bibinfo {author} {\bibfnamefont
  {Z.}~\bibnamefont {Bian}}, \bibinfo {author} {\bibfnamefont {W.}~\bibnamefont
  {Yi}}, \ and\ \bibinfo {author} {\bibfnamefont {P.}~\bibnamefont {Xue}},\
  }\href@noop {} {\bibfield  {journal} {\bibinfo  {journal} {Phys. Rev. Lett.}\
  }\textbf {\bibinfo {volume} {122}},\ \bibinfo {pages} {020501} (\bibinfo
  {year} {2019})}\BibitemShut {NoStop}%
\bibitem [{\citenamefont {Lin}\ \emph {et~al.}(2022{\natexlab{a}})\citenamefont
  {Lin}, \citenamefont {Li}, \citenamefont {Xiao}, \citenamefont {Wang},
  \citenamefont {Yi},\ and\ \citenamefont {Xue}}]{lin2022topological}%
  \BibitemOpen
  \bibfield  {author} {\bibinfo {author} {\bibfnamefont {Q.}~\bibnamefont
  {Lin}}, \bibinfo {author} {\bibfnamefont {T.}~\bibnamefont {Li}}, \bibinfo
  {author} {\bibfnamefont {L.}~\bibnamefont {Xiao}}, \bibinfo {author}
  {\bibfnamefont {K.}~\bibnamefont {Wang}}, \bibinfo {author} {\bibfnamefont
  {W.}~\bibnamefont {Yi}}, \ and\ \bibinfo {author} {\bibfnamefont
  {P.}~\bibnamefont {Xue}},\ }\href@noop {} {\bibfield  {journal} {\bibinfo
  {journal} {Phys. Rev. Lett.}\ }\textbf {\bibinfo {volume} {129}},\ \bibinfo
  {pages} {113601} (\bibinfo {year} {2022}{\natexlab{a}})}\BibitemShut
  {NoStop}%
\bibitem [{\citenamefont {Longhi}(2019)}]{longhi2019probing}%
  \BibitemOpen
  \bibfield  {author} {\bibinfo {author} {\bibfnamefont {S.}~\bibnamefont
  {Longhi}},\ }\href@noop {} {\bibfield  {journal} {\bibinfo  {journal} {Phys.
  Rev. Res.}\ }\textbf {\bibinfo {volume} {1}},\ \bibinfo {pages} {023013}
  (\bibinfo {year} {2019})}\BibitemShut {NoStop}%
\bibitem [{\citenamefont {Xiao}\ \emph {et~al.}(2020)\citenamefont {Xiao},
  \citenamefont {Deng}, \citenamefont {Wang}, \citenamefont {Zhu},
  \citenamefont {Wang}, \citenamefont {Yi},\ and\ \citenamefont
  {Xue}}]{xiao2020non}%
  \BibitemOpen
  \bibfield  {author} {\bibinfo {author} {\bibfnamefont {L.}~\bibnamefont
  {Xiao}}, \bibinfo {author} {\bibfnamefont {T.}~\bibnamefont {Deng}}, \bibinfo
  {author} {\bibfnamefont {K.}~\bibnamefont {Wang}}, \bibinfo {author}
  {\bibfnamefont {G.}~\bibnamefont {Zhu}}, \bibinfo {author} {\bibfnamefont
  {Z.}~\bibnamefont {Wang}}, \bibinfo {author} {\bibfnamefont {W.}~\bibnamefont
  {Yi}}, \ and\ \bibinfo {author} {\bibfnamefont {P.}~\bibnamefont {Xue}},\
  }\href@noop {} {\bibfield  {journal} {\bibinfo  {journal} {Nat. Phys.}\
  }\textbf {\bibinfo {volume} {16}},\ \bibinfo {pages} {761} (\bibinfo {year}
  {2020})}\BibitemShut {NoStop}%
\bibitem [{\citenamefont {Lin}\ \emph {et~al.}(2022{\natexlab{b}})\citenamefont
  {Lin}, \citenamefont {Li}, \citenamefont {Xiao}, \citenamefont {Wang},
  \citenamefont {Yi},\ and\ \citenamefont {Xue}}]{lin2022observation}%
  \BibitemOpen
  \bibfield  {author} {\bibinfo {author} {\bibfnamefont {Q.}~\bibnamefont
  {Lin}}, \bibinfo {author} {\bibfnamefont {T.}~\bibnamefont {Li}}, \bibinfo
  {author} {\bibfnamefont {L.}~\bibnamefont {Xiao}}, \bibinfo {author}
  {\bibfnamefont {K.}~\bibnamefont {Wang}}, \bibinfo {author} {\bibfnamefont
  {W.}~\bibnamefont {Yi}}, \ and\ \bibinfo {author} {\bibfnamefont
  {P.}~\bibnamefont {Xue}},\ }\href@noop {} {\bibfield  {journal} {\bibinfo
  {journal} {Nat. Commun.}\ }\textbf {\bibinfo {volume} {13}},\ \bibinfo
  {pages} {3229} (\bibinfo {year} {2022}{\natexlab{b}})}\BibitemShut {NoStop}%
\bibitem [{\citenamefont {Kawabata}\ \emph {et~al.}(2023)\citenamefont
  {Kawabata}, \citenamefont {Numasawa},\ and\ \citenamefont
  {Ryu}}]{kawabata2023entanglement}%
  \BibitemOpen
  \bibfield  {author} {\bibinfo {author} {\bibfnamefont {K.}~\bibnamefont
  {Kawabata}}, \bibinfo {author} {\bibfnamefont {T.}~\bibnamefont {Numasawa}},
  \ and\ \bibinfo {author} {\bibfnamefont {S.}~\bibnamefont {Ryu}},\
  }\href@noop {} {\bibfield  {journal} {\bibinfo  {journal} {Phys. Rev. X}\
  }\textbf {\bibinfo {volume} {13}},\ \bibinfo {pages} {021007} (\bibinfo
  {year} {2023})}\BibitemShut {NoStop}%
\bibitem [{\citenamefont {Weidemann}\ \emph {et~al.}(2020)\citenamefont
  {Weidemann}, \citenamefont {Kremer}, \citenamefont {Helbig}, \citenamefont
  {Hofmann}, \citenamefont {Stegmaier}, \citenamefont {Greiter}, \citenamefont
  {Thomale},\ and\ \citenamefont {Szameit}}]{weidemann2020topological}%
  \BibitemOpen
  \bibfield  {author} {\bibinfo {author} {\bibfnamefont {S.}~\bibnamefont
  {Weidemann}}, \bibinfo {author} {\bibfnamefont {M.}~\bibnamefont {Kremer}},
  \bibinfo {author} {\bibfnamefont {T.}~\bibnamefont {Helbig}}, \bibinfo
  {author} {\bibfnamefont {T.}~\bibnamefont {Hofmann}}, \bibinfo {author}
  {\bibfnamefont {A.}~\bibnamefont {Stegmaier}}, \bibinfo {author}
  {\bibfnamefont {M.}~\bibnamefont {Greiter}}, \bibinfo {author} {\bibfnamefont
  {R.}~\bibnamefont {Thomale}}, \ and\ \bibinfo {author} {\bibfnamefont
  {A.}~\bibnamefont {Szameit}},\ }\href@noop {} {\bibfield  {journal} {\bibinfo
   {journal} {Science}\ }\textbf {\bibinfo {volume} {368}},\ \bibinfo {pages}
  {311} (\bibinfo {year} {2020})}\BibitemShut {NoStop}%
\bibitem [{\citenamefont {McDonald}\ and\ \citenamefont
  {Clerk}(2020)}]{mcdonald2020exponentially}%
  \BibitemOpen
  \bibfield  {author} {\bibinfo {author} {\bibfnamefont {A.}~\bibnamefont
  {McDonald}}\ and\ \bibinfo {author} {\bibfnamefont {A.~A.}\ \bibnamefont
  {Clerk}},\ }\href@noop {} {\bibfield  {journal} {\bibinfo  {journal} {Nat.
  Commun.}\ }\textbf {\bibinfo {volume} {11}},\ \bibinfo {pages} {5382}
  (\bibinfo {year} {2020})}\BibitemShut {NoStop}%
\bibitem [{\citenamefont {Wanjura}\ \emph {et~al.}(2020)\citenamefont
  {Wanjura}, \citenamefont {Brunelli},\ and\ \citenamefont
  {Nunnenkamp}}]{wanjura2020topological}%
  \BibitemOpen
  \bibfield  {author} {\bibinfo {author} {\bibfnamefont {C.~C.}\ \bibnamefont
  {Wanjura}}, \bibinfo {author} {\bibfnamefont {M.}~\bibnamefont {Brunelli}}, \
  and\ \bibinfo {author} {\bibfnamefont {A.}~\bibnamefont {Nunnenkamp}},\
  }\href@noop {} {\bibfield  {journal} {\bibinfo  {journal} {Nat. Commun.}\
  }\textbf {\bibinfo {volume} {11}},\ \bibinfo {pages} {3149} (\bibinfo {year}
  {2020})}\BibitemShut {NoStop}%
\bibitem [{\citenamefont {Chen}\ \emph {et~al.}(2022)\citenamefont {Chen},
  \citenamefont {Zhou}, \citenamefont {Chen},\ and\ \citenamefont
  {Ye}}]{chen2022quantum}%
  \BibitemOpen
  \bibfield  {author} {\bibinfo {author} {\bibfnamefont {L.~M.}\ \bibnamefont
  {Chen}}, \bibinfo {author} {\bibfnamefont {Y.}~\bibnamefont {Zhou}}, \bibinfo
  {author} {\bibfnamefont {S.~A.}\ \bibnamefont {Chen}}, \ and\ \bibinfo
  {author} {\bibfnamefont {P.}~\bibnamefont {Ye}},\ }\href@noop {} {\bibfield
  {journal} {\bibinfo  {journal} {Phys. Rev. B}\ }\textbf {\bibinfo {volume}
  {105}},\ \bibinfo {pages} {L121115} (\bibinfo {year} {2022})}\BibitemShut
  {NoStop}%
\bibitem [{\citenamefont {Gao}\ \emph {et~al.}(2024)\citenamefont {Gao},
  \citenamefont {Sheng}, \citenamefont {Zhao}, \citenamefont {He},
  \citenamefont {Lu}, \citenamefont {Chen}, \citenamefont {Ding}, \citenamefont
  {Zhu},\ and\ \citenamefont {Liu}}]{gao2024quantum}%
  \BibitemOpen
  \bibfield  {author} {\bibinfo {author} {\bibfnamefont {M.}~\bibnamefont
  {Gao}}, \bibinfo {author} {\bibfnamefont {C.}~\bibnamefont {Sheng}}, \bibinfo
  {author} {\bibfnamefont {Y.}~\bibnamefont {Zhao}}, \bibinfo {author}
  {\bibfnamefont {R.}~\bibnamefont {He}}, \bibinfo {author} {\bibfnamefont
  {L.}~\bibnamefont {Lu}}, \bibinfo {author} {\bibfnamefont {W.}~\bibnamefont
  {Chen}}, \bibinfo {author} {\bibfnamefont {K.}~\bibnamefont {Ding}}, \bibinfo
  {author} {\bibfnamefont {S.}~\bibnamefont {Zhu}}, \ and\ \bibinfo {author}
  {\bibfnamefont {H.}~\bibnamefont {Liu}},\ }\href@noop {} {\bibfield
  {journal} {\bibinfo  {journal} {Phys. Rev. B}\ }\textbf {\bibinfo {volume}
  {110}},\ \bibinfo {pages} {094308} (\bibinfo {year} {2024})}\BibitemShut
  {NoStop}%
\bibitem [{\citenamefont {Flurin}\ \emph {et~al.}(2017)\citenamefont {Flurin},
  \citenamefont {Ramasesh}, \citenamefont {Hacohen~Gourgy}, \citenamefont
  {Martin}, \citenamefont {Yao},\ and\ \citenamefont
  {Siddiqi}}]{flurin2017observing}%
  \BibitemOpen
  \bibfield  {author} {\bibinfo {author} {\bibfnamefont {E.}~\bibnamefont
  {Flurin}}, \bibinfo {author} {\bibfnamefont {V.~V.}\ \bibnamefont
  {Ramasesh}}, \bibinfo {author} {\bibfnamefont {S.}~\bibnamefont
  {Hacohen~Gourgy}}, \bibinfo {author} {\bibfnamefont {L.~S.}\ \bibnamefont
  {Martin}}, \bibinfo {author} {\bibfnamefont {N.~Y.}\ \bibnamefont {Yao}}, \
  and\ \bibinfo {author} {\bibfnamefont {I.}~\bibnamefont {Siddiqi}},\
  }\href@noop {} {\bibfield  {journal} {\bibinfo  {journal} {Phys. Rev. X}\
  }\textbf {\bibinfo {volume} {7}},\ \bibinfo {pages} {031023} (\bibinfo {year}
  {2017})}\BibitemShut {NoStop}%
\bibitem [{\citenamefont {Ramasesh}\ \emph {et~al.}(2017)\citenamefont
  {Ramasesh}, \citenamefont {Flurin}, \citenamefont {Rudner}, \citenamefont
  {Siddiqi},\ and\ \citenamefont {Yao}}]{ramasesh2017direct}%
  \BibitemOpen
  \bibfield  {author} {\bibinfo {author} {\bibfnamefont {V.~V.}\ \bibnamefont
  {Ramasesh}}, \bibinfo {author} {\bibfnamefont {E.}~\bibnamefont {Flurin}},
  \bibinfo {author} {\bibfnamefont {M.}~\bibnamefont {Rudner}}, \bibinfo
  {author} {\bibfnamefont {I.}~\bibnamefont {Siddiqi}}, \ and\ \bibinfo
  {author} {\bibfnamefont {N.~Y.}\ \bibnamefont {Yao}},\ }\href@noop {}
  {\bibfield  {journal} {\bibinfo  {journal} {Phys. Rev. Lett.}\ }\textbf
  {\bibinfo {volume} {118}},\ \bibinfo {pages} {130501} (\bibinfo {year}
  {2017})}\BibitemShut {NoStop}%
\bibitem [{\citenamefont {Ryan}\ \emph {et~al.}(2005)\citenamefont {Ryan},
  \citenamefont {Laforest}, \citenamefont {Boileau},\ and\ \citenamefont
  {Laflamme}}]{ryan2005experimental}%
  \BibitemOpen
  \bibfield  {author} {\bibinfo {author} {\bibfnamefont {C.~A.}\ \bibnamefont
  {Ryan}}, \bibinfo {author} {\bibfnamefont {M.}~\bibnamefont {Laforest}},
  \bibinfo {author} {\bibfnamefont {J.~C.}\ \bibnamefont {Boileau}}, \ and\
  \bibinfo {author} {\bibfnamefont {R.}~\bibnamefont {Laflamme}},\ }\href@noop
  {} {\bibfield  {journal} {\bibinfo  {journal} {Phys. Rev. A}\ }\textbf
  {\bibinfo {volume} {72}},\ \bibinfo {pages} {062317} (\bibinfo {year}
  {2005})}\BibitemShut {NoStop}%
\bibitem [{\citenamefont {Karski}\ \emph {et~al.}(2009)\citenamefont {Karski},
  \citenamefont {F{\"o}rster}, \citenamefont {Choi}, \citenamefont {Steffen},
  \citenamefont {Alt}, \citenamefont {Meschede},\ and\ \citenamefont
  {Widera}}]{karski2009quantum}%
  \BibitemOpen
  \bibfield  {author} {\bibinfo {author} {\bibfnamefont {M.}~\bibnamefont
  {Karski}}, \bibinfo {author} {\bibfnamefont {L.}~\bibnamefont {F{\"o}rster}},
  \bibinfo {author} {\bibfnamefont {J.-M.}\ \bibnamefont {Choi}}, \bibinfo
  {author} {\bibfnamefont {A.}~\bibnamefont {Steffen}}, \bibinfo {author}
  {\bibfnamefont {W.}~\bibnamefont {Alt}}, \bibinfo {author} {\bibfnamefont
  {D.}~\bibnamefont {Meschede}}, \ and\ \bibinfo {author} {\bibfnamefont
  {A.}~\bibnamefont {Widera}},\ }\href@noop {} {\bibfield  {journal} {\bibinfo
  {journal} {Science}\ }\textbf {\bibinfo {volume} {325}},\ \bibinfo {pages}
  {174} (\bibinfo {year} {2009})}\BibitemShut {NoStop}%
\bibitem [{\citenamefont {Schmitz}\ \emph {et~al.}(2009)\citenamefont
  {Schmitz}, \citenamefont {Matjeschk}, \citenamefont {Schneider},
  \citenamefont {Glueckert}, \citenamefont {Enderlein}, \citenamefont {Huber},\
  and\ \citenamefont {Schaetz}}]{schmitz2009quantum}%
  \BibitemOpen
  \bibfield  {author} {\bibinfo {author} {\bibfnamefont {H.}~\bibnamefont
  {Schmitz}}, \bibinfo {author} {\bibfnamefont {R.}~\bibnamefont {Matjeschk}},
  \bibinfo {author} {\bibfnamefont {C.}~\bibnamefont {Schneider}}, \bibinfo
  {author} {\bibfnamefont {J.}~\bibnamefont {Glueckert}}, \bibinfo {author}
  {\bibfnamefont {M.}~\bibnamefont {Enderlein}}, \bibinfo {author}
  {\bibfnamefont {T.}~\bibnamefont {Huber}}, \ and\ \bibinfo {author}
  {\bibfnamefont {T.}~\bibnamefont {Schaetz}},\ }\href@noop {} {\bibfield
  {journal} {\bibinfo  {journal} {Phys. Rev. Lett.}\ }\textbf {\bibinfo
  {volume} {103}},\ \bibinfo {pages} {090504} (\bibinfo {year}
  {2009})}\BibitemShut {NoStop}%
\bibitem [{\citenamefont {Z{\"a}hringer}\ \emph {et~al.}(2010)\citenamefont
  {Z{\"a}hringer}, \citenamefont {Kirchmair}, \citenamefont {Gerritsma},
  \citenamefont {Solano}, \citenamefont {Blatt},\ and\ \citenamefont
  {Roos}}]{zahringer2010realization}%
  \BibitemOpen
  \bibfield  {author} {\bibinfo {author} {\bibfnamefont {F.}~\bibnamefont
  {Z{\"a}hringer}}, \bibinfo {author} {\bibfnamefont {G.}~\bibnamefont
  {Kirchmair}}, \bibinfo {author} {\bibfnamefont {R.}~\bibnamefont
  {Gerritsma}}, \bibinfo {author} {\bibfnamefont {E.}~\bibnamefont {Solano}},
  \bibinfo {author} {\bibfnamefont {R.}~\bibnamefont {Blatt}}, \ and\ \bibinfo
  {author} {\bibfnamefont {C.~F.}\ \bibnamefont {Roos}},\ }\href@noop {}
  {\bibfield  {journal} {\bibinfo  {journal} {Phys. Rev. Lett.}\ }\textbf
  {\bibinfo {volume} {104}},\ \bibinfo {pages} {100503} (\bibinfo {year}
  {2010})}\BibitemShut {NoStop}%
\bibitem [{\citenamefont {Sansoni}\ \emph {et~al.}(2012)\citenamefont
  {Sansoni}, \citenamefont {Sciarrino}, \citenamefont {Vallone}, \citenamefont
  {Mataloni}, \citenamefont {Crespi}, \citenamefont {Ramponi},\ and\
  \citenamefont {Osellame}}]{sansoni2012two}%
  \BibitemOpen
  \bibfield  {author} {\bibinfo {author} {\bibfnamefont {L.}~\bibnamefont
  {Sansoni}}, \bibinfo {author} {\bibfnamefont {F.}~\bibnamefont {Sciarrino}},
  \bibinfo {author} {\bibfnamefont {G.}~\bibnamefont {Vallone}}, \bibinfo
  {author} {\bibfnamefont {P.}~\bibnamefont {Mataloni}}, \bibinfo {author}
  {\bibfnamefont {A.}~\bibnamefont {Crespi}}, \bibinfo {author} {\bibfnamefont
  {R.}~\bibnamefont {Ramponi}}, \ and\ \bibinfo {author} {\bibfnamefont
  {R.}~\bibnamefont {Osellame}},\ }\href@noop {} {\bibfield  {journal}
  {\bibinfo  {journal} {Phys. Rev. Lett.}\ }\textbf {\bibinfo {volume} {108}},\
  \bibinfo {pages} {010502} (\bibinfo {year} {2012})}\BibitemShut {NoStop}%
\bibitem [{\citenamefont {Crespi}\ \emph {et~al.}(2013)\citenamefont {Crespi},
  \citenamefont {Osellame}, \citenamefont {Ramponi}, \citenamefont
  {Giovannetti}, \citenamefont {Fazio}, \citenamefont {Sansoni}, \citenamefont
  {De~Nicola}, \citenamefont {Sciarrino},\ and\ \citenamefont
  {Mataloni}}]{crespi2013anderson}%
  \BibitemOpen
  \bibfield  {author} {\bibinfo {author} {\bibfnamefont {A.}~\bibnamefont
  {Crespi}}, \bibinfo {author} {\bibfnamefont {R.}~\bibnamefont {Osellame}},
  \bibinfo {author} {\bibfnamefont {R.}~\bibnamefont {Ramponi}}, \bibinfo
  {author} {\bibfnamefont {V.}~\bibnamefont {Giovannetti}}, \bibinfo {author}
  {\bibfnamefont {R.}~\bibnamefont {Fazio}}, \bibinfo {author} {\bibfnamefont
  {L.}~\bibnamefont {Sansoni}}, \bibinfo {author} {\bibfnamefont
  {F.}~\bibnamefont {De~Nicola}}, \bibinfo {author} {\bibfnamefont
  {F.}~\bibnamefont {Sciarrino}}, \ and\ \bibinfo {author} {\bibfnamefont
  {P.}~\bibnamefont {Mataloni}},\ }\href@noop {} {\bibfield  {journal}
  {\bibinfo  {journal} {Nat. Photonics}\ }\textbf {\bibinfo {volume} {7}},\
  \bibinfo {pages} {322} (\bibinfo {year} {2013})}\BibitemShut {NoStop}%
\bibitem [{\citenamefont {Goyal}\ \emph {et~al.}(2013)\citenamefont {Goyal},
  \citenamefont {Roux}, \citenamefont {Forbes},\ and\ \citenamefont
  {Konrad}}]{goyal2013implementing}%
  \BibitemOpen
  \bibfield  {author} {\bibinfo {author} {\bibfnamefont {S.~K.}\ \bibnamefont
  {Goyal}}, \bibinfo {author} {\bibfnamefont {F.~S.}\ \bibnamefont {Roux}},
  \bibinfo {author} {\bibfnamefont {A.}~\bibnamefont {Forbes}}, \ and\ \bibinfo
  {author} {\bibfnamefont {T.}~\bibnamefont {Konrad}},\ }\href@noop {}
  {\bibfield  {journal} {\bibinfo  {journal} {Phys. Rev. Lett.}\ }\textbf
  {\bibinfo {volume} {110}},\ \bibinfo {pages} {263602} (\bibinfo {year}
  {2013})}\BibitemShut {NoStop}%
\bibitem [{\citenamefont {Giordani}\ \emph {et~al.}(2019)\citenamefont
  {Giordani}, \citenamefont {Polino}, \citenamefont {Emiliani}, \citenamefont
  {Suprano}, \citenamefont {Innocenti}, \citenamefont {Majury}, \citenamefont
  {Marrucci}, \citenamefont {Paternostro}, \citenamefont {Ferraro},
  \citenamefont {Spagnolo} \emph {et~al.}}]{giordani2019experimental}%
  \BibitemOpen
  \bibfield  {author} {\bibinfo {author} {\bibfnamefont {T.}~\bibnamefont
  {Giordani}}, \bibinfo {author} {\bibfnamefont {E.}~\bibnamefont {Polino}},
  \bibinfo {author} {\bibfnamefont {S.}~\bibnamefont {Emiliani}}, \bibinfo
  {author} {\bibfnamefont {A.}~\bibnamefont {Suprano}}, \bibinfo {author}
  {\bibfnamefont {L.}~\bibnamefont {Innocenti}}, \bibinfo {author}
  {\bibfnamefont {H.}~\bibnamefont {Majury}}, \bibinfo {author} {\bibfnamefont
  {L.}~\bibnamefont {Marrucci}}, \bibinfo {author} {\bibfnamefont
  {M.}~\bibnamefont {Paternostro}}, \bibinfo {author} {\bibfnamefont
  {A.}~\bibnamefont {Ferraro}}, \bibinfo {author} {\bibfnamefont
  {N.}~\bibnamefont {Spagnolo}},  \emph {et~al.},\ }\href@noop {} {\bibfield
  {journal} {\bibinfo  {journal} {Phys. Rev. Lett.}\ }\textbf {\bibinfo
  {volume} {122}},\ \bibinfo {pages} {020503} (\bibinfo {year}
  {2019})}\BibitemShut {NoStop}%
\bibitem [{\citenamefont {Schreiber}\ \emph {et~al.}(2010)\citenamefont
  {Schreiber}, \citenamefont {Cassemiro}, \citenamefont {Poto{\v{c}}ek},
  \citenamefont {G{\'a}bris}, \citenamefont {Mosley}, \citenamefont
  {Andersson}, \citenamefont {Jex},\ and\ \citenamefont
  {Silberhorn}}]{schreiber2010photons}%
  \BibitemOpen
  \bibfield  {author} {\bibinfo {author} {\bibfnamefont {A.}~\bibnamefont
  {Schreiber}}, \bibinfo {author} {\bibfnamefont {K.~N.}\ \bibnamefont
  {Cassemiro}}, \bibinfo {author} {\bibfnamefont {V.}~\bibnamefont
  {Poto{\v{c}}ek}}, \bibinfo {author} {\bibfnamefont {A.}~\bibnamefont
  {G{\'a}bris}}, \bibinfo {author} {\bibfnamefont {P.~J.}\ \bibnamefont
  {Mosley}}, \bibinfo {author} {\bibfnamefont {E.}~\bibnamefont {Andersson}},
  \bibinfo {author} {\bibfnamefont {I.}~\bibnamefont {Jex}}, \ and\ \bibinfo
  {author} {\bibfnamefont {C.}~\bibnamefont {Silberhorn}},\ }\href@noop {}
  {\bibfield  {journal} {\bibinfo  {journal} {Phys. Rev. Lett.}\ }\textbf
  {\bibinfo {volume} {104}},\ \bibinfo {pages} {050502} (\bibinfo {year}
  {2010})}\BibitemShut {NoStop}%
\bibitem [{\citenamefont {Schreiber}\ \emph {et~al.}(2011)\citenamefont
  {Schreiber}, \citenamefont {Cassemiro}, \citenamefont {Poto{\v{c}}ek},
  \citenamefont {G{\'a}bris}, \citenamefont {Jex},\ and\ \citenamefont
  {Silberhorn}}]{schreiber2011decoherence}%
  \BibitemOpen
  \bibfield  {author} {\bibinfo {author} {\bibfnamefont {A.}~\bibnamefont
  {Schreiber}}, \bibinfo {author} {\bibfnamefont {K.}~\bibnamefont
  {Cassemiro}}, \bibinfo {author} {\bibfnamefont {V.}~\bibnamefont
  {Poto{\v{c}}ek}}, \bibinfo {author} {\bibfnamefont {A.}~\bibnamefont
  {G{\'a}bris}}, \bibinfo {author} {\bibfnamefont {I.}~\bibnamefont {Jex}}, \
  and\ \bibinfo {author} {\bibfnamefont {C.}~\bibnamefont {Silberhorn}},\
  }\href@noop {} {\bibfield  {journal} {\bibinfo  {journal} {Phys. Rev. Lett.}\
  }\textbf {\bibinfo {volume} {106}},\ \bibinfo {pages} {180403} (\bibinfo
  {year} {2011})}\BibitemShut {NoStop}%
\bibitem [{\citenamefont {Schreiber}\ \emph {et~al.}(2012)\citenamefont
  {Schreiber}, \citenamefont {G{\'a}bris}, \citenamefont {Rohde}, \citenamefont
  {Laiho}, \citenamefont {{\v{S}}tefa{\v{n}}{\'a}k}, \citenamefont
  {Poto{\v{c}}ek}, \citenamefont {Hamilton}, \citenamefont {Jex},\ and\
  \citenamefont {Silberhorn}}]{schreiber20122d}%
  \BibitemOpen
  \bibfield  {author} {\bibinfo {author} {\bibfnamefont {A.}~\bibnamefont
  {Schreiber}}, \bibinfo {author} {\bibfnamefont {A.}~\bibnamefont
  {G{\'a}bris}}, \bibinfo {author} {\bibfnamefont {P.~P.}\ \bibnamefont
  {Rohde}}, \bibinfo {author} {\bibfnamefont {K.}~\bibnamefont {Laiho}},
  \bibinfo {author} {\bibfnamefont {M.}~\bibnamefont
  {{\v{S}}tefa{\v{n}}{\'a}k}}, \bibinfo {author} {\bibfnamefont
  {V.}~\bibnamefont {Poto{\v{c}}ek}}, \bibinfo {author} {\bibfnamefont
  {C.}~\bibnamefont {Hamilton}}, \bibinfo {author} {\bibfnamefont
  {I.}~\bibnamefont {Jex}}, \ and\ \bibinfo {author} {\bibfnamefont
  {C.}~\bibnamefont {Silberhorn}},\ }\href@noop {} {\bibfield  {journal}
  {\bibinfo  {journal} {Science}\ }\textbf {\bibinfo {volume} {336}},\ \bibinfo
  {pages} {55} (\bibinfo {year} {2012})}\BibitemShut {NoStop}%
\bibitem [{\citenamefont {Lin}\ \emph {et~al.}(2023)\citenamefont {Lin},
  \citenamefont {Yi},\ and\ \citenamefont {Xue}}]{lin2023manipulating}%
  \BibitemOpen
  \bibfield  {author} {\bibinfo {author} {\bibfnamefont {Q.}~\bibnamefont
  {Lin}}, \bibinfo {author} {\bibfnamefont {W.}~\bibnamefont {Yi}}, \ and\
  \bibinfo {author} {\bibfnamefont {P.}~\bibnamefont {Xue}},\ }\href@noop {}
  {\bibfield  {journal} {\bibinfo  {journal} {Nat. Commun.}\ }\textbf {\bibinfo
  {volume} {14}},\ \bibinfo {pages} {6283} (\bibinfo {year}
  {2023})}\BibitemShut {NoStop}%
\bibitem [{\citenamefont {Xue}\ \emph {et~al.}(2015)\citenamefont {Xue},
  \citenamefont {Zhang}, \citenamefont {Qin}, \citenamefont {Zhan},
  \citenamefont {Bian}, \citenamefont {Li},\ and\ \citenamefont
  {Sanders}}]{xue2015experimental}%
  \BibitemOpen
  \bibfield  {author} {\bibinfo {author} {\bibfnamefont {P.}~\bibnamefont
  {Xue}}, \bibinfo {author} {\bibfnamefont {R.}~\bibnamefont {Zhang}}, \bibinfo
  {author} {\bibfnamefont {H.}~\bibnamefont {Qin}}, \bibinfo {author}
  {\bibfnamefont {X.}~\bibnamefont {Zhan}}, \bibinfo {author} {\bibfnamefont
  {Z.}~\bibnamefont {Bian}}, \bibinfo {author} {\bibfnamefont {J.}~\bibnamefont
  {Li}}, \ and\ \bibinfo {author} {\bibfnamefont {B.~C.}\ \bibnamefont
  {Sanders}},\ }\href@noop {} {\bibfield  {journal} {\bibinfo  {journal} {Phys.
  Rev. Lett.}\ }\textbf {\bibinfo {volume} {114}},\ \bibinfo {pages} {140502}
  (\bibinfo {year} {2015})}\BibitemShut {NoStop}%
\bibitem [{\citenamefont {Zhan}\ \emph {et~al.}(2017)\citenamefont {Zhan},
  \citenamefont {Xiao}, \citenamefont {Bian}, \citenamefont {Wang},
  \citenamefont {Qiu}, \citenamefont {Sanders}, \citenamefont {Yi},\ and\
  \citenamefont {Xue}}]{zhan2017detecting}%
  \BibitemOpen
  \bibfield  {author} {\bibinfo {author} {\bibfnamefont {X.}~\bibnamefont
  {Zhan}}, \bibinfo {author} {\bibfnamefont {L.}~\bibnamefont {Xiao}}, \bibinfo
  {author} {\bibfnamefont {Z.}~\bibnamefont {Bian}}, \bibinfo {author}
  {\bibfnamefont {K.}~\bibnamefont {Wang}}, \bibinfo {author} {\bibfnamefont
  {X.}~\bibnamefont {Qiu}}, \bibinfo {author} {\bibfnamefont {B.~C.}\
  \bibnamefont {Sanders}}, \bibinfo {author} {\bibfnamefont {W.}~\bibnamefont
  {Yi}}, \ and\ \bibinfo {author} {\bibfnamefont {P.}~\bibnamefont {Xue}},\
  }\href@noop {} {\bibfield  {journal} {\bibinfo  {journal} {Phys. Rev. Lett.}\
  }\textbf {\bibinfo {volume} {119}},\ \bibinfo {pages} {130501} (\bibinfo
  {year} {2017})}\BibitemShut {NoStop}%
\bibitem [{\citenamefont {Longhi}(2023)}]{longhi2023phase}%
  \BibitemOpen
  \bibfield  {author} {\bibinfo {author} {\bibfnamefont {S.}~\bibnamefont
  {Longhi}},\ }\href@noop {} {\bibfield  {journal} {\bibinfo  {journal} {Phys.
  Rev. B}\ }\textbf {\bibinfo {volume} {108}},\ \bibinfo {pages} {075121}
  (\bibinfo {year} {2023})}\BibitemShut {NoStop}%
\bibitem [{\citenamefont {Yao}\ and\ \citenamefont {Wang}(2018)}]{yao2018edge}%
  \BibitemOpen
  \bibfield  {author} {\bibinfo {author} {\bibfnamefont {S.}~\bibnamefont
  {Yao}}\ and\ \bibinfo {author} {\bibfnamefont {Z.}~\bibnamefont {Wang}},\
  }\href@noop {} {\bibfield  {journal} {\bibinfo  {journal} {Phys. Rev. Lett.}\
  }\textbf {\bibinfo {volume} {121}},\ \bibinfo {pages} {086803} (\bibinfo
  {year} {2018})}\BibitemShut {NoStop}%
\bibitem [{\citenamefont {Yokomizo}\ and\ \citenamefont
  {Murakami}(2019)}]{yokomizo2019non}%
  \BibitemOpen
  \bibfield  {author} {\bibinfo {author} {\bibfnamefont {K.}~\bibnamefont
  {Yokomizo}}\ and\ \bibinfo {author} {\bibfnamefont {S.}~\bibnamefont
  {Murakami}},\ }\href@noop {} {\bibfield  {journal} {\bibinfo  {journal}
  {Phys. Rev. Lett.}\ }\textbf {\bibinfo {volume} {123}},\ \bibinfo {pages}
  {066404} (\bibinfo {year} {2019})}\BibitemShut {NoStop}%
\bibitem [{\citenamefont {Imura}\ and\ \citenamefont
  {Takane}(2019)}]{imura2019generalized}%
  \BibitemOpen
  \bibfield  {author} {\bibinfo {author} {\bibfnamefont {K.}~\bibnamefont
  {Imura}}\ and\ \bibinfo {author} {\bibfnamefont {Y.}~\bibnamefont {Takane}},\
  }\href@noop {} {\bibfield  {journal} {\bibinfo  {journal} {Phys. Rev. B}\
  }\textbf {\bibinfo {volume} {100}},\ \bibinfo {pages} {165430} (\bibinfo
  {year} {2019})}\BibitemShut {NoStop}%
\bibitem [{\citenamefont {Li}\ \emph {et~al.}(2025)\citenamefont {Li},
  \citenamefont {Zhang}, \citenamefont {Kou}, \citenamefont {Xiao},
  \citenamefont {Jia}, \citenamefont {Li},\ and\ \citenamefont
  {Mei}}]{li2025observation}%
  \BibitemOpen
  \bibfield  {author} {\bibinfo {author} {\bibfnamefont {Y.}~\bibnamefont
  {Li}}, \bibinfo {author} {\bibfnamefont {J.}~\bibnamefont {Zhang}}, \bibinfo
  {author} {\bibfnamefont {Y.}~\bibnamefont {Kou}}, \bibinfo {author}
  {\bibfnamefont {L.}~\bibnamefont {Xiao}}, \bibinfo {author} {\bibfnamefont
  {S.}~\bibnamefont {Jia}}, \bibinfo {author} {\bibfnamefont {L.}~\bibnamefont
  {Li}}, \ and\ \bibinfo {author} {\bibfnamefont {F.}~\bibnamefont {Mei}},\
  }\href@noop {} {\bibfield  {journal} {\bibinfo  {journal} {arXiv:2504.18063}\
  } (\bibinfo {year} {2025})}\BibitemShut {NoStop}%
\bibitem [{\citenamefont {Hatano}\ and\ \citenamefont
  {Nelson}(1996)}]{hatano1996localization}%
  \BibitemOpen
  \bibfield  {author} {\bibinfo {author} {\bibfnamefont {N.}~\bibnamefont
  {Hatano}}\ and\ \bibinfo {author} {\bibfnamefont {D.~R.}\ \bibnamefont
  {Nelson}},\ }\href@noop {} {\bibfield  {journal} {\bibinfo  {journal} {Phys.
  Rev. Lett.}\ }\textbf {\bibinfo {volume} {77}},\ \bibinfo {pages} {570}
  (\bibinfo {year} {1996})}\BibitemShut {NoStop}%
\bibitem [{\citenamefont {Evers}\ and\ \citenamefont
  {Mirlin}(2000)}]{evers2000fluctuations}%
  \BibitemOpen
  \bibfield  {author} {\bibinfo {author} {\bibfnamefont {F.}~\bibnamefont
  {Evers}}\ and\ \bibinfo {author} {\bibfnamefont {A.}~\bibnamefont {Mirlin}},\
  }\href@noop {} {\bibfield  {journal} {\bibinfo  {journal} {Phys. Rev. Lett.}\
  }\textbf {\bibinfo {volume} {84}},\ \bibinfo {pages} {3690} (\bibinfo {year}
  {2000})}\BibitemShut {NoStop}%
\bibitem [{\citenamefont {Buarque}\ and\ \citenamefont
  {Dias}(2019)}]{buarque2019aperiodic}%
  \BibitemOpen
  \bibfield  {author} {\bibinfo {author} {\bibfnamefont {A.}~\bibnamefont
  {Buarque}}\ and\ \bibinfo {author} {\bibfnamefont {W.~d.~S.}\ \bibnamefont
  {Dias}},\ }\href@noop {} {\bibfield  {journal} {\bibinfo  {journal} {Phys.
  Rev. E}\ }\textbf {\bibinfo {volume} {100}},\ \bibinfo {pages} {032106}
  (\bibinfo {year} {2019})}\BibitemShut {NoStop}%
\end{thebibliography}%
\end{document}